\documentclass[12pt]{article}
\usepackage{amsmath,amsfonts,epsf}
\usepackage{amssymb}
\usepackage{graphicx}
\usepackage{grffile}
\input epsf

\makeatletter\@addtoreset{equation}{section}\makeatother

\def\be{\begin{equation}}
\def\ee{\end{equation}}
\def\bea{\begin{eqnarray}}
\def\eea{\end{eqnarray}}
\def\ie{\begin{equation}\begin{aligned}}
\def\fe{\end{aligned}\end{equation}}

\newcommand{\m}{\mu}
\newcommand{\n}{\nu}
\newcommand{\af}{\alpha}
\newcommand{\p}{\nabla}

\newcommand{\A}{{\alpha}}
\newcommand{\B}{{\beta}}
\newcommand{\C}{{\gamma}}
\newcommand{\D}{{\delta}}

\makeatletter\@addtoreset{equation}{section}\makeatother

\newcommand{\vev}[1]{{\left< {#1} \right>}}

\newcommand{\cN}{{\mathcal N}}
\newcommand{\cO}{{\mathcal O}}

\newcommand{\ul}{\underline}
\newcommand{\vp}{{\vec p}}
\newcommand{\vY}{{\vec Y}}
\newcommand{\ty}{{\tilde y}}
\newcommand{\tz}{{\tilde z}}

\renewcommand{\title}[1]{\vbox{\center\LARGE{#1}}\vspace{5mm}}
\renewcommand{\author}[1]{\vbox{\center#1}\vspace{5mm}}
\newcommand{\address}[1]{\vbox{\center\em#1}}
\newcommand{\email}[1]{\vbox{\center\tt#1}\vspace{5mm}}

\begin{document}

\unitlength = .8mm
\begin{titlepage}
\begin{center}
\hfill \\
\hfill \\
\vskip 1cm

\title{Higher Spin Gravity with Matter in $AdS_3$\\ and Its CFT Dual}

\author{Chi-Ming Chang$^{a}$ and
Xi Yin$^{b}$}

\address{
Jefferson Physical Laboratory, Harvard University,\\
Cambridge, MA 02138 USA}

\email{$^a$cmchang@physics.harvard.edu,
$^b$xiyin@fas.harvard.edu}

\end{center}

\abstract{
We study Vasiliev's system of higher spin gauge fields coupled to massive scalars in $AdS_3$, and compute the tree level two and three point functions. These are compared to the large $N$ limit of the $W_N$ minimal model, and nontrivial agreements are found. We propose a modified version of the conjecture of Gaberdiel and Gopakumar, under which the bulk theory is perturbatively dual to a subsector of the CFT that closes on the sphere.
}

\vfill

\end{titlepage}

\eject \tableofcontents

\section{Introduction}

The AdS/CFT correspondence \cite{Maldacena:1997re} has given us a tremendous amount of insight in quantum gravity through its duality with large $N$ gauge theories. Progress does not come easily, however. The regime in which the bulk theory reduces to semi-classical gravity is typically dual to a gauge theory in the strong 't Hooft coupling regime, and is difficult to solve. In the opposite limit, where the gauge theory is weakly coupled, the bulk theory is typically in a very stringy regime, involving strings in AdS whose radius is very small in string units (though large in Planck units, as long as $N$ is large). With a few exceptions, such as the purely NS-NS background of $AdS_3$ \cite{Giveon:1998ns}, in which case the dual CFT is singular \cite{Larsen:1999uk, Maldacena:2000hw}, generally the bulk string theory involves Ramond-Ramond fluxes; even the free string spectrum is difficult to solve, and the full string field theory appears to be out of reach at the moment.

A particularly simple class of conjectured AdS/CFT dualities \cite{Klebanov:2002ja, Sezgin:2002rt, Gaberdiel:2010pz} avoids these difficulties. These involve boundary CFTs whose numbers of degrees of freedom scales like $N$ rather than $N^2$. In the $AdS_4/CFT_3$ conjecture of \cite{Klebanov:2002ja}, the boundary theory is given by the critical $O(N)$ vector model. Such a duality can be extended to Chern-Simons-matter theories with vector matter representations \cite{GMPTY}. In the $AdS_3/CFT_2$ conjecture of \cite{Gaberdiel:2010pz}, the boundary theory is the $W_N$ minimal model, which can be realized as the coset model
\ie\label{coseta}
{SU(N)_k\oplus SU(N)_1\over SU(N)_{k+1}}.
\fe
In these examples, the CFT is either exactly solvable or has a simple $1/N$ expansion that can be  computed straightforwardly order by order. The dual bulk theories, however, are higher spin extensions of gravity, involving an infinite tower\footnote{While a {\it pure} higher spin gauge theory in $AdS_3$ involving spins up to $N$ can be formulated in terms of $SL(N,{\mathbb R})\times SL(N,{\mathbb R})$ Chern-Simons theory, it is not known how to couple this theory to scalar matter fields. The construction of \cite{Vasiliev:1999ba} requires an infinite set of gauge fields of spins $s=2,3,\cdots,\infty$. This is the system conjectured to be dual to the $W_N$ minimal model in \cite{Gaberdiel:2010pz}.  While the dynamical mechanism that renders the set of spins finite in the interacting theory has not yet been understood, this seeming mismatch is not visible at any given order in perturbation theory.} of higher spin gauge fields. In the case of \cite{Gaberdiel:2010pz}, additional massive scalar matter fields are coupled to the higher spin gauge fields. It is likely that these higher spin gauge theories are UV complete (at least perturbatively) theories that contain gravity, due to the large number of gauge symmetries, and are interesting toy models for quantum gravity. However, they do not reduce to semi-classical gravity in any limit. Note that the higher spin symmetry can be broken by AdS boundary conditions \cite{Klebanov:2002ja, Giombi:2011ya}, but this breaking is controlled by the coupling constant of the theory and is in some sense rather mild.

The goal of the current paper is to understand the conjectured duality of \cite{Gaberdiel:2010pz} at the interacting level, in particular, to the second order in perturbation theory. In fact, a careful examination of the spectrum of the linearized Vasiliev system leads us to propose a modification of the conjecture of \cite{Gaberdiel:2010pz}. A key insight of \cite{Gaberdiel:2010pz} is that, in the large $N$ limit of the coset model (\ref{coseta}), $\lambda=N/(N+k)$ plays the role of the 't Hooft coupling, and the basic primaries labelled by representations $(\Box;0)$ and $(0;\Box)$ (as well as the conjugate representations) have finite scaling dimensions $\Delta_+$ and $\Delta_-$ in the 't Hooft limit, and are conjectured to be dual to massive scalars in the bulk. We will consider a version of Vasiliev's system that involve a gauge field of spin $s$ for $s=2,3,\cdots,\infty$, coupled to two real massive scalar fields.
We propose that it is dual to a {\it subsector} of the $W_N$ minimal model, generated by the $W_N$ currents together with two basic primary operators of dimension $\Delta_+$, labelled by $(\Box;0)$ and $(\overline\Box;0)$, or two basic primaries of dimension $\Delta_-$ labelled by $(0;\Box)$ and $(0;\overline\Box)$, depending on the boundary condition imposed on the bulk scalar. We will refer to these two subsectors as the $\Delta_+$ subsector and the $\Delta_-$ subsector, respectively. Each subsector has closed OPEs, and hence consistent $n$-point functions on the sphere, in the sense that they only factorize through operators within in the same subsector. This identification is natural by comparing the bulk fields and boundary operators, and also avoids the puzzle with ``light states" in the 't Hooft limit of the coset model.\footnote{The ``light states" are the primaries labelled by a pair of identical representations, $(R;R)$, whose dimension scales like $1/N$ in the large $N$ limit. While the contribution of such states to the partition function is argued in \cite{Gaberdiel:2010pz} to decouple in the strict infinite $N$ limit, they show up in OPEs of basic primaries when $1/N$ corrections are taken into account.} However, it suggests that the bulk Vasiliev system is non-perturbatively incomplete, though makes sense to all order in perturbation theory. It may be possible to enlarge Vasiliev's system to obtain a higher spin-matter theory that is dual to the full $W_N$ minimal model, but such a bulk theory would be subject to the strange feature of having a large number of light states. We will not address this possibility in the current paper. There is, on the other hand, a minimal truncation of Vasiliev system, where one keeps only the even spin fields and one out of the two real massive scalars. We conjecture that this system is dual to the orthogonal group version of the $W_N$ minimal model.\footnote{The 't Hooft limit of this class of CFTs are recently studied in \cite{Ahn:2011pv}.}

The main nontrivial check of our proposal is a comparison of the tree level three-point functions involving two scalars and one higher spin field in the bulk, and the 't Hooft limit of the corresponding three point function in the dual CFT. In order to carry out such a computation, we first solve for the boundary to bulk propagators of Vasiliev's master fields, and then expand the nonlinear equations of motion to second order in perturbation theory and compute the three point function. We encounter subtleties with gauge ambiguity and boundary condition on the higher spin fields, and will find explicit formulae for the gauge field propagators obeying the boundary condition of \cite{Henneaux:2010xg}. While one may expect that, in principle, such three point functions are determined by symmetries and Ward identities, the implementation of the latter is not so trivial on the CFT side. For instance, we do not know a simple way to carry out the $1/N$ expansion of the coset model, and must calculate correlators exactly at finite $N$ first, and then take the 't Hooft limit. For various quantities of interest in the CFT, analytic formulae for general spins are often difficult to obtain, and instead one computes case by case for the first few spins. The results have a nontrivial dependence on the 't Hooft coupling $\lambda$, which is mapped to a deformation parameter $\nu$ in the bulk theory. The case in which the bulk theory is the simplest, namely the $\nu=0$ ``undeformed" theory, is mapped to $\lambda=1/2$. In this paper, most of our computation is performed within the $\nu=0$ theory, and is compared to the $\lambda=1/2$ case of the $W_N$ minimal model. In Appendix C we give some formulae useful for the deformed bulk theory with nonzero $\nu$, though the analogous computation of correlators in the deformed theory is left to future work. 

More precisely, we compute correlators of the form $\langle {\overline{\cal O}} {\cal O} J^{(s)}\rangle$ at tree level in the $\nu=0$ undeformed bulk theory. These three-point functions are fixed by conformal symmetry up to the overall coefficient; the latter is computed unambiguously as a function of the spin $s$. The result is then compared to the three point functions in the $W_N$ minimal model, in the large $N$ limit, at 't Hooft coupling $\lambda=1/2$. We test the conjectured duality using the explicit expression for the spin 3 current in the coset construction, and found perfect agreement. 

We begin with a brief review of the three-dimensional Vasiliev's system in section 2. In section 3 we describe the linearized spectrum of the bulk theory, as well as propagators and boundary conditions, while leaving technical details to Appendix A. Some useful formulae for the deformed bulk theory (i.e. with nonzero $\nu$) are given in Appendices C. In section 4, we work to second order in perturbation theory and compute the three point functions of interest. The details of these derivations are given in Appendix B. Our proposal of the dualities and a test on the three point functions are presented in section 5. We conclude in section 6.

\section{A brief review of Vasiliev's system in $AdS_3$}

Throughout this paper, we will consider the Vasiliev system in $AdS_3$, which consists of one higher spin gauge field for each spin $s=2,3,4,\cdots$, coupled to a pair of real massive scalar fields. We will often work explicitly with the Poincar\'e coordinates of $AdS_3$, with $x^\m=(z,x^i)$, $i=1,2$, and the metric $ds^2={1\over z^2}(dz^2+dx^i dx^i)$. Following Vasiliev, we introduce the auxiliary bosonic twistor variables $y_\A,z_\A$, where $\A=1,2$ is a spinorial index, as well as the Grassmannian variables $\psi_i$, $i=1,2,$ which obey $\{\psi_i,\psi_j\} = 2 \delta_{ij}$.\footnote{Note that while the equations of motion treats $\psi_1$ and $\psi_2$ on equal footing, the choice of vacuum will not. The $\psi_i$'s can be thought of as purely a bookkeeping device.} The master fields are: $W$ a 1-form in the spacetime parameterized by $x^\mu$, $S$ a 1-form in the auxiliary $z^\A$-space, and $B$ a scalar field. All of them are functions of $x^\mu, y_\A, z_\A$, as well as $\psi_i$,\footnote{In Vasiliev's original papers, the master fields depend on the additional Grassmannian variables $k,\rho$. This will be discussed in Appendix C. We will refer it as the ``extended Vasiliev system", the Vasiliev system we present here is obtained by making a projection $(1+k)/2$ on all fields, and effectively eliminating $k,\rho$.}
\ie
&W = W_\mu(x|y,z,\psi_i) dx^\mu,
\\
&S = S_\A(x|y,z,\psi_i) dz^\A,
\\
&B = B(x|y,z,\psi_i).
\fe
These fields are subject to a large set of gauge symmetries. The infinitesimal gauge transformation is parameterized by a function $\epsilon(x|y,z,\psi)$,
\ie
&\delta W = d_x\epsilon + [ W, \epsilon ]_*,
\\
&\delta S =d_z\epsilon + [ S, \epsilon]_*,
\\
&\delta B = [ B, \epsilon]_*.
\fe
One further imposes a truncation so that $W, B$ are even functions of $(y,z)$ whereas $S_\A$ is odd in $(y,z)$ (so that the 1-form $S$ is even under $(y,z,dz)\mapsto (-y,-z,-dz)$). The gauge parameter $\epsilon$ is then restricted to be an even function of $(y,z)$ as well.
One introduces a star-product $*$ on functions of $(y,z)$, defined by
\ie
f(y,z)*g(y,z)=\int d^2ud^2v e^{uv}f(y+u,z+u)g(y+v,z-v).
\fe
Here and throughout this paper, the spinors are contracted as $uv=u^\A v_\A=-v^\A u_\A=-vu$ and $u\sigma v=u^\A \sigma_\A{}^\B v_\B$ for a matrix $\sigma$. The integration measure $d^2u d^2 v$ above is normalized such that $f*1=f$. The Grassmannian variables $\psi_i$ commute with $y_\A, z_\A$ and do not participate in the $*$ product. Under the star-product, the auxiliary variables $y_\A$ generate the three dimensional higher spin algebra $hs(1,1)$ \cite{Vasiliev:1991}\footnote{We will also consider $hs(\lambda)$ the one parameter deformation of $hs(1,1)$ in Appendix C.}, which is an associative algebra, whose general element can be represented by a even analytic function of in $y_\A$. In particular, $hs(1,1)$ has a subalgebra $sl(2)$ whose generator can be written as $T_{\A\B}=y_{(\A}*y_{\B)}$. An inner product on this algebra is defined as $(A,B)=A(y)*B(y)\big|_{y=0}$.
%\footnote{The factor $i$ here is due to a different convention of our star product compared to \cite{Vasiliev:1991}. The inner product on the star product will only be needed later in recovering the Chern-Simons form of the higher spin action, in computing the two point functions.}

We define an involution $\iota$ on the star algebra as follows: $\iota(y^\A)=i y^\A$, $\iota(z^\A)=-i z^\A$, $\iota(dz^\A)=-i dz^\A$, and the action of $\iota$ reverses the order of all products (including the multiplication of $\psi_i$'s); in particular, $\iota(\psi_1\psi_2)=\psi_2\psi_1=-\psi_1\psi_2$.
The master fields $W,S,B$ are then subject to the reality condition\footnote{Such a reality condition is necessary because, as we will see later, the physical components of the $B$ master field are of the form $\psi_2 C_{even}+\psi_2\psi_1 C_{odd}$ where $C_{even}$ is a real scalar and $C_{odd}$ is a purely imaginary scalar field.}
\ie\label{real}
\iota(W)^*=-W,~~ \iota(S)^*=-S,~~\text{and}~~ \iota(B)^*=B,
\fe
where the superscript $*$ stands for taking the complex conjugate on the component fields while leaving the auxiliary variables $y^\A,z^\A,\psi_i$ untouched.

Vasiliev's equations of motion are now written as
\ie\label{Vasiliev's eq}
&d_x W + W*W = 0,
\\
&d_x S + d_z W + \{W,S\}_* = 0,
\\
&d_zS + S*S = B*K dz^2,
\\
&d_x B + [W,B]_* = 0,
\\
&d_z B + [S,B]_* = 0.
\fe
Here $d_x$ and $d_z$ denote the exterior derivative in spacetime coordinates $x^\mu$ and the auxiliary variables $z^\A$ respectively. $K\equiv e^{zy}$ is known as the Kleinian. It has the properties
\ie
&K*K=1,~~~~
K*f(y,z) = K f(z,y),~~~~
f(y,z)*K=K f(-z,-y).
\fe
A few comments on (\ref{Vasiliev's eq}) are in order. The third equation in (\ref{Vasiliev's eq}) can be thought of as the definition of the scalar master field $B$. The fourth equation is equivalent to a Bianchi identity for the field strength of the connection ${\cal A}=W+S$, which follows from the second and third equation. The last equation, however, is an independent equation for $B$.\footnote{This is different from the four-dimensional version of Vasiliev's system, which involves a similar set of equations.}

Note that the equations of motion (\ref{Vasiliev's eq}) are preserved under the involution $\iota$, if one sends $(W,S,B)$ to $(-W, -S, B)$ at the same time.
In particular, Vasiliev's system can be further truncated down to what we refer to as the ``minimal Vasiliev's system". The latter is defined by projecting the master fields onto the $\iota$-invariant components, namely
\ie\label{trunc}
\iota(W)=-W,~~ \iota(S)=-S,~~\text{and}~~ \iota(B)=B.
\fe
We will see later that the minimal Vasiliev's system contains only the even spin gauge fields and a single matter scalar. Though, in most of this paper, we will be considering the untruncated Vasiliev's system, where gauge spins of all spins greater than or equal to 2 are included.

%In general, $W_\m(x|y,z,\psi)$ contains both integer and half-integer higher spin $s\ge 3/2$ gauge fields as the coefficients of the its expansion of the auxiliary variables $y^\A$ and $z^\A$, and $B(x|y,z,\psi)$ contains a massive scalar and a massive fermion. But we can project out all the fermionic fields by restricting $W_\m(x|y,z,\psi)$ and $B(x|y,z,\psi)$ being even function of the auxiliary variable $y^\A$ and $z^\A$, and it follows that $S_\A(x|y,z,\psi)$ is an odd function of $y^\A$ and $z^\A$. 

The equations (\ref{Vasiliev's eq}) are formulated in a background independent manner. 
To formulate the perturbation theory, one begins by choosing a vacuum solution, and identifies the physical propagating degrees of freedom by linearizing the equations around the vacuum solution. One may then proceed to higher orders in perturbation theory and study interactions in this background. It turns out that the system (\ref{Vasiliev's eq}) admits a 1-parameter family of distinct $AdS_3$ vacua, labeled by a real parameter $\nu$. In fact, the parameter $\nu$ appears in a non-dynamical, auxiliary component of $B$, and thus the 1-parameter family of $AdS_3$ vacua are not connected by physical deformations, but should rather be thought of as different theories in $AdS_3$. In this paper, we will focus on the simplest, ``undeformed" theory, corresponding to the $\nu=0$ vacuum. The deformed vacua/theories ($\nu\not=0$) are discussed in Appendix C. The perturbation theory, and in particular the study of three point functions, of the {\it deformed} theory is left to future work.

The undeformed $AdS_3$ vacuum solution is given by
\ie\label{undvac}
B = 0,~~~~S = 0,~~~~W = W_0 \equiv w_0(x|y) + \psi_1 e_0(x|y),
\fe
where $W_0$ is a flat connection satisfying $d_x W_0 + W_0*W_0 = 0$. With $W_0(x|y,\psi_1)$ chosen to be a quadratic function of $y$, the flatness condition is classically equivalent to the Chern-Simons formulation of Einstein's equation with negative cosmological constant in three dimensions. In other words, the equations of motion is obeyed if the 1-forms $e_0,w_0$ are chosen as the dreibein and spin connection for $AdS_3$, contracted with $y^\A$ in spinorial notation. In Poincar\'e coordinates $x^\mu=(z, x^i)$, they can be written as
\ie
& w_0(x|y) \equiv w_0^{\A\B}(x) y_\A y_\B =  - {y \sigma^{\m z} y\over 8 z}dx^\m, ~~~~ e_0(x|y) \equiv
e_0^{\A\B}(x) y_\A y_\B = - {y \sigma^\m y\over 8z} dx^\m .
\fe
Our convention for $e_0$ is such that
\ie
(e_0^\mu)_{\A\B} (e_{0\mu})^{\C\D} = -{1\over 64} (\delta_\A^\C \delta_\B^\D + \delta_\A^\D \delta_\B^\C),
~~~~~ (e_0^\mu)_{\A\B} (e_{0\nu})^{\A\B} = -{1\over 32} \delta^\mu_\nu.
\fe
Expanding around this vacuum solution, we will write $W=W_0+\widehat W$, and the equations of motion in its perturbative form as
\ie\label{eomperturb}
&D_0 \widehat W = - \widehat W*\widehat W,
\\
&D_0 S + d_z \widehat W = -\{\widehat W,S\}_*,
\\
&d_z S - B*Kdz^2 = -S*S,
\\
&d_z B = -[S,B]_*,
\\
&D_0 B = -[\widehat W,B]_*,
\fe
where we have defined $D_0\equiv d_x +[W_0,\cdot]_*$. By choosing a $z_\A$-dependent gauge function, one can always go to a gauge in which $S|_{z_\A=0} = 0$. The physical degrees of freedom are entirely contained in the $z_\A$-independent part of the master fields, whereas the $z_\A$-dependence are determined via the equations of motion.
It is then useful to decompose $W,B$ as
\ie\label{decomposeWB}
&W(x|y,z,\psi)=W_0+\Omega(x|y,\psi)+W'(x|y,z,\psi)\\
&B(x|y,z,\psi)=C(x|y,\psi)+B'(x|y,z,\psi)
\fe
where $\Omega$ and $C$ are the restriction of $\widehat W$ and $B$ to $z_\A=0$, respectively, while $W'$ and $B'$ obey $W'\big|_{z_\A=0}=B'\big|_{z_\A=0}=0$. We will see that $\Omega$ and $C$ contain the higher spin gauge fields and two real scalar fields, whereas $W'$ and $B'$ are auxiliary fields. At the linearized level, the equations (\ref{eomperturb}) reduce to
\bea
&&D_0 \Omega^{(1)} =  - \{ W_0,  W'^{(1)} \}_*|_{z=0},\label{Omega}\\
&&d_z W'^{(1)} = -D_0 S^{(1)},\label{W'}\\
&&d_z S^{(1)} = C^{(1)}*K dz^2,\label{S}\\
&&B'^{(1)} = 0, \label{B'}\\
&& D_0C^{(1)} =0,\label{C}
\eea
where the superscript $(n)$ labels the order of the component of the respective field in the perturbative expansion. These equations will be analyzed in detail in the next section as well as in Appendix A. We will then proceed to the quadratic order and study the cubic coupling and three point functions in section 4.

Let us note that the system of equations (\ref{Vasiliev's eq}) and the $AdS_3$ vacuum (\ref{undvac}) are invariant under a global $U(1)$ symmetry,
\ie\label{uone}
W\to e^{i\theta \psi_1} W e^{-i\theta \psi_1},~~~~
S\to e^{i\theta \psi_1} S e^{-i\theta \psi_1},~~~~
B\to e^{i\theta \psi_1} B e^{-i\theta \psi_1}.
\fe
This $U(1)$ rotates the phase of the complex scalar matter field, while leaving the higher spin fields invariant. Note that (\ref{uone}) preserves the reality condition (\ref{real}). While it is a symmetry of the classical theory, and is expected to be a perturbative symmetry of the quantum theory, it should be broken non-perturbatively (or alternatively, become gauged), as anticipated in any quantum theory of gravity \cite{Banks:2010zn,Hellerman:2010fv}. In the proposed dual CFT, the $U(1)$ rotates the basic primaries $(\Box;0)$ and $(\overline\Box;0)$ with opposite phases. As far as correlators of a fixed number of basic primaries are concerned, in the large $N$ limit, this $U(1)$ is effectively a symmetry of the theory, since any correlation function that violates the $U(1)$ vanishes by the fusion rule. This $U(1)$ is obviously broken when $N$ basic primaries are inserted, as the tensor product of $N$ fundamental representations of $SU(N)$ contains a singlet.

%In the next subsection, we will solve for the boundary-to-bulk propagator for the master field $C^{(1)}$ using (\ref{C}). In subsection 2.3, we will study the equations (\ref{Omega}), (\ref{W'}), and (\ref{S}), especially (\ref{Omega}), which will give us the linearized equations of the higher spin gauge fields. We will further rewrite the linearized higher spin equations in terms of the equations of the metric-like field or frame-like field. In subsection 2.4, we will solve for the boundary-to-bulk propagator in the modified de Donder gauge using the equation of metric-like field. However, in subsection 2.5, we will show that when specializing to spin 2 such boundary-to-bulk propagator explicitly violates the Brown-Henneaux boundary condition. We will further propose a generalized  Brown-Henneaux boundary condition for higher spin gauge fields, and study the asymptotic behavior of higher spin gauge fields under this boundary condition. It turn out that just by knowing the asymptotic behavior is enough for us to compute the higher spin two point function $\vev{J_s(x_1)J_s(x_2)}$ and the three point function of one higher spin gauge field and two scalar fields $\vev{\cO(x_1)\cO(x_2)J_s(x_3)}$. We will perform these computations in subsection 2.6 and in section 3.

\section{Propagators and two point functions}

\subsection{The physical fields and propagators}

In this subsection we will describe the physical degrees of freedom in the linearized master fields, as well as their propagators. The details of the derivations starting from Vasiliev's equation are given in Appendix A. 

\subsubsection{The scalar matter field}

The linearized scalar master field $C^{(1)}(x|y,\psi)$ can be decomposed as
\ie\label{Cdam}
C^{(1)}(x|y,\psi_i) = C^{(1)}_{aux}(x|y,\psi_1) + \psi_2 C_{mat}^{(1)}(x|y,\psi_1).
\fe
$C^{(1)}_{aux}$ is purely auxiliary; the only solution to its equation of motion is a constant, which parameterizes a family of $AdS_3$ vacua. We will set $C^{(1)}_{aux}=0$ for now. $C^{(1)}_{mat}$ can be expanded in $y$ as
\ie\label{Cdy}
C_{mat}^{(1)} = \sum C_{mat}^{(1),n}(x|y,\psi_1) = \sum C_{mat}^{(1),n}{}_{\A_1\cdots\A_n}(x|\psi_1) y^{\A_1}\cdots y^{\A_n}.
\fe 
It follows from $D_0 (\psi_2 C^{(1)}_{mat})=0$ that the bottom component $C^{(1),0}_{mat}(x|\psi_1)$ obeys the usual Klein-Gordon equation for a massive scalar field in $AdS_3$,
\ie\label{Klein-Gordon}
\left(\nabla^\mu \partial_\mu - m^2 \right) C^{(1),0}_{mat}(x|\psi_1) = 0,~~~m^2 = -{3\over 4}.
\fe
Expanding further in $\psi_1$, $C_{mat}^{(1),0}(x|\psi_1) = C_{even}(x) + \psi_1 C_{odd}(x)$ contain a pair of real scalars of mass squared $m^2=-{3\over 4}$ in $AdS$ units. Due to the reality condition (\ref{real}), $C_{even}$ is real whereas $C_{odd}$ is a purely imaginary scalar field.  They can be paired up to a complex massive scalar as $C_{even}+C_{odd}$, with $C_{even}-C_{odd}$ its complex conjugate. Under the global $U(1)$ symmetry (\ref{uone}), $C_{even}\pm C_{odd}$ transform as
\ie
C_{even}\pm C_{odd}\rightarrow e^{\pm i \theta}\left(C_{even}\pm C_{odd}\right).
\fe

In the dual boundary CFT, this complex scalar corresponds to a complex scalar operator of dimension $\Delta_+$ or $\Delta_-$, depending on the choice of boundary condition. Here
\ie
\Delta_\pm = 1\pm {1\over 2} = {3\over 2}~{\rm or}~{1\over 2}.
\fe
The higher components $C_{mat}^{(1),n}$ are expressed in terms of derivatives of $C_{mat}^{(1),0}$ through the equation of motion.

In the $\nu$-deformed vacua, $C_{mat}^{(1)}$ still describes a pair of real massive scalar fields, with mass squared $m^2 = -{3\over 4}+{\nu (\nu \pm 2)\over 4}$, where the $\pm$ sign depends on a choice of projection. This is discussed in Appendix C.

The boundary-to-bulk propagator for the scalar is $C^{mat,0} = K(\vec x,z)^\Delta$ for $\Delta = 3/2$ or $\Delta = 1/2$, where $K(\vec x,z)\equiv {z\over \vec x^2 + z^2}$, $\vec x=(x^1, x^2)$. It is convenient to introduce another auxiliary variable $\widetilde \psi_1$, satisfying $\widetilde{\psi}_1^2=1$, to label the two different boundary conditions, so that $\Delta = 1 + \widetilde\psi_1 /2$. With the $\delta$-function source on $C_{even}$ component:
\ie\label{source}
C_{mat}^{(1)}(\vec x,z\rightarrow 0|y,\psi_1)=2\pi \widetilde\psi_1 z^{1-{\widetilde\psi_1\over 2}}\delta^2(x)
\fe
turned on on the boundary, the boundary-to-bulk propagator for the master field $C_{mat}^{(1)}(x|y,\psi_1)$ is then given by
\ie\label{b-t-b scalar}
C_{mat}^{(1)}(x|y,\psi_1)=\left(1+\psi_1{1+\widetilde \psi_1\over 2} y\Sigma y\right)e^{{\psi_1\over 2}y\Sigma y}K^{1+{\widetilde\psi_1\over 2}},
\fe
where $\Sigma\equiv \sigma^z - {2z\over x^2}\sigma^\m x^\m $. We can also turn on the source on $C_{odd}$ component:
\ie
C_{mat}^{(1)}(\vec x,z\rightarrow 0|y,\psi_1)=2\pi \psi_1\widetilde\psi_1 z^{1-{\widetilde\psi_1\over 2}}\delta^2(x)
\fe
on the boundary. The boundary-to-bulk propagator will be just (\ref{b-t-b scalar}) times $\psi_1$.

Under the action of the involution $\iota$, $C_{even}$ is invariant whereas $C_{odd}$ changes sign. Hence only $C_{even}$ survives the minimal truncation (\ref{trunc}). Thus, the ``minimal Vasiliev system" contains only a single real scalar scalar, which is dual to a real scalar operator in the boundary CFT. Note that in writing the boundary-to-bulk propagator (\ref{b-t-b scalar}), we have chosen to turn on a source for $C_{even}$ only, and the result is invariant under the projection by $\iota$. 

\subsubsection{The higher spin fields}

The higher spin gauge fields, as well as some auxiliary fields, are contained in $\Omega(x|y,\psi)$, which may be decomposed in the form
\ie\label{Odhssc}
\Omega^{(1)}(x|y,\psi_i) = \Omega^{hs}(x|y,\psi_1) + \psi_2 \Omega^{sc}(x|y,\psi_1).
\fe
As the notations suggest, $\Omega^{hs}$ contain the higher spin gauge fields in $AdS_3$, while $\Omega^{sc}$ are in fact auxiliary fields determined by the scalar matter fields. The linearized equations take the form
\ie\label{linom}
D_0 \Omega^{hs} = 0,~~~\tilde D_0 \Omega^{sc} = -\psi_2 \{ W_0, \psi_2 W^{mat}\}_*|_{z=0}.
\fe
where we have defined
\ie
\tilde D_0 \equiv d_x + [w_0,\cdot]_* - \psi_1 \{ e_0,\cdot \}_*.
\fe
It is demonstrated in Appendix A.2 that up to gauge transformations, $\Omega^{sc}$ have no propagating degrees of freedom and are determined entirely in terms of $C_{mat}$. $\Omega^{hs}$, on the other hand, obeys the (linearized) Chern-Simons equation with higher spin algebra ${\rm hs}(1,1)\oplus {\rm hs}(1,1)$. They are related to the metric-like higher spin fields, which are usually written in terms of traceless symmetric tensors, in the following way.

First, expand $\Omega^{hs}_{\A\B} \equiv \Omega^{hs}_\mu (e_0^\mu)_{\A\B}$ in $y$ as
\ie
\Omega_{\A\B}^{hs}(x|y,\psi_1) = \sum \Omega^{hs,(n)}_{\A\B}(x|y,\psi_1) =  \sum \Omega^{hs,n}_{\A\B|\A_1\cdots\A_n}(x|\psi_1) y^{\A_1}\cdots y^{\A_n} ,
\fe
and then express the components in terms of symmetric traceless tensors (in spinorial notation) as
\ie
\Omega^{hs,(n)}_{\A\B|\A_1\cdots\A_n}(x|\psi_1) = \chi^{n,+}_{\A\B\A_1\cdots\A_n} + \epsilon_{(\A_1(\underline{\A}}\chi^{n,0}_{\underline{\B})\A_2\cdots\A_n)} + \epsilon_{(\underline{\A}(\A_1}\epsilon_{\underline{\B})\A_2} \chi^{n,-}_{\A_3\cdots\A_n)},
\fe
or equivalently,
\ie
& \Omega^{hs,(n)}_{\A\B}(x|y,\psi_1) = {1\over (n+2)(n+1)}\partial_\A \partial_\B \chi_n^+(x|y,\psi_1)
+ {1\over n} y_{(\A}\partial_{\B)} \chi_n^0(x|y,\psi_1) + y_\A y_\B \chi_n^-(x|y,\psi_1).
\fe
Here $\chi_n^+(x|y,\psi_1)$ is defined as $\chi^{n,+}_{\A_1\cdots\A_{n+2}}$ contracted with $y^\A$'s, and similarly for $\chi_n^0(x|y,\psi_1)$ and $\chi_n^-(x|y,\psi_1)$. Next, we expand in $\psi_1$, and write
\ie
\chi_n^{\pm/0} = \chi^{n,\pm/0}_{even} + \psi_1 \chi^{n,\pm/0}_{odd}.
\fe
It turns out that $\chi_{even}$ are determined in terms of (derivatives of) $\chi_{odd}$ through the equation of motion. Furthermore, $\chi_{odd}^{n,0}$ can be gauged away entirely. The residual gauge symmetry on $\chi_{odd}^{n,\pm}(y)$ takes the form
\ie\label{gaugechi}
& \delta \chi_{odd}^{n,+}(y) = - \nabla^+\lambda_{odd}^n(y),
\\
& \delta \chi_{odd}^{n,-}(y) = - {1\over n(n+1)} \nabla^-\lambda_{odd}^n(y),
\fe
where $\lambda_{odd}^n(y)$ is related to the gauge parameter $\epsilon$ by $\epsilon = \psi_1 \lambda_{odd}^n$. $\nabla^\pm$ are defined here as
\ie
\nabla^+ \equiv (y e_0^\mu y) \nabla_\mu,~~~\nabla^- \equiv (\partial_y e_0^\mu \partial_y) \nabla_\mu,
\fe
where $\nabla_\mu$ acts on a tensor $(\cdots)_{\A_1\A_2\cdots}$ as the spin-covariant derivative. Under the $\iota$-action, only the even spin fields are invariant. Hence, the ``minimal" Vasiliev's system only contains higher spin gauge fields with even spins, and its dual boundary CFT contains only even spin currents.

In the metric-like formulation, the spin-$s$ gauge field is described by a rank $s$ double traceless symmetric tensor $\Phi_{\mu_1\cdots\mu_s}$. It may be decomposed into irreducible representations of the Lorentz group as
\ie\label{Phidxichi}
\Phi_{\mu_1\cdots\mu_s} = \xi_{\mu_1\cdots\mu_s} + g_{(\mu_1\mu_2} \chi_{\mu_3\cdots\mu_s)},
\fe
where $\xi$ and $\chi$ are traceless symmetric tensors of rank $s$ and $s-2$, respectively. With the identification
\ie\label{idchi}
\chi_{odd}^{2s-2,+} = \xi^{(s)},~~~~ \chi_{odd}^{2s-2,-} = -{2s-3\over 32(s-1)}\chi^{(s)},
\fe
where $\xi^{(s)}$ is defined as $\xi_{\mu_1\cdots\mu_s}$ contracted with $(e_0^\m)_{\A\B}y^\A y^\B$, and similarly for $\chi^{(s)}$, the Chern-Simons form of the equations of motion can be shown to be equivalent to the Fronsdal form of the equation on $\Phi$,
\ie\label{fransdal}
&(\Box-m^2)\Phi_{\m_1\cdots\m_s}-s \p_{(\ul{\m_1}}\p^\m\Phi_{\m\ul{\m_2\cdots\m_s})}+{1\over 2}s(s-1)\p_{(\ul\m_1}\p_{\ul\m_2}\Phi^{\m}{}_{\m\ul{\m_3\cdots\m_s})}
\\
&~~~~-s(s-1)g_{(\ul{\m_1\m_2}}\Phi^{\m}{}_{\m\ul{\m_3\cdots\m_s})}=0,
\fe
which is invariant under the gauge transformation:
\ie\label{fransdal gauge trans}
\delta \Phi_{\m_1\cdots \m_s} = \nabla_{(\m_1}\eta_{\m_2\cdots\m_s)},
\fe
where $\eta_{\m_2\cdots\m_s}$ is a symmetric traceless gauge parameter. The gauge transformation (\ref{fransdal gauge trans}) is also equivalent to (\ref{gaugechi}) under the identification (\ref{idchi}).

In three dimensions, the higher spin gauge fields do not have bulk propagating degrees of freedom. In $AdS_3$, just as in the more familiar case of gravitons ($s=2$), there are boundary excitations of the higher spin fields, corresponding to field configurations that cannot be gauged away by gauge transformations that vanish on the boundary of the AdS spacetime. A careful analysis of the gauge conditions is necessary in order to talk about boundary-to-bulk propagators and bulk-to-bulk propagators. We will first consider Metsaev's modified de Donder gauge \cite{Metsaev:2009}, which is convenient for solving higher spin propagators in AdS in general dimensions. We will see, however, that the propagators found in this gauge violates (the higher spin generalization of) Brown-Henneaux boundary condition, and are not directly applicable to the computation of boundary correlators. Nonetheless, this gauge should be useful in doing loop computations in the bulk. We will then proceed to find the appropriate boundary-to-bulk propagators that obey Brown-Henneaux boundary condition, which allows for computations of boundary correlators.

\subsection{Propagators in modified de Donder gauge}

The modified de Donder gauge was introduced by Metsaev in \cite{Metsaev:2009}. This gauge has the advantage that the equations of motion for different components of free higher spin gauge fields decouple, and hence the solutions can be obtained easily. The implementation of the gauge condition, on the other hand, is a bit complicated. It can be described as follows. Start with the double traceless symmetric $\Phi^{s}_{\mu_1\cdots\mu_s}$ which obeys the Fronsdal equation in $AdS_3$. Write $\Phi^{s}_{A_1\cdots A_s} = \Phi^{s}_{\mu_1\cdots\mu_s} e^{\mu_1}_{A_1}\cdots e^{\mu_s}_{A_s}$ where $A_i$ are local Lorentz frame indices. Define a generating function/field
\ie\label{PhiY}
\Phi^s(x|Y) = \Phi^s_{A_1\cdots A_s} Y^{A_1} \cdots Y^{A_s},
\fe
where $Y^A=(Y^z, Y^1, Y^2)$ are auxiliary vector variables (analogous to the twistor variables $y^\A$ introduced previously). One then performs a linear transformation on $\Phi^s(x|Y)$,
\ie
\phi(x|Y) = z^{-{1\over 2}} {\cal N} \Pi^{\phi\Phi} \Phi^s(x|Y),
\fe
where $z$ is the Poincar\'e radial coordinate, ${\cal N}$ is an operator that acts as a separate normalization factor on each component of $\Phi(x|Y)$ of given degree in $Y^z$ and $\vec Y=(Y^1,Y^2)$, and $\Pi^{\phi\Phi}$ involves derivatives on $Y^z$ and $\vec Y$. See Appendix A.3 for the definition of these operators. The resulting generating field $\phi(x|Y)$ is double traceless with respect to the directions parallel to the boundary, namely
\ie\label{dtphi}
\left( {\partial^2\over \partial \vec Y^2}\right)^2 \phi(x|Y) = 0.
\fe
The modified de Donder gauge is defined by a gauge condition of the form
\ie\label{Cgauge}
\overline{C} \phi(x|Y) = 0,
\fe
where $\overline{C}$ is an operator involving up to two derivatives on $\vec Y$ and one spacetime derivative. The key point is that, in this case, the Fronsdal equation for $\Phi^s$ is re-expressed in terms of equations on $\phi(x|Y)$ as
\ie
\left[ \Box + \partial_z^2 - {(r-{1\over 2})(r-{3\over 2})\over z^2}  \right] \phi_r(x|\vec Y)=0,
\fe
where $\phi_r(x|\vec Y)$ are the components of $\phi(x|Y)$ expanded in $Y^z$,
\ie\label{phiexpand}
\phi(x|Y) = \sum_{r=0}^s (Y^z)^{s-r} \phi_r(x|\vec Y).
\fe
The equation of motion is then straightforwardly solved in momentum space. Note that the gauge condition (\ref{Cgauge}) relates the different components $\phi_r(x|\vec Y)$. After solving $\phi(x|Y)$, one can translate it back into $\Phi^s(x|Y)$, and further into the frame-like fields $\chi_{odd}^{(s),\pm}$. The result for the boundary-to-bulk propagator of $\chi_{odd}^{(s),\pm}$ due to a chiral spin-$s$ current $J^{(s)}_{++\cdots +}$ source inserted at $\vec x=0$ is given in momentum space explicitly by (up to the overall normalization factor)
\ie\label{chibh}
\chi^{(s),+}_{odd}(\vp,z|y)=& \sum^s_{r=0}{i^{r}}{s\choose r}p^{r-1} { (p^+)^{s-r}} (y^1)^{s+r} (y^2)^{s-r} zK_{r-1}(z|\vp|),
\\
\chi^{(s),-}_{odd}(\vp,z|y) =& {s\over 2(2s-1)}\sum^s_{r=0} i^{r}{s-2\choose r} {p^{r-1} (p^+)^{s-r}}(y^1)^{s+r-2} (y^2)^{s-r-2}zK_{r-1}(z|\vp|).
\fe
The details of the derivation is given in Appendix A.3. These propagators, however, do not obey the higher spin analog \cite{Henneaux:2010xg, Campoleoni:2010zq} of Brown-Henneaux boundary condition \cite{Brown:1986nw}, which should be imposed in order for the dual CFT to have the appropriate higher spin symmetry. In fact, we know that any solution to the linearized higher spin equations in $AdS_3$ must be a pure gauge in the bulk. The key to finding the appropriate boundary-to-bulk propagator is then to find the appropriate gauge transformation near the boundary. In the next subsection, we will see that such a gauge transformation takes a rather simple form. The bulk-to-bulk propagators in the modified de Donder gauge may still prove useful for loop computations in the bulk, which we hope to revisit in the future.

\subsection{The asymptotic boundary condition}

Let us begin with the spin 2 case, and consider the Brown-Henneaux boundary condition \cite{Brown:1986nw} on metric fluctuations. In the $Y$-algebra language, a spin 2 tensor field sourced by a positively polarized stress-energy tensor insertion on the boundary, at $\vec x=0$, that obeys Brown-Henneaux boundary condition is given by
\ie\label{Brown-Henneaux}
\Phi^2(x|Y)\sim\delta^2(\vec x)(Y^+)^2+(\text{subleading contact terms}) + {z^2\over (x^-)^4}(Y^-)^2.
\fe
On the RHS we only indicated the leading order terms in the $z\to 0$ limit; their coefficients are not specified. The boundary-to-bulk propagators in the modified de Donder gauge, derived in the previous subsection, does not obey this boundary condition. It suffices to examine the spin 2 case. In position space, the graviton boundary to bulk propagator in the modified de Donder gauge (for a positively polarized source) is
\ie
\Phi^2(Y)&={2i \over \pi} Y^z Y^+   {x^+ z \over (x^2+z^2)^2} - {i \over \pi} (Y^+)^{2}  {z^{2}\over (x^2+z^2)^2} +{  i \over \pi} Y^+Y^-  {(x^+)^{2}\over (x^2+z^2)^2}.
\fe
In the limit $z\rightarrow 0$, it goes like
\ie
\Phi^2(Y)\sim \delta^2(x) (Y^+)^2 + (\text{subleading contact terms})+{Y^- Y^+ \over (x^-)^{2}},
\fe
which clearly violates the boundary behavior of (\ref{Brown-Henneaux}).

Similarly, the higher spin gauge fields are subject to the an analog of the Brown-Henneaux boundary conditions \cite{Henneaux:2010xg,Campoleoni:2010zq}. For general spin $s$, the boundary condition is such that the boundary-to-bulk propagator for a positive polarized spin-$s$ source is
\ie\label{generalized Brown-Henneaux}
\Phi^s(x|Y)\sim z^{2-s}\delta^2(\vec x)(Y^+)^s+(\text{subleading contact terms})+{(Y^-)^s z^s\over (x^-)^{2s}},
\fe
where the coefficient are again not specified. Let us examine this boundary condition (\ref{generalized Brown-Henneaux}) in more detail. In three dimension, similarly to gravitons, the higher spin gauge fields do not have any propagating degrees of freedom in the bulk. In other words, any solution to the equation of motion can be (locally) written in a pure gauge form, $\Phi^s(x|Y) = Y^A D^A \eta^s(x|Y)$. However, the gauge parameter $\eta^s(x|Y)$ may have nonzero higher spin charge, the latter is given by a boundary integral, and the higher spin gauge field $\Phi^s(x|Y)$ would not be gauge equivalence to zero. As proposed in \cite{Henneaux:2010xg}, the boundary behavior of the gauge parameter $\eta^s(x|Y)$ can be fixed by demanding the gauge field $\Phi^s(x|Y)$ obeys the boundary conditions (\ref{generalized Brown-Henneaux}). With some effort, we find the appropriate gauge parameter $\eta^s(x|Y)$ near the boundary:
\ie\label{boundary gauge parameter eta}
\eta^s(x|Y)&= \sum^{s-1}_{u=0}\sum^{2s-2u-1}_{r=1}\sum^{u}_{v=0}{(-1)^{r+u}\over (2u)!}{u\choose v}\left(\prod^{2u-1}_{j=0}(r+j)\right)\left(\prod^u_{j=1}{2j-1\over 2s-2j-1}\right)
\\
&~~~~~~~~\times(Y^3)^{2v+r-1}(Y^-)^{u-v}(Y^+)^{s-r-v-u}{z^{2u+r-s}\over (x^-)^{2u+r}} + \cO(z^{s+1}),
\fe
and the corresponding gauge field
\ie\label{boundary gauge field}
\Phi^s(x|Y)&=Y^A D^A \eta^s(x|Y)
\\
&=2\pi z^{2-s}\delta^2(x)(Y^+)^s+(\text{subleading contact terms})
\\
&~~~~+(-1)^s (2s-1){(Y^-)^s z^s\over (x^-)^{2s}} + \cO(z^{s+1}).
\fe
Notice that the leading analytic term on the RHS of (\ref{boundary gauge field}) is proportional to the two point function of the boundary higher spin currents. Since the gauge parameter is a traceless tensor, i.e. $\partial_Y^2 \eta_s(Y)=0$, we can substitute $Y^A=e^A_{\A\B}y^\A y^\B$ in (\ref{boundary gauge parameter eta}) and obtain, modulo an overall normalization coefficient, the gauge parameter in the (spinorial) $y$-algebra language (see (\ref{gaugechi})):
\ie\label{boundary gauge parameter lambda}
\lambda^s(y)&= -4\sum^{2s-1}_{r=1}(y^1)^{2s-r-1}(y^2)^{r-1}{z^{r-s}\over (x^-)^{r}} + \cO(z^{s+1}).
\fe
For later use, we also compute the boundary-to-bulk propagators for the generating function of frame-like fields, $\chi^{(s),\pm/0}_{odd}$ and $\chi^{(s),\pm/0}_{even}$ using (\ref{gauge lambda}) and (\ref{odd to even}), and  compute $\Omega^{hs,(s)}_{11}$ and $\Omega^{hs,(s)}_{22}$ using (\ref{chi to omega}). They are
\ie
&\chi^{(s),+}_{odd}=2\pi (y^1)^{2s}z^{2-s}\delta^2(x) +(\text{subleading contact terms})+{(2s-1)(y^2)^{2s}z^s\over  (x^-)^{2s}}+ \cO(z^{s+1}),
\\
&\chi^{(s),0}_{odd}=0,
\\
&\chi^{(s),-}_{odd} =(\text{contact terms of the order $z^{4-2s}$ and higher})+\cO(z^{s+1}),
\fe
and
\ie
& \chi_{even}^{(s),+} =-2\pi (y^1)^{2s}z^{2-s}\delta^2(x) +(\text{subleading contact terms})-{(2s-1)(y^2)^{2s}z^s\over  (x^-)^{2s}}+ \cO(z^{s+1}),
\\
& \chi_{even}^{(s),0} =(\text{contact terms of the order $z^{3-2s}$ and higher})+\cO(z^{s+1}),
\\
& \chi_{even}^{(s),-} =(\text{contact terms of the order $z^{4-2s}$ and higher})+\cO(z^{s+1}),
\fe
as well as
\ie\label{RHBC Omega}
&\Omega^{hs,(s)}_{11}(y)=-2(1-\psi_1)\pi (y^1)^{2s-2}z^{2-s}\delta^2(x)+(\text{subleading contact terms})+\cO(z^{s+1}),
\\
&\Omega^{hs,(s)}_{22}(y)=(\text{contact terms of the order $z^{4-s}$ and higher})-(1-\psi_1){(2s-1)(y^2)^{2s-2}z^s\over  (x^-)^{2s}}+\cO(z^{s+1}).
\fe
Notice that the leading contact term in $\Omega^{hs,(s)}_{11}$ is proportional to $(1-\psi_1)$; in other words, we have imposed the Dirichlet boundary condition on the component $(1-\psi_1)\Omega^{hs,(s)}_{11}$. Similarly, for the negative polarized higher spin gauge field, we impose the Dirichlet boundary condition on the component $(1+\psi_1)\Omega^{hs,(s)}_{22}$.

\subsection{Higher spin two point function}

With these formulae at hand, we can now compute the two point function of the higher spin currents on the boundary. The linearized higher spin equation $D_0\Omega^{hs}=0$ can be obtained from the quadratic part of a Chern-Simons type action:
\ie\label{csbulk}
S_{hs}=-\int d\psi_1 \int \left(\Omega^{hs},d\Omega^{hs}+2W_0*\Omega^{hs}\right).
\fe
We decompose the higher spin gauge field as
\ie
\Omega^{hs}=\Omega^{hs}_z dz + \Omega^{hs}_+ dx^+ +\Omega^{hs}_- dx^-.
\fe 
Modulo the equation of motion, the variation of the action (\ref{csbulk}) is
\ie
\delta S_{hs} = -\int d\psi_1 \int dx^{+}dx^{-}{1\over z^2}\left[\left(\Omega^{hs}_{+},\delta \Omega^{hs}_{-}\right) -\left(\Omega^{hs}_{-},\delta \Omega^{hs}_{+}\right) \right],
\fe
which, however, is non-vanishing under the boundary condition (\ref{RHBC Omega}). To cancel it, we add a boundary term to the action:
\ie\label{hs b action}
S_{hs,b}=-\int d\psi_1 \int dx^{+}dx^{-}{1\over z^2}\psi_1\left(\Omega^{hs}_{+}, \Omega^{hs}_{-}\right),
\fe
whose variation is
\ie
\delta S_{hs,b}=-\int d\psi_1 \int dx^{+}dx^{-}{1\over z^2}\psi_1\left[\left(\Omega^{hs}_{+},\delta \Omega^{hs}_{-}\right)+\left(\Omega^{hs}_{-}, \delta\Omega^{hs}_{+}\right)\right].
\fe
Hence, the variation of the total action $S_{hs}+S_{hs,b}$ is
\ie
\delta S_{hs} + \delta S_{hs,b}=- \int d\psi_1 \int dx^{+}dx^{-}{1\over z^2}\left[(1+\psi_1)\left(\Omega^{hs}_{+},\delta \Omega^{hs}_{-}\right) -(1-\psi_1)\left(\Omega^{hs}_{-},\delta \Omega^{hs}_{+} \right)\right].
\fe
which indeed vanishes under the boundary condition (\ref{RHBC Omega}), or equivalently the Dirichlet boundary condition on the components $(1-\psi_1)\Omega^{hs}_{+}$ and $(1+\psi_1)\Omega^{hs}_{-}$.

Since the bulk action (\ref{csbulk}) vanishes on-shell, the only contribution to the two-point function comes from the boundary term (\ref{hs b action}). Evaluating the boundary integral (\ref{hs b action}) using the higher spin boundary-to-bulk propagators, we obtain the two point function of higher spin currents:
\ie
\vev{J_s(x_1)J_s(x_2)}&=\int d^2 x{1\over z^2}4\pi (\partial_{y^2})^{2s-2}z^{2-s}\delta^2(x-x_1){(2s-1)(y^2)^{2s-2}z^s\over  (x^- - x^-_2)^{2s}}
\\
&=4\pi {(2s-1)! \over  (x^-_{12})^{2s}}.
\fe
This is indeed the structure expected from conformal invariance. 
%To make the two point function coefficients positive definite, we may normalize the higher spin currents by defining $j_s=iJ_s$ for $s$ odd and $j_s=J_s$ for $s$ even. Then, we have
%\ie
%\vev{j_s(x_1)j_s(x_2)}=4\pi (2s-1)(2s-2)! {1\over  (x^-_{12})^{2s}}.
%\fe

\section{Three point functions}

\subsection{The second order equation for the scalars}

To extract the cubic couplings in the bulk Lagrangian, or the three point correlation function of boundary operators, we need to express the master fields in terms of the physical fields and expand the equations of motion to quadratic order. For the purpose of studying three point functions involving the scalars, it suffices to work with the equations for the master field $B$, to the second order. They are
\ie
&d_z B^{(2)} = -[S^{(1)},B^{(1)}]_*,
\\
&D_0 B^{(2)} = -[ W^{(1)},B^{(1)}]_*.
\fe
Decomposing $W^{(1)},B^{(1)},B^{(2)}$ as in (\ref{decomposeWB}), and restricting the second equation at $z=0$, we obtain
\ie
&d_z B'^{(2)} = -[S^{(1)},\psi_2 C^{(1)}_{mat}]_*,
\\
&D_0C^{(2)}=-[W_0,B'^{(2)}]_*\big|_{z=0}-[W'^{(1)},\psi_2C^{(1)}_{mat}]_*\big|_{z=0}
\\
&~~~~~~~~~~~~~-[\Omega^{hs},\psi_2C^{(1)}_{mat}]_*-[\psi_2\Omega^{sc},\psi_2C^{(1)}_{mat}]_*.
\fe
We remind the reader that $C^{(1)}=C^{(1)}_{aux}+\psi_2 C^{(1)}_{mat}$ and $\Omega^{(1)}= \Omega^{hs}+\psi_2 \Omega^{sc}$, and we have set $C^{(1)}_{aux}=0$. The $S^{(1)}$ and $W'^{(1)}$ are linear in $\psi_2$, and the first equation implies $B'^{(2)}$ is independent of $\psi_2$. Decomposing $C^{(2)}$ in a similar way as $C^{(2)}(x|y,\psi)=C^{(2)}_{aux}(x|y,\psi_1)+\psi_2 C^{(2)}_{mat}(x|y,\psi_1)$, we obtain the second order equation for the scalars:
\ie
D_0 \psi_2 C^{(2)}_{mat}=-[\Omega^{hs},\psi_2C^{(1)}_{mat}]_*,
\fe
or more explicitly
\ie\label{2nd scalar equation}
D_0\psi_2C^{(2)}_{mat}=-\psi_2[\Omega^{even},C^{(1)}_{mat}]_*+\psi_2\psi_1\{\Omega^{odd},C^{(1)}_{mat}\}_*,
\fe
where $\Omega^{even}$ and $\Omega^{odd}$ are the components in the decomposition $\Omega^{hs}=\Omega^{even}+\psi_1\Omega^{odd}$.

We further decompose $C^{(2)}_{mat}$ as $C^{(2)}_{mat}(y)=\sum^{\infty}_{n=0} C^{(2),n}_{mat}{}_{\A_1\cdots\A_n}y^{\A_1}\cdots y^{\A_n}$, and specialize (\ref{2nd scalar equation}) to the case $n=0,2$.
\ie
&\partial_\m C^{(2),0}_{mat}{}-4\psi_1(e_{0\m})^{\af\beta}C^{(2),2}_{mat}{}_{\af\beta}=U^0_\m,
\\
&\nabla_\m C^{(2),2}_{mat}{}_{\af\beta}-2\psi_1(e_{0\m})_{\af\beta}C^{(2),0}_{mat}{}-24\psi_1(e_{0\m})^{\gamma\delta}C^{(2),4}_{mat}{}_{\gamma\delta\af\beta}=U^2_{\m|\af\beta},
\fe
where $U^{0}_{\m}$ and $U^{2}_{\m|\A_1\A_2}$ are the first two coefficient of the $y$-expansion of the RHS of (\ref{2nd scalar equation}). After some simple manipulations, it follows that
\ie\label{2nd scalar}
(\Box-m^2) C^{(2),0}_{mat}=\nabla_\m U^{0,\m} + 4\psi_1 (e^\m_0)^{\af\beta}U^{2}_{\m|\af\beta}.
\fe
The RHS is calculated in terms of the first order fields in Appendix B.2. The resulting the second order equation for the scalars can be written in the form
\ie\label{2orderse}
(\Box-m^2) C^{(2),0}_{mat}=\sum^\infty_{s=2}C^{(1),2s-2}_{mat}(\partial_y)\Xi_s(y),
\fe
where $\Xi_s(y)$ is expressed in terms of the higher spin fields as
\ie
\Xi_s(y) &= 8 \left[  \chi_{odd}^{(s),+}(y) + (2s-2)(2s-1) \chi_{odd}^{(s),-}(y) \right]
\\
&~~+ 32\psi_1 \left[ {1\over (2s-1)} \nabla^- \chi_{odd}^{(s),+}(y) - ( 2s-2) \nabla^+ \chi_{odd}^{(s),-}(y) \right].
\fe

\subsection{The three point function}

The boundary-to-bulk propagator for the higher spin gauge field satisfying the generalized Brown-Henneaux boundary condition (\ref{generalized Brown-Henneaux}) is determined by the boundary behavior of the gauge transformation (\ref{boundary gauge parameter lambda}). The latter is enough for us to compute the three point function of one higher spin gauge field and two scalars. Suppose the cubic action of a higher spin gauge field and two scalars is of the form as the higher spin gauge field couples to the higher spin current, i.e.
\ie\label{action1}
\int  {d^2x}\left({dz\over z^3}\right)\Phi^s_{\m_1\cdots\m_s}T_s^{\m_1\cdots\m_s}
\fe
 where the higher spin current $T_s^{\m_1\cdots\m_s}$ is a quadratic function of the scalar and its derivatives. Since the boundary to bulk propagator for high spin gauge field can be written in a ``pure gauge" form: $\Phi^s_{\m_1\cdots\m_s}= \nabla_{(\ul{\m_1}}\eta^s_{\ul{\m_2\cdots\m_s})}$, and the higher spin current is conserved: $\nabla_\m T_s^{\m\m_1\cdots\m_{s-1}}=0$, we have
\ie\label{intbypt}
&\int  {d^2x}\left({dz\over z^3}\right)\nabla_{\m_1}\eta^s_{\m_2\cdots\m_s}T_s^{\m_1\cdots\m_s}
\\
&=\int  {d^2x}dz\partial_{\m_1}\left({1\over z^3}\eta^s_{\m_2\cdots\m_s}T_s^{\m_1\cdots\m_s}\right)
\\
&=-\lim_{z\rightarrow0} {1\over z^3}\int  {d^2x}\,\eta^s_{\m_2\cdots\m_s}T_s^{z\m_2\cdots\m_s},
\fe
which only depends on the boundary behavior of the gauge parameter at $z\rightarrow 0$.

The RHS of the second order equation (\ref{2orderse}) gives the variation of the cubic action with respect to the scalar up to some possible boundary terms.
\ie\label{vaction}
\delta S=\int d\psi_1\int {d^2x dz\over z^3}\, \psi_1\delta C^{(1),0}_{mat}\sum^\infty_{s=2}C^{(1),2s-2}_{mat}(\partial_y)\Xi_s(y).
\fe
While it is possible to recover the cubic part of the action from (\ref{vaction}), in the form (\ref{action1}), we will not need it for the computation of the three point function. The tree level three point function is computed by varying the bulk action with respect to three sources inserted on the boundary, and so it suffices to work with (\ref{vaction}) directly, by evaluating it on the boundary-to-bulk propagators for the higher spin gauge field and scalars.
This computation is performed explicitly in Appendix B.3. The resulting three point function of one higher spin current and two scalars is:
\ie\label{thtmp}
\vev{\overline{\cO}(x_1)\cO(x_2)J_s(x_3)}=-4\pi (s+\widetilde\psi_1(s-1))\Gamma(s) {1\over |x_{12}|^{2+\widetilde\psi_1}}\left({x_{12}^-\over x^-_{13} x^-_{23}}\right)^s.
\fe
Here ${\cal O}$ and $\overline{\cal O}$ are dual to $C_{even}+C_{odd}$ and $C_{even}-C_{odd}$ respectively. They have scaling dimension $\Delta_+={3\over 2}$ or $\Delta_-={1\over 2}$ depending on the choice of boundary condition, corresponding to $\widetilde\psi_1=1$ or $\widetilde\psi_1=-1$.
The position dependent factor on the RHS of (\ref{thtmp}) is fixed by conformal symmetry. The only nontrivial data here are contained in the overall coefficient, which is unambiguous given the normalization of the currents. These will be compared to representations of the $W_N$ algebra in the 't Hooft limit in the next section.

\section{The dual CFT}

\subsection{The proposal}

It has been proposed in \cite{Gaberdiel:2010pz} that Vasiliev's higher spin-matter system (more precisely, a version of this theory with four real massive scalars) is dual to the $W_N$ minimal model, which can be realized by the coset model
\ie
{SU(N)_k\oplus SU(N)_1\over SU(N)_{k+1}}.
\fe
This CFT has a 't Hooft-like scaling limit, in which $N$ is taken to be large while keeping the 't Hooft coupling
\ie
\lambda = {N\over N+k}
\fe
to be fixed. In the infinite $N$ limit, $\lambda$ becomes a continuous parameter, in the range $0<\lambda <1$. It is proposed that $\lambda$ is mapped to the parameter $\nu$ that label $AdS_3$ vacua, with the identification $\lambda ={1\over 2} (1\pm \nu)$. The undeformed, $\nu=0$ vacuum we have been considering so far would be mapped to the $\lambda=1/2$ case. In the 't Hooft limit, ``basic primaries" of (left plus right) scaling dimension $\Delta_\pm = {1\pm \lambda}$ are mapped to the massive scalars in the bulk, whereas all other primaries are found in the OPEs of the basic primaries, their duals interpreted as bound states in the bulk.

A puzzle with this proposal is the existence of low lying primary operators in the coset CFT, whose dimension scale like $1/N$ and form a discretuum in the 't Hooft limit. This has been further addressed in \cite{GGHR}. It is unclear how to interpret the dual of such states in the bulk.

Here we put forward a different proposal, namely that the Vasiliev higher spin-matter system, involving only two real massive scalars in the bulk, is dual to a subsector of the $W_N$ minimal model, generated by the two basic primaries of either dimension $\Delta_+$ or dimension $\Delta_-$, depending on the boundary condition for the bulk scalar field. This subsector has closed OPE and is consistent as a CFT on the sphere, though not on Riemann surfaces of nonzero genus, as it is not modular invariant. Hence, we believe that the bulk Vasiliev's system is nonperturbatively incomplete, though makes sense perturbatively to all orders in its coupling constant (i.e. $1/N$).

In a similar manner, we further propose that the ``minimal" Valisiev's system, obtained via the truncation to fields invariant under the $\iota$-involution (\ref{trunc}), is dual to a subsector of the orthogonal group version of the coset model,\footnote{The bulk gauge group of the minimal Vasiliev theory, in the Chern-Simons language, when truncated to a finite (even) spin $N$, is $Sp(N,{\mathbb R})\times Sp(N,{\mathbb R})$. In mapping representations of the higher spin algebra in the bulk to primaries labeled by representations of the affine Lie algebra of the minimal model, a transpose on the Young tableaux is involved \cite{GGHR}. This suggests that the dual minimal model is based on $SO$ rather than $Sp$ coset. We thank T. Hartman for pointing this out. Note also that the analogous $Sp$ coset construction would not give a $W_N$ minimal model; its primaries are generally not labelled simply by a pair of representations, but a triple of representations \cite{Bouwknegt:1992wg}. }
\ie
{SO(N)_k\oplus SO(N)_1\over SO(N)_{k+1}}.
\fe
Because $SO(N)$ has only even degree Casimir invariants, the coset model contains only the even spin currents. The real scalar in the ``minimal" Valisiev's system is dual to one of the real basic primary operators, either $(\Box;0)$ or $(0;\Box)$, depending on the choice of boundary condition for the bulk scalar.

\subsection{$W_N$ currents and primaries}

Let $K^a(z)$ be the currents of the $SU(N)_k$ current algebra, and $J^a(z)$ the currents of $SU(N)_1$. Our convention for the group generators of $SU(N)$ is such that
\ie
{\rm Tr}(T^a T^b) = -\delta^{ab}
\fe
where ${\rm Tr}$ is taken in the fundamental representation. The cubic symmetric tensor is defined to be
\ie
d^{abc} = -i {\rm Tr} (\{T^a, T^b\} T^c).
\fe
The $SU(N)_k$ currents, for instance, are normalized with the OPE
\ie
K^a(z) K^b(0) \sim -{k\over z^2} \delta^{ab} + f^{abc} {K^c(0)\over z},
\fe
where $f^{abc} = -{\rm Tr}([T^a, T^b] T^c)$. The spin-2 current, i.e. the stress-energy tensor of the coset model constructed out of the Sugawara tensors, is given by
\ie
&T(z) = W^2(z) \\
&= -{1\over 2(N+k)}:K^a K^a: - {1\over 2(N+1)} :J^a J^a: + {1\over 2(N+k+1)}:(K^a+J^a)(K^a+J^a):
\fe
The spin-3 current $W^3$, in the 't Hooft limit, is written as
\ie
W^3(z) = d_{abc} \left[ {3\lambda^2\over (1-\lambda)(2-\lambda)} :K^a K^b J^c: - {3\lambda\over 1-\lambda} :K^a J^b J^c: + :J^a J^b J^c: \right].
\fe
The normalization is such that the two point function of $W^3$ is given by
\ie
\langle W^3(z) W^3(0)\rangle = -6{(1+\lambda)(2+\lambda)\over (1-\lambda)(2-\lambda)} N^5 + (1/N~{\rm corrections}).
\fe
One may also construct higher spin-$s$ currents out of the product of $s$ $K^a$ and $J^a$'s, subject to the constraint that $W^s$ is primary with respect to the diagonal $SU(N)_{k+1}$. This is rather cumbersome, which we shall not attempt here. Nonetheless, we will perform one unambiguous check with the spin-3 current.

Let us now turn to the primary operators with respect to the $W_N$ algebra. These are labelled by three representations of $SU(N)$, $(\rho,\mu;\nu)$; here $\rho, \mu, \nu$ are the height weight vectors of the respective representations, subject to the condition that the sum of the Dynkin labels is less than or equal to the level, and the constraint that $\rho+\mu-\nu$ lies in the root lattice of $SU(N)$. Further, it follows from the second $SU(N)$ being at level 1 that $\mu$ is uniquely determined given $\rho$ and $\nu$. Following the notation of \cite{Gaberdiel:2010pz}, the primaries are labeled by $(\rho;\nu)$. We consider the diagonal modular invariant, by pairing up identical representations on the left and right moving sectors. The basic primaries are:
\ie
&{\cal O}_+ = (\Box;0)\otimes (\Box;0),~~~~~\overline{\cal O}_+ = (\overline\Box;0)\otimes (\overline\Box;0),\\
&{\cal O}_- = (0;\Box)\otimes (0;\Box),~~~~~\overline{\cal O}_- = (0;\overline\Box)\otimes (0;\overline\Box).
\fe
In the 't Hooft limit, ${\cal O}_\pm$ (and $\overline{\cal O}_\pm$) have conformal weight $h_\pm=\bar h_\pm = {1\pm \lambda\over 2}$.

Our proposal is that with the $\Delta_+$ boundary condition, the two real massive scalars in the bulk, combined into a complex scalar $C_{even}+C_{odd}$, is dual to ${\cal O}_+$, while its complex conjugate $C_{even}-C_{odd}$ is dual to $\overline{\cal O}_+$. According to the fusion rule, the OPEs of ${\cal O}_+$ and $\overline{\cal O}_+$ involve only primaries labeled by the representations of the form $(R;0)$. In particular, the operators ${\cal O}_-, \overline{\cal O}_-$ and the low lying primaries of the form $(R;R)$ do not appear in the OPEs of ${\cal O}_+$ and $\overline{\cal O}_+$. Thus, this subsector of the CFT closes on the sphere.

Alternatively, with $\Delta_-$ boundary condition imposed on the bulk scalar, we propose the dual to the be subsector generated by ${\cal O}_-$ and $\overline{\cal O}_-$. 

\subsection{A test on the three point function}

The spin-3 current acts on the basic primaries ${\cal O}_\pm$ as
\ie
&W^3_0|{\cal O}_-\rangle = C_\Box |{\cal O}_-\rangle,
\\
&W^3_0|{\cal O}_+\rangle = - C_\Box {(1+\lambda)(2+\lambda)\over (1-\lambda)(2-\lambda)} |{\cal O}_+\rangle,
\fe
where $C_\Box$ is the cubic Casimir for the fundamental representation, given by
\ie
&C_\Box|\Box\rangle = d_{abc}J_0^a J_0^b J_0^c |\Box\rangle,~~~~C_\Box=iN^2
\fe
in our convention. The three point function $\langle{\cal O}_\Delta(z_1) {\cal O}_\Delta(z_2)W^s(z_3)\rangle$ is determined by conformal symmetry to be of the form
\ie
{A(s) \over |z_{12}|^{2\Delta}} \left( {z_{12}\over z_{13}z_{23}} \right)^s.
\fe
We will write $\langle{\overline{\cO}}_\Delta{\cal O}_\Delta W^s\rangle \equiv A(s)$ for the coefficient. It follows from the action of $W^3_0$ on the primary states that
\ie
\langle{\overline{\cO}}_+{\cal O}_+ W^3\rangle = - iN^2 {(1+\lambda)(2+\lambda)\over (1-\lambda)(2-\lambda)}, ~~~~~ \langle{\overline{\cO}}_-{\cal O}_- W^3\rangle = iN^2.
\fe
If we define $J^{(s)}$ to be the spin-$s$ current with normalized two-point function, namely $\langle J^{(s)}(z) J^{(s)}(0)\rangle = z^{-2s}$ (this fixes $J^{(s)}$ up to a sign), then we have
\ie\label{oojcft}
&\langle{\overline{\cO}}_+{\cal O}_+ J^{(2)}\rangle = N^{-{1\over 2}} \sqrt{1+\lambda\over 2(1-\lambda)}, ~~~~ \langle{\overline{\cO}}_-{\cal O}_- J^{(2)}\rangle = N^{-{1\over 2}}\sqrt{1-\lambda\over 2(1+\lambda)},
\\
&\langle{\overline{\cO}}_+{\cal O}_+ J^{(3)}\rangle = N^{-{1\over 2}} \sqrt{(1+\lambda)(2+\lambda)\over 6(1-\lambda)(2-\lambda)}, ~~~~ \langle{\overline{\cO}}_-{\cal O}_- J^{(3)}\rangle = -N^{-{1\over 2}}\sqrt{(1-\lambda)(2-\lambda)\over 6(1+\lambda)(2+\lambda)}.
\fe

From the bulk, we have computed the three point function $\langle {\overline{\cO}}{\cal O}J^{(s)}\rangle$ in the undeformed theory, with the result (after normalizing the spin-$s$ current)
\ie\label{oojbulk}
\langle {\overline{\cO}}_+{\cal O}_+ J^{(s)} \rangle = g \Gamma(s) \sqrt{2s-1\over \Gamma(2s-1)},~~~~
\langle {\overline{\cO}}_-{\cal O}_- J^{(s)} \rangle =  (-)^s g {\Gamma(s) \over \sqrt{\Gamma(2s)}}.
\fe
Here $g$ is the overall coupling constant of the bulk theory. 
%Note that a priori, its identification with $1/\sqrt{N}$ of the dual CFT may come with a different normalization for $\Delta_+$ and $\Delta_-$ boundary conditions. 
This should be compared with the CFT at $\lambda=1/2$. With the identification
\ie
g = {1\over \sqrt{N}},
\fe
we see that (\ref{oojbulk}) precisely agrees with (\ref{oojcft}) at $\lambda=1/2$. (\ref{oojbulk}) then further makes predictions for the three point functions $\langle {\overline{\cal O}}{\cal O}J^{(s)}\rangle$ of spin $s\geq 4$ in the $W_N$ coset CFT, in the 't Hooft limit at $\lambda=1/2$, which remains to be computed directly on the CFT side. Further, it would be very interesting to compute these three point functions in the {\it deformed} bulk theory, i.e. the $AdS_3$ vacua with nonzero $\nu$, which should be mapped to the CFT with 't Hooft parameter away from $\lambda=1/2$. We hope to report on this in future works.

\section{Concluding remarks}

In this paper, we have developed the perturbation theory of Vasiliev's higher spin-matter system in $AdS_3$, to the second order. This allowed us to compute the bulk tree level three point functions, in the undeformed $\nu=0$ vacuum. The result passed a nontrivial test that involves the explicit expression for the spin-3 current in the $W_N$ minimal model (at the special value of 't Hooft coupling $\lambda=1/2$). Our result from the bulk also makes predictions on three point functions involving currents of spin $s\geq 4$ which in principle can be straightforwardly computed (though tedious) in the coset CFT, by constructing the $W_N$ currents out of the spin 1 affine currents, and then taking the 't Hooft limit. 

A natural next step is to move away from the undeformed, $\nu=0$ vacuum, and consider the deformed bulk theory, which should be dual to the CFT away from $\lambda=1/2$. In Appendix C, we have derived the boundary to bulk propagator for the scalar master field in the deformed theory. The computation of correlators using these expressions could be complicated, though at least one can work order by order expanding in $\nu$, which amounts to expanding in $\lambda-{1\over 2}$ in the dual CFT.

Next, one would like to go beyond leading order in $1/N$. The basic primaries in the $W_N$ minimal model have exact scaling dimensions
\ie
& \Delta_+ = 2h(\Box;0) = {N-1\over N} (1+{N+1\over N}\lambda),\\
& \Delta_- = 2h(0;\Box) = {N-1\over N} (1-{N+1\over N+\lambda}\lambda).
\fe
Identifying $\Delta_\pm = 1\pm \sqrt{1+m_\pm^2}$, we see that the renormalized mass of the bulk scalar with the two different boundary conditions are
\ie
& m_+^2 = -\left[\big(1+{\lambda\over N}\big)^2-\lambda^2\right] \left(1-{1\over N^2}\right),
\\
& m_-^2 =  -(1-\lambda^2)\left(1+ {\lambda\over N}\right)^{-2} \left(1-{1\over N^2}\right).
\fe
The bulk scalar propagator depend on the boundary condition ($\Delta_+$ or $\Delta_-$), which presumably leads to the different  renormalized masses $m_+$ and $m_-$ through loop corrections. The difference between $m_+$ and $m_-$, say at order $1/N$, or one-loop in the bulk, can in principle be understood \cite{Hartman:2006,Giombi:2011ya} in terms of the tree level four-point functions, by factorizing the difference in the bulk propagators for the two boundary conditions into the product of boundary-to-bulk propagators. To compute either $m_-^2$ or $m_+^2$ form the bulk, however, requires performing a genuine one-loop computation in Vasiliev's theory. The precise relation between the bulk deformation parameter $\nu$ and the 't Hooft coupling $\lambda$ of the boundary CFT, beyond the leading order in $1/N$, is presumably also regularization dependent.

We proposed that Vasiliev's system is dual to not the entire $W_N$ minimal model CFT, but only a subsector of it, generated by the basic primaries ${\cal O}_+, \overline{\cal O}_+$ and the $W_N$ currents, or the subsector generated by ${\cal O}_-, \overline{\cal O}_-$ and the $W_N$ currents, depending on whether $\Delta_+$ or $\Delta_-$ boundary condition is imposed on the two bulk scalars. These two subsectors close on their OPEs, and lead to consistent $n$-point functions on the sphere. However, they are not modular invariant. From the perspective of the bulk higher spin gravity theory, modular invariance is expected to be restored by gravitational instantons (analytic continuation of BTZ black holes), which are non-perturbative. At the level of perturbation theory, it is consistent that the bulk theory is dual to a subsector of a modular invariant CFT. The duality we are proposing is analogous to the statement that pure gravity in $AdS_3$, at the level of perturbation theory, is dual to the subsector of a CFT involving only Virasoro descendants of the vacuum, i.e. operators made out of products of stress-energy tensors. The latter lead to a consistent set of $n$-point functions on the sphere, though do not give modular invariant genus one partition functions by themselves.

If our proposal is correct, then it suggests that Vasiliev's system is non-perturbatively incomplete, though makes sense to all orders in perturbation theory. One may suspect that solitons, in particular black hole solutions, should be included and could make the theory modular invariant. However, we are not aware of a modular invariant completion of the $\Delta_+$ or $\Delta_-$ subsector of $W_N$ minimal model that requires adding only states/operators whose dimensions scale with $N$ (and are large in the large $N$ limit). The $W_N$ minimal model itself would amount to adding not only states of dimension of order 1, but also a large number of light states whose dimensions go like $1/N$, which seems pathological from the perspective of the bulk theory.

It is clearly of great interest, still, to understand the bulk theory dual to the full $W_N$ minimal model, since the latter is non-perturbative defined and exactly solvable. It is shown in \cite{GGHR} that the descendants of the light states give rise to bound states of the basic primaries, while the light states themselves become null in the infinite $N$ limit. It is unclear how to understand this from the bulk. A possibility is that additional {\it massless} scalars should be added in the bulk theory, with the non-standard boundary condition (so that they are dual to operators of dimension 0 rather than 2, classically). It would be an interesting challenge to construct such a theory in $AdS_3$.

\subsection*{Acknowledgments}

We are grateful to R. Loganayagam and Wei Song for collaborations at the initial stage of this work, to Suvrat Raju, Rajesh Ropakumar and especially Tom Hartman for very useful discussions, and to the authors of \cite{GGHR} for sharing a draft of their paper. X.Y. would like to thank the organizers of Indian Strings Meeting 2011, Tata Institute of Fundamental Research, Berkeley Center for Theoretical Physics, Fields Institute, and Perimeter Institute for their hospitality during the course of this work. C.C. would like to thank the organizers of Taiwan String Theory Workshop 2011 for their hospitality during the course of this work. We would like to thank the organizers of the higher spin workshop at Simons Center for Geometry and Physics for providing the opportunities for many stimulating discussions.
This work is supported in part by the Fundamental Laws Initiative Fund at Harvard University, and by NSF Award PHY-0847457.

\appendix

\section{Linearizing Vasiliev's equations}

\subsection{Derivation of the scalar boundary to bulk propagator}
In this subsection, we study the linearized equations (\ref{C}), and solve for the boundary-to-bulk propagator for the master field $C^{(1)}$. 

Decomposing the $C^{(1)}$ as in (\ref{Cdam}) the equation $D_0 C^{(1)} =0$ is written as
\ie\label{CaCmeq}
&d_x C^{(1)}_{aux}+4(w_0^{\af\beta}y_\af{\partial \over\partial y^\beta}+\psi_1e_0^{\af\beta}y_\af{\partial \over\partial y^\beta})C^{(1)}_{aux}=0\\
&d_x C^{(1)}_{mat}+4w_0^{\af\beta}y_\af{\partial \over\partial y^\beta}C^{(1)}_{mat}-2\psi_1e_0^{\af\beta}(y_\af y_\beta+{\partial^2\over\partial y^\af\partial y^\beta})C^{(1)}_{mat}=0
\fe
Expand $C^{(1)}_{mat/aux}(x|y,\psi_i)$ as in (\ref{Cdy}), we write the first equation of (\ref{CaCmeq}) as
\ie
\partial_\m C_{aux}^{(1),n}{}_{\af_1\cdots\af_n}-4n(w_{0\m})_{(\underline\A_1}{}^{\beta}C_{aux}^{(1),n}{}_{\beta\underline{\af_2\cdots\af_{n}})}-4n\psi_1(e_{0\m})_{(\underline\A_1}{}^{\beta}C_{aux}^{(1),n}{}_{\beta\underline{\af_2\cdots\af_{n}})}=0.
\fe
Contracting this equation with $(e^\m_0)_{\gamma\delta}$, and symmetrizing the indices $(\gamma\delta\af_1\cdots\af_n)$, we get
\ie
\nabla_{(\underline{\gamma\delta}}C_{aux}^{(1),n}{}_{\underline{\af_1\cdots\af_n})}=0~~\text{with}~~\nabla_{\A\B} = e^\mu_{\A\B}\nabla_\mu,
\fe
which means that $C^{(1)}_{aux}$ carries no propagating degree of freedom. We can simply set $C^{(1)}_{aux}=0$.

The second equation of (\ref{CaCmeq}) can be written as
\ie
&\partial_\m C^{(1),n}_{mat}{}_{\af_1\cdots\af_n}-4n(w_{0\m})_{(\underline\A_1}{}^{\beta}C^{(1),n}_{mat}{}_{\beta\underline{\af_2\cdots\af_{n}})}
\\
&~~-2\psi_1(e_{0\m})_{(\underline{\A_1\A_2}}C_{mat}^{(1),n-2}{}_{\underline{\af_3\cdots\af_{n}})}-2(n+2)(n+1)\psi_1(e_{0\m})^{\af\beta}C^{(1),n+2}_{mat}{}_{\A\B\A_1\cdots\A_{n}}=0.
\fe
Or contracting this equation with $(e_0^\m)_{\A\B}$ gives
\ie\label{seqn}
&\nabla_{\A\B} C_{mat}^{(1),n}{}_{\af_1\cdots\af_n}+{1\over 16}\psi_1\epsilon_{(\A(\ul{\A_1}}\epsilon_{\B)\ul{\A_2}} C_{mat}^{(1),n-2}{}_{\underline{\af_3\cdots\af_{n}})}
\\
&~~~~+{1\over 16}(n+2)(n+1)\psi_1C_{mat}^{(1),n+2}{}_{\A\B\A_1\cdots\A_{n}}=0.
\fe
This equation is in a reducible representation of the permutation group of permuting the indices. To simplify the equation, we decompose it into irreducible representations by contracting with the tensor $\epsilon^{\A\B}$ or symmetrizing all the indices. First, contracting (\ref{seqn}) with $\epsilon^{\A\A_1}$ gives
\ie\label{seqna}
\nabla^{\A}{}_{\B} C_{mat}^{(1),n}{}_{\af\A_2\cdots\af_n}-{n+1\over 16n}\psi_1\epsilon_{\B(\ul{\A_2}} C_{mat}^{(1),n-2}{}_{\underline{\af_3\cdots\af_{n}})}=0.
\fe
Contracting (\ref{seqna}) with $\epsilon^{\B\A_2}$ gives
\ie\label{seqnb}
\nabla^{\A\B} C_{mat}^{(1),n}{}_{\A\B\af_3\cdots\af_n}+{n+1\over 16(n-1)}\psi_1 C_{mat}^{(1),n-2}{}_{\af_3\cdots\af_{n}}=0.
\fe
Next, we want to symmetrize the indices of equations (\ref{seqn}), (\ref{seqna}), and (\ref{seqnb}). It is convenient to reintroduce the auxiliary $y^\A$-variable. By contracting the indices of the equations (\ref{seqn}), (\ref{seqna}), and (\ref{seqnb}) with the $y^{\A}$'s which automatically symmetrizes all the indices, we obtain
\ie\label{eqnscalar}
&\nabla^+ C_{mat}^{(1),n}(y)-{1\over 16}(n+2)(n+1)\psi_1C_{mat}^{(1),n+2}(y)=0,
\\
&\nabla^0 C_{mat}^{(1),n}(y)=0,
\\
&\nabla^{-} C_{mat}^{(1),n}{}(y)-{1\over 16}(n+1)n\psi_1 C_{mat}^{(1),n-2}(y)=0,
\fe
where
\ie
C_{mat}^{(1),n}(y)=C_{mat}^{(1),n}{}_{\af_1\cdots\af_n}y^{\A_1}\cdots y^{\A_n}
\fe
which is the degree $n$ homogeneous polynomial in the Taylar expansion of the matter field $C^{mat}(y)$, and we define the operators
\ie
\nabla^+ = (y{\slash\!\!\!\nabla}y),~~~~\nabla^0 = (y{\slash\!\!\!\nabla} \partial_y), ~~~~
\nabla^- = (\partial_y{\slash\!\!\!\nabla}\partial_y) .
\fe
They obey commutation relations
\ie\label{y-algebra}
& [\nabla^0,\nabla^\pm] = 0,\\
& [\nabla^+,\nabla^-] = {{\cal N}+1\over 16} \Box_{AdS} - {{\cal N}({\cal N}+2)({\cal N}+1)\over 64},\\
& (\nabla^0)^2 = \nabla^+\nabla^- + {{\cal N}^2\over 64}\Box_{AdS} + {{\cal N}^2({\cal N}+2)\over 128}.
\fe
with ${\cal N} = y\partial_y$ and $\Box_{AdS}\equiv -32\nabla_{\A\B}\nabla^{\A\B}$ where $\nabla_{\A\B}$ is defined to act covariantly both on explicit spinor indices as well as on indices contracted with $y^\A$. Iterating the first equation of (\ref{eqnscalar}), we get
\ie\label{iterate}
C_{mat}^{(1),2s}(y)={1\over (2s)!}(16\psi_1\nabla^+)^s C_{mat}^{(1),0}.
\fe
Since $C^{(1)}_{mat}(y)$ is an even function in $y^\A$, it is totally determined by its lowest component $C_{mat}^{(1),0}$ via the above relation. After some simple manipulations of (\ref{eqnscalar}) using (\ref{y-algebra}), we derive
\ie
\Box_{AdS} C_{mat}^{(1),n}&=-{1\over 4}(3+n(n+2))C_{mat}^{(1),n}.
\fe
For $n=0$, the equation gives the usual Klein-Gordon equation on $AdS_3$, (\ref{Klein-Gordon}).
The higher components $C_{mat}^{(1),n}$ are determined by $C_{mat}^{(1),0}$ through the linearized equations of motion.

The equation (\ref{Klein-Gordon}) is solved by scalar boundary to bulk propagator $C^{mat,0} = K(x,z)^\Delta$ for $\Delta = 3/2$ or $\Delta = 1/2$, where $K(x,z)\equiv {z\over x^2 + z^2}$. It is convenient to introduce another auxiliary variable $\tilde \psi_1$, satisfying $\tilde \psi_1^2=1$, to label the different boundary conditions, so that $\Delta = 1 + \tilde\psi_1 /2$. The $(\nabla^+)^s$ acting on $K^\Delta$ is
\ie\label{nsK}
(\nabla^+)^s K^\Delta={1\over 8^s}\left(\prod^s_{j=1}(\Delta +j-1)\right)(y\Sigma y)^s K^\Delta,
\fe
and using (\ref{iterate}), we obtain
\ie
C_{mat}^{(1)}(y)=\left(1+\psi_1{1+\tilde \psi_1\over 2} y\Sigma y\right)e^{{\psi_1\over 2}y\Sigma y}K^{1+{\tilde\psi_1\over 2}},
\fe
where $\Sigma=\sigma^z - {2z\over x^2}\sigma^\m x^\m $.

\subsection{The linearized higher spin equations}
In this subsection, we study the linearized equations (\ref{Omega}),(\ref{W'}),(\ref{S}), and rewrite them as the (linearized) Chern-Simons equation and Fronsdal equation by eliminating all the auxiliary degrees of freedom.

The (\ref{W'}) and (\ref{S}) imply that $W'$ is solved in terms of $S$ and further in terms of $C^{(1)}_{mat}$; hence, in particular, it is linear in $\psi_2$. Decomposing $\Omega^{(1)}$ as in (\ref{Odhssc}), the linearized equations are written in (\ref{linom}).

The linearized gauge transformations act by
\ie
&\delta W^{(1)} = d_x \epsilon + [W_0,\epsilon]_*,
\\
&\delta S^{(1)} = d_z \epsilon.
\fe
Let us restrict to gauge transformations that leave $S^{(1)}$ invariant, namely $\epsilon =\lambda(x|y,\psi_1)+\psi_2 \rho(x|y,\psi_1)$, where $\lambda(x|y,\psi_1)$ and $\rho(x|y,\psi_1)$ transform $\Omega^{hs}$ and $\Omega^{sc}$ independently at the linearized level. Their actions are
\ie\label{hs gt}
&\delta \Omega^{sc} = d_x \rho + \psi_2 [W_0, \psi_2\rho]_* = \nabla_x \rho - \psi_1\{e_0, \rho \}_*,
\\
&\delta \Omega^{hs} = d_x \lambda + [W_0, \lambda]_* = \nabla_x \lambda + \psi_1[e_0, \lambda ]_*.
\fe

We show that $\Omega^{sc}$ contains no dynamical degrees of freedom. First consider the homogeneous part of the equation,
\ie
\tilde D_0 \Omega^{sc} = 0,
\fe
or more explicitly,
\ie
\nabla_x  \Omega^{sc}(x|y,\psi_1) - \psi_1 e_0(x|y) \wedge_*  \Omega^{sc}(x|y,\psi_1) + \psi_1  \Omega^{sc}(x|y,\psi_1) \wedge_*  e_0(x|y) = 0.
\fe
We have emphasized the wedge product between 1-forms, so the last terms involve the $*$-anti-commutator of the components of $e_0$ and $\Omega^{sc}$. Expand $\Omega^{sc}$ as
\ie
\Omega^{sc}(x|y,\psi_1) = dx^\mu \sum_{n=0}^\infty \Omega^{sc,n}_{\mu|\A_1\cdots\A_n}(x|\psi_1) y^{\A_1} \cdots y^{\A_n}.
\fe
In components, the homogeneous equation for $\Omega^{sc}$ is written as
\ie
\nabla_{[\mu} \Omega^{sc,n}_{\nu]|\A_1\cdots\A_n} -2 \psi_1 (e_{0[\mu})_{(\underline{\A_1\A_2}} \Omega^{sc,n-2}_{\nu] | \underline{\A_3\cdots\A_n})} - 2(n+2)(n+1) \psi_1 (e_{0[\mu})^{\A\B} \Omega^{sc,n+2}_{\nu]| \A\B\A_1\cdots\A_n} = 0.
\fe
Converting $\mu,\nu$ into spinor indices, we obtain
\ie\label{eqnaa}
\nabla_{(\A}{}^\C \Omega^{sc,n}_{\B)\C|\A_1\cdots\A_n}-2 \psi_1 e_{\A}{}^\C{}_{|(\underline{\A_1\A_2}} \Omega^{sc,n-2}_{\B)\C | \underline{\A_3\cdots\A_n})} - 2(n+2)(n+1) \psi_1 e_{(\A}{}^{\C|\D\tau} \Omega^{sc,n+2}_{\B)\C| \D\tau\A_1\cdots\A_n} = 0.
\fe
where
\ie
e_{\A\B|\C\D} \equiv (e_0^\mu)_{\A\B} (e_{0\mu})_{\C\D} = -{1\over 64} (\epsilon_{\A\C}\epsilon_{\B\D} + \epsilon_{\A\D}\epsilon_{\B\C}).
\fe
We can write (\ref{eqnaa}) as
\ie\label{eqnaf}
\nabla_{(\A}{}^\C \Omega^{sc,n}_{\B)\C|\A_1\cdots\A_n}-{1\over 16} \psi_1 \epsilon_{(\A(\underline{\A_1}} \Omega^{sc,n-2}_{\B)\underline{\A_2} | \underline{\A_3\cdots\A_n})} +{1\over 16}(n+2)(n+1) \psi_1 \epsilon^{\C\D} \Omega^{sc,n+2}_{\C(\A|\B) \D\A_1\cdots\A_n} = 0.
\fe
In components, the gauge transformation (\ref{hs gt}) for $\Omega^{sc}$ can be written as
\ie
\delta\Omega^{sc,n}_{\m|\A_1\cdots\A_n} = \nabla_\m \rho^n_{\A_1\cdots\A_n} - 2\psi_1 (e_\m)_{(\A_1\A_2}\rho^{n-2}_{\A_3\cdots\A_n)} - 2(n+2)(n+1)\psi_1 (e_\m)^{\A\B}\rho^{n+2}_{\A\B\A_1\cdots\A_n},
\fe
or
\ie
\delta\Omega^{sc,n}_{\A\B|\A_1\cdots\A_n} = \nabla_{\A\B} \rho^n_{\A_1\cdots\A_n} + {1\over 16}\psi_1 \epsilon_{(\A(\ul{\A_1}}\epsilon_{\B)\ul{\A_2}}\rho^{n-2}_{\ul{\A_3\cdots\A_n})} + {1\over 16}(n+2)(n+1)\psi_1 \rho^{n+2}_{\A\B\A_1\cdots\A_n}.
\fe
Decomposing $\Omega^{sc,(n)}_{\A\B|\A_1\cdots\A_n}$ as
\ie
\Omega^{sc,(n)}_{\A\B|\A_1\cdots\A_n} = \zeta^{n,+}_{\A\B\A_1\cdots\A_n} + \epsilon_{(\A_1(\underline{\A}}\zeta^{n,0}_{\underline{\B})\A_2\cdots\A_n)} + \epsilon_{(\underline{\A}(\A_1}\epsilon_{\underline{\B})\A_2} \zeta^{n,-}_{\A_3\cdots\A_n)},
\fe
we find that $\zeta^{n,+}$ and $\zeta^{n,-}$ can be gauged away by $\rho^{n+2}$ and $\rho^{n-2}$. Furthermore, by symmetrizing $(\A\B\A_1\cdots\A_m)$ of (\ref{eqnaf}), $\zeta^{n,0}$ can be fully determined by $\zeta^{n,+}$ and $\zeta^{n,-}$.

Now let us turn to the higher spin fields, $\Omega^{hs}$. Their linearized equations are written more explicitly as
\ie
\nabla_x \Omega^{hs} + e_0\wedge_* \Omega^{hs} + \Omega^{hs}\wedge_* e_0 = 0,
\fe
or in components,
\ie
\nabla_{[\mu} \Omega^{hs,n}_{\nu] | \A_1\cdots\A_n} - 4n\psi_1 (e_{0[\mu})_{(\underline{\A_1}}{}^{\B} \Omega^{hs,n}_{\nu]|\B\underline{\A_2\cdots\A_n})} = 0.
\fe
Replacing $[\mu\nu]$ with spinor indices, we can write it as
\ie
\nabla_{(\A}{}^\C \Omega^{hs,n}_{\B)\C | \A_1\cdots\A_n} - 4n\psi_1 e_{(\A}{}^\C{}_{|(\underline{\A_1}}{}^{\D} \Omega^{hs,n}_{\B)\C|\D\underline{\A_2\cdots\A_n})} = 0,
\fe
or
\ie\label{eqnab}
\nabla_{(\A}{}^\C \Omega^{hs,n}_{\B)\C | \A_1\cdots\A_n} + {1\over 16}n\psi_1 \epsilon_{(\underline{\A_1}(\A} \Omega^{hs,n}{}_{\B)}{}^\C{}_{|\C\underline{\A_2\cdots\A_n})} - {1\over 16} n \psi_1 \Omega^{hs,n}_{(\A(\underline{\A_1}|\B)\underline{\A_2\cdots\A_n)}} = 0.
\fe
Let us decompose $\Omega^{hs,(n)}_{\A\B|\A_1\cdots\A_n} $ into the irreducible representation of the permutation group of permuting the indices as
\ie
\Omega^{hs,(n)}_{\A\B|\A_1\cdots\A_n} = \chi^{n,+}_{\A\B\A_1\cdots\A_n} + \epsilon_{(\A_1(\underline{\A}}\chi^{n,0}_{\underline{\B})\A_2\cdots\A_n)} + \epsilon_{(\underline{\A}(\A_1}\epsilon_{\underline{\B})\A_2} \chi^{n,-}_{\A_3\cdots\A_n)}.
\fe
Conversely,
\ie
& \Omega^{hs,n}_{(\A\B|\A_1\cdots\A_n)} = \chi^{n,+}_{\A\B\A_1\cdots\A_n},
\\
& \Omega^{hs,n}{}_{(\A_1}{}^\C{}_{|\C\A_2\cdots\A_n)} = {n+2\over 2n} \chi^{n,0}_{\A_1\cdots\A_n},
\\
& \Omega^{hs,n}{}^{\C\D}{}_{|\C\D\A_1\cdots\A_{n-2}} = {n+1\over n-1} \chi^{n,-}_{\A_1\cdots\A_{n-2}}.
\fe
Next, we want to also decompose the equation (\ref{eqnab}) into the irreducible representation of the permutation group. Symmetrizing all indices $(\A\B\A_1\cdots\A_n)$ in (\ref{eqnab}) gives
\ie\label{chip}
\nabla_{(\A_1}{}^\C \chi^{n,+}_{\A_2\cdots\A_{n+2})\C} - {1\over 2}\nabla_{(\A_1\A_2}\chi^{n,0}_{\A_3\cdots\A_{n+2})}  - {1\over 16} n \psi_1 \chi^{n,+}_{\A_1\cdots\A_{n+2}} = 0.
\fe
On the other hand, contracting (\ref{eqnab}) with $\epsilon^{\A\A_1}$ gives
\ie\label{eqnac}
& \nabla_\A{}^\C \Omega_{\B\C|}{}^\A{}_{\A_2\cdots\A_n} + \nabla_\B{}^\C \Omega_{\A\C|}{}^\A{}_{\A_2\cdots\A_n} \\
& - {\psi_1\over 16} \left[ (n+3) \Omega_\B{}^\C{}_{|\C\A_2\cdots\A_n} 
+(n-1) \epsilon_{(\underline{\A_2}\B} \Omega^{\C\D}{}_{|\C\D\underline{\A_3\cdots\A_n})}
+(n-1) \Omega_{\A(\underline{\A_2}|\B}{}^\A{}_{\underline{\A_3\cdots\A_n})}\right] = 0.
\fe
Now symmetrizing $(\B\A_2\cdots\A_n)$ gives
\ie\label{chi0}
-\nabla^{\C\D} \chi^{n,+}_{\C\D\A_1\cdots\A_n} -{2\over n} \nabla_{(\A_1}{}^\C\chi^{n,0}_{\A_2\cdots\A_n)\C}
+ {n+2\over n} \nabla_{(\A_1\A_2} \chi^{n,-}_{\A_3\cdots\A_n)} - {n+2\over 8n} \psi_1 \chi^{n,0}_{\A_1\cdots\A_n} = 0.
\fe
Alternatively, contract (\ref{eqnac}) with $\epsilon^{\B\A_2}$ gives
\ie\label{chin}
{n+2\over n} \nabla^{\C\D} \chi^{n,0}_{\C\D\A_1\cdots \A_{n-2}} - {2(n+1)(n-2)\over n(n-1)} \nabla^\C{}_{(\A_1}
\chi^{n,-}_{\A_2\cdots\A_{n-2})\C} + {(n+2)(n+1)\over 8(n-1)} \psi_1 \chi^{n,-}_{\A_1\cdots\A_{n-2}} = 0.
\fe

As in the previous subsection, we reintroduce the auxiliary variable $y^\A$, and define
\ie
&\chi_n^{+}(y) = \chi^{n,+}_{\A_1\cdots \A_{n+2}} y^{\A_1}\cdots y^{\A_{n+2}},\\
&\chi_n^{0}(y) = \chi^{n,0}_{\A_1\cdots \A_{n}} y^{\A_1}\cdots y^{\A_{n}},\\
&\chi_n^{-}(y) = \chi^{n,-}_{\A_1\cdots \A_{n-2}} y^{\A_1}\cdots y^{\A_{n-2}},\\
\fe
and so
\ie\label{chi to omega}
& \Omega^{hs,(n)}_{\A\B}(y) = {1\over (n+2)(n+1)}\partial_\A \partial_\B \chi_n^+(y)
+ {1\over n} y_{(\A}\partial_{\B)} \chi_n^0(y) + y_\A y_\B \chi_n^-(y).
\fe
The three equations derived previously for $\chi$, (\ref{chip}), (\ref{chi0}), and (\ref{chin}), can now be written as 
\ie\label{eqnchi}
&{1\over n+2}\nabla^0 \chi_n^+(y) + {1\over 2} \nabla^+ \chi_n^0(y)
-{n\over 16}\psi_1 \chi_n^+(y) = 0,
\\
& {1\over (n+2)(n+1)}\nabla^- \chi_n^+(y) - {2\over n^2} \nabla^0 \chi_n^0(y) - {n+2\over n}\nabla^+ \chi_n^-(y) - {n+2\over 8n}\psi_1 \chi_n^0(y) = 0,
\\
& -{n+2\over n^2(n-1)} \nabla^- \chi_n^0(y) - {2(n+1)\over n(n-1)}\nabla^0 \chi_n^-(y) + {(n+2)(n+1)\over 8(n-1)} \psi_1 \chi_n^-(y) = 0.
\fe
Now expand $\chi^{\pm/0}_n$ in $\psi_1$,
\ie
\chi^{\pm/0}_n = \chi^{n,\pm/0}_{even} + \psi_1 \chi^{n,\pm/0}_{odd}.
\fe
We can now solve $\chi_{even}$ in terms of $\chi_{odd}$:
\ie
& \chi_{even}^{n,+}(y)  = {16\over n}\left[{1\over n+2}\nabla^0 \chi_{odd}^{n,+}(y) + {1\over 2} \nabla^+ \chi_{odd}^{n,0}(y)\right],
\\
& \chi_{even}^{n,0}(y) = {8\over n+2} \left[ {n\over (n+2)(n+1)}\nabla^- \chi_{odd}^{n,+}(y) -{2\over n} \nabla^0 \chi_{odd}^{n,0}(y) - (n+2) \nabla^+ \chi_{odd}^{n,-}(y)\right] ,
\\
& \chi_{even}^{n,-}(y) = {8\over n}\left[ {1\over n(n+1)} \nabla^- \chi_{odd}^{n,0}(y) + {2\over n+2} \nabla^0 \chi_{odd}^{n,-}(y) \right].
\fe
At this point, it is convenient to use part of the gauge symmetry to gauge away $\chi_{odd}^0$ completely (we will show this in the later part of this subsection), and then write
\ie\label{odd to even}
& \chi_{even}^{n,+}(y)  = {16\over n(n+2)} \nabla^0 \chi_{odd}^{n,+}(y),
\\
& \chi_{even}^{n,0}(y) = {8\over n+2} \left[ {n\over (n+2)(n+1)}\nabla^- \chi_{odd}^{n,+}(y)  - (n+2) \nabla^+ \chi_{odd}^{n,-}(y)\right] ,
\\
& \chi_{even}^{n,-}(y) ={16\over n(n+2)} \nabla^0 \chi_{odd}^{n,-}(y) .
\fe
Plugging back in (\ref{eqnchi}) (with $\chi^0_{odd}=0$), we obtain (the second equation is automatically satisfied because of the second equation of (\ref{y-algebra}))
\ie\label{hs equation frame}
&{16\over n(n+2)^2}(\nabla^0)^2 \chi_{odd}^{n,+}(y) + {4n\over (n+2)^2(n+1)}\nabla^+\nabla^- \chi_{odd}^{n,+}(y) - 4 (\nabla^+)^2 \chi_{odd}^{n,-}(y)
-{n\over 16}\chi_{odd}^{n,+}(y) = 0,
\\
& - {8\over (n+2)(n+1)n}(\nabla^-)^2 \chi_{odd}^{n,+}(y)  + {8(n+2)\over n^2} \nabla^- \nabla^+ \chi_{odd}^{n,-}(y) - {32(n+1)\over n^2(n+2)}(\nabla^0)^2 \chi_{odd}^{n,-}(y) \\
&~~~~+ {(n+2)(n+1)\over 8} \chi_{odd}^{n,-}(y) = 0.
\fe
By using (\ref{y-algebra}), we rewrite (\ref{hs equation frame}) as
\ie\label{hs equation y}
&\Box_{AdS} \chi_{odd}^{n,+}(y) + {2n+8-n^2 \over 4} \chi_{odd}^{n,+}(y) + {16\over (n+1)} \nabla^+\nabla^- \chi_{odd}^{n,+}(y) - 16n (\nabla^+)^2 \chi_{odd}^{n,-}(y) = 0,
\\
&\Box_{AdS}  \chi_{odd}^{n,-}(y) - {(n^2+2n+4)\over 4} \chi_{odd}^{n,-}(y) -  {8\over n} \nabla^+ \nabla^-  \chi_{odd}^{n,-}(y)  + {8\over (n+1)n^2}(\nabla^-)^2 \chi_{odd}^{n,+}(y)= 0.
\fe

Now let us examine the gauge transformations on $\chi^\pm$. The gauge transformation on the components of $\Omega^{hs,n}$ is
\ie
\delta \Omega^{hs,n}_{\A\B|\A_1\cdots\A_n} = \nabla_{\A\B} \lambda^n_{\A_1\cdots\A_n}
- {n\over 16}\psi_1 \epsilon_{(\A_1(\underline{\A}} \lambda^n_{\underline{\B})\A_2\cdots\A_n)}.
\fe
In terms of $\chi^{\pm,0}$, we have
\ie
& \delta \chi^{n,+}_{\A_1\cdots\A_{n+2}} = \nabla_{(\A_1\A_2} \lambda^n_{\A_3\cdots\A_{n+2})},\\
&  \delta \chi^{n,0}_{\A_1\cdots\A_n} ={2n\over n+2} \nabla_{(\A_1}{}^\C \lambda^n_{\A_2\cdots\A_n)\C}
+ {n\over 16}\psi_1 \lambda^n_{\A_1\cdots\A_n},
\\
& \delta\chi^{n,-}_{\A_1\cdots\A_{n-2}} = {n-1\over n+1} \nabla^{\C\D} \lambda^n_{\C\D\A_1\cdots\A_{n-2}} .
\fe
Expanding $\lambda^n$ as $\lambda^n=\lambda^n_{even}+\psi_1\lambda^n_{odd}$, we can use $\lambda^{n}_{even}$ to set $\chi^{n,0}_{odd}=0$, and $\chi_{odd}^{n,+},\chi_{odd}^{n,-}$ transform under gauge transformation generated by the residual gauge parameter $\lambda_{odd}^{n}$ as
\ie\label{gauge lambda}
& \delta \chi_{odd}^{n,+}(y) = - \nabla^+\lambda_{odd}(y),
\\
& \delta \chi_{odd}^{n,-}(y) = - {1\over n(n+1)} \nabla^-\lambda_{odd}(y).
\fe

It is very useful to rewrite the equations of motion in the metric-like formulation. In the metric like formulation, we have the metric like field $\Phi_{\m_1\cdots\m_s}$ which is totally symmetric and satisfies the double traceless condition:
\ie\label{dbtl}
\Phi^{\m\n}{}_{\m\n\m_5\cdots \m_s}=0.
\fe
$\Phi_{\m_1\cdots\m_s}$ satisfies the Fronsdal equation (\ref{fransdal}), and transforms under the gauge transformation as (\ref{fransdal gauge trans}).

We show that the Fronsdal equation (\ref{fransdal}) and the frame-like equation (\ref{hs equation frame}) are equivalent. Let us decompose $\Phi_{\m_1\cdots\m_s}$ into the irreducible representation of the Lorentz group as in (\ref{Phidxichi}). Plugging this in to (\ref{fransdal}), we obtain
\ie\label{fransdal 1}
&(\Box-m^2)\xi_{\m_1\cdots\m_s}+(\Box-m^2)g_{(\ul{\m_1\m_2}}\chi_{\ul{\m_3\cdots\m_s})}-s\p_{(\ul{\m_1}}\p^\m\xi_{\m\ul{\m_2\cdots\m_s})}
\\
&+(2s-3)\p_{(\ul{\m_1}}\p_{\ul{\m_2}}\chi_{\ul{\m_3\cdots\m_s})} - (s-2) g_{(\ul{\m_1\m_2}}\p_{\ul{\m_3}}\p^\m\chi_{\m\ul{\m_4\cdots\m_s})}\\
&-2 (2s-1) g_{(\ul{\m_1\m_2}}\chi_{\ul{\m_3\cdots\m_s})}=0.
\fe
Contracting this with $g^{\m_1\m_2}$, we get
\ie\label{fransdal trace}
&(2s-1)(\Box-m^2)\chi_{\m_3\cdots\m_s}- s(s-1)\p^\m\p^\n\xi_{\m\n\m_3\cdots\m_s}+(2s-3)\Box\chi_{\m_3\cdots\m_s}
\\
&+(2s-3)(s-2)\p^{\m}\p_{(\ul{\m_3}}\chi_{\m\ul{\m_4\cdots\m_s})}-2(s-2)\p_{(\ul{\m_3}}\p^{\m}\chi_{\m\ul{\m_4\cdots\m_s})}
\\
&-(s-2)(s-3)g_{(\ul{\m_3\m_4}}\p^\m\p^\n\chi_{\m\n\ul{\m_5\cdots\m_s})}-2(2s-1)^2\chi_{\m_3\cdots\m_s}=0.
\fe
By using the formula
\ie
\p^{\m}\p_{(\ul{\m_3}}\chi_{\m\ul{\m_4\cdots\m_s})}=\p_{(\ul{\m_3}}\p^{\m}\chi_{\m\ul{\m_4\cdots\m_s})}-(s-1)\chi_{\m_3\cdots\m_s},
\fe
we can simplify (\ref{fransdal trace}) as
\ie\label{fransdal trace 1}
&(2s-1)(\Box-m^2)\chi_{\m_3\cdots\m_s}- s(s-1) \p^\m\p^\n\xi_{\m\n\m_3\cdots\m_s}+(d+2s-5)\Box\chi_{\m_3\cdots\m_s}
\\
&+(2s-5)(s-2) \p_{(\ul{\m_3}}\p^{\m}\chi_{\m\ul{\m_4\cdots\m_s})}-(2s-3)(s-2)(s-1)\chi_{\m_3\cdots\m_s}
\\
&-2(2s-1)^2\chi_{\m_3\cdots\m_s}-(s-2)(s-3)g_{(\ul{\m_3\m_4}}\p^\m\p^\n\chi_{\m\n\ul{\m_5\cdots\m_s})}=0.
\fe
Defining
\ie
&\xi^s(y)=y^{\af_1}\cdots y^{\af_{2s}}(e^{\m_1}_0){}_{\af_1\af_2}\cdots (e^{\m_s}_0){}_{\af_{2s-1}\af_2s}\xi_{\m_1\cdots\m_s},
\\
&\chi^s(y)=y^{\af_1}\cdots y^{\af_{2s}}(e^{\m_1}_0){}_{\af_1\af_2}\cdots (e^{\m_{s-2}}_0){}_{\af_{2s-5}\af_{2s-4}}\chi_{\m_1\cdots\m_{s-2}},
\fe
we can write (\ref{fransdal 1}) and (\ref{fransdal trace 1}) as
\ie\label{fransdal y}
&\Box_{AdS}\xi^s-s(s-3)\xi^s+{16\over 2s-1}\p^+ \p^-\xi^s + (2s-3)(\p^+ )^2\chi^s=0,
\\
&\Box_{AdS}\chi^s-(s^2-s+1)\chi^s-{4\over s-1} \p^+ \p^- \chi^s
-{64\over (2s-1)(s-1)(2s-3)} (\p^-)^2\xi^s=0.
\fe
We can then identify (\ref{hs equation y}) and (\ref{fransdal y}) by
\ie\label{identification}
 \chi^{2s-2,+}_{odd}=\xi^s,~~~~\chi^{2s-2,-}_{odd}=-{2s-3\over 32(s-1)}\chi^s.
 \fe
Later, we will also write $\chi^{2s-2,\pm}_{odd}$ as $\chi^{(s),\pm}_{odd}$ for convenience.
 
Let us also analyze the gauge transformation. Plugging (\ref{Phidxichi}) into (\ref{fransdal gauge trans}), we have
\ie
\delta \xi_{\m_1\cdots\m_s}+g_{(\ul{\m_1\m_2}}\delta\chi_{\ul{\m_3\cdots\m_s})} = \nabla_{(\m_1}\eta_{\m_2\cdots\m_s)}.
\fe
Contracting this with $g^{\m_1\m_2}$, we obtain
\ie
\delta\chi_{\m_3\cdots\m_s} = {s-1\over 2s-1}\nabla^{\m}\eta_{\m\m_3\cdots\m_s}.
\fe
It follows that
\ie\label{gaugell}
&\delta \xi^s(y) = \nabla^+ \eta^s(y),
\\
&\delta \chi^s(y) = -{16\over (2s-1)(2s-3)} \nabla^-\eta^s(y).
\fe
The gauge transformations (\ref{gauge lambda}) and (\ref{gaugell}) are also equivalent by the identification (\ref{identification}).

\subsection{Derivation of higher spin boundary-to-bulk propagator in modified de Donder gauge}

The Fronsdal equation (\ref{fransdal}) can be easily solved in the modified de Donder gauge proposed by Metsaev in \cite{Metsaev:2009}. As in (\ref{Odhssc}), we define the generating function $\Phi^s(x|Y)$ of the metric-like higher spin gauge field $\Phi^s_{\m_1\cdots\m_s}$. The field $\Phi^s(x|Y)$ is related to $\chi^{2s-2,+}$ and $\chi^{2s-2,+}$ by
\ie\label{relation}
&\chi^{2s-2,+}_{odd}(y)=\xi^s(y)=\Phi^s(Y)\big|_{Y^A\rightarrow e^A{}_{\af\beta}y^\af y^\beta},
\\
&\chi^{2s-2,-}_{odd}(y)=-{2s-3\over 32(s-1)}\chi^s(y)=-{2s-3\over 64(2s-1)(s-1)}{\partial^2\Phi^s(Y)\over\partial Y^2}\big|_{Y^A\rightarrow e^A{}_{\af\beta}y^\af y^\beta}.
\fe
Using the variable $Y^A$, we can rewrite the Fronsdal equation (\ref{fransdal}), the gauge transformation (\ref{fransdal gauge trans}), and the double traceless condition (\ref{dbtl}) as
\ie
&\left(\Box_{AdS}-s(s-3)-Y^AD^A{\partial\over\partial Y^B}D^B\right.
\\
&~~~~~~~~\left.+{1\over 2}Y^AD^AY^BD^B{\partial\over\partial Y^C}{\partial\over\partial Y^C}-Y^AY^A{\partial\over\partial Y^B}{\partial\over\partial Y^B}\right)\Phi^s(x|Y)=0,
\\
&\delta \Phi^s(x|Y) = Y^A D^A \eta^s(x|Y),
\\
&\left({\partial^2\over\partial Y^2}\right)^2\Phi^s(x|Y)=0,
\fe
where $D^A$ is the covariant derivative acting both on explicit frame indices as well as on indices contracted with $Y^A$; in particular $\Box_{AdS}=D^A D_A$. As proposed by Metsaev \cite{Metsaev:2009}, one then perform a linear transformation:
\be 
\phi(x|Y)=  z^{-\frac{1}{2}}  \cN\Pi^{\phi\Phi}\Phi^s(x|Y),
\ee
and the inverse of it is
\be\label{transformphi}
\Phi^s(x|Y) = z^{\frac{1}{2}} \Pi^{\Phi\phi} \cN \phi(x|Y),
\ee
where the various operators are defined as
\ie
& \cN \equiv \left(\frac{2^{N_z} \Gamma( N_\vY +
N_z - \frac{1}{2}) \Gamma( 2N_\vY - 1)}{\Gamma(N_\vY -
\frac{1}{2})\Gamma(2N_\vY + N_z - 1)}\right)^{1/2},
\\
&\Pi^{\phi\Phi} \equiv  \Pi_\vY + \vec{Y}^2 \frac{1}{4(N_{\vY} + 1)}
\Pi_\vY \left({\partial^2\over \partial \vY^2} + \frac{N_\vY + 1}{N_\vY }{\partial^2\over\partial Y^{z2}}\right),
\\
&\Pi^{\Phi\phi} \equiv  \Pi_{Y} +  Y^2 \frac{1}{2(2 N_Y + 3)}\Pi_{Y} \left({\partial^2\over \partial \vY^2} - \frac{2}{2 N_Y + 1}{\partial^2\over \partial Y^{z2}}\right),
\\
& \Pi_\vY \equiv
\Pi(\vY,0,N_\vY,{\partial\over \partial \vY},0,2),~~~~~~~~\Pi_{Y} \equiv
\Pi(\vY,Y^z,N_Y,{\partial\over\partial\vY},{\partial\over\partial Y^z},3),
\\
&\Pi(\vY,Y^z, A,{\partial\over \partial \vY},{\partial\over \partial Y^z}, B)\equiv \sum_{n=0}^\infty (Y^2)^n \frac{(-)^n \Gamma(A +
\frac{B-2}{2} + n)}{4^n n! \Gamma(A + \frac{B-2}{2} +
2n)}\left({\partial^2\over\partial Y^2}\right)^n,
\\
& N_\vY =\vY\cdot {\partial\over \partial \vY},~~~~~~~~N_z
=Y^z{\partial\over \partial Y^z},~~~~~~~~ \ N_Y \equiv N_\vY + N_z.
\fe

The modified de Donder gauge condition written in terms of the field $\phi(x|Y)$ is:
\ie\label{gauge}
\bar C\phi(x|Y)=0,
\fe
where
\ie
&\bar{C} \equiv{\partial\over\partial\vY}\cdot{\vec\partial} - {1\over 2} \vY\cdot{\vec\partial} {\partial^2\over\partial\vY^2} + {1\over 2} e_1 {\partial^2\over\partial\vY^2} - {\bar e}_1\Pi',
\\
&\Pi' \equiv 1 -\vY^2\frac{1}{4(N_\vY
+1)}{\partial^2\over\partial\vY^2},
\\
&e_1 = e_{1,1} \left( \partial_z + \frac{2s -3
-2N_z}{2z}\right),
\\
& {\bar e}_1 = \Bigl(\partial_z - \frac{2s - 3
-2N_z}{2z}\Bigr) {\bar e}_{1,1},
\\
& e_{1,1} = Y^z f,~~~~~~~~ {\bar e}_{1,1} = f
{\partial\over\partial Y^z},
\\
& f \equiv \varepsilon
\Bigl(\frac{2s-2-N_z}{2s-2-2N_z}\Bigr)^{1/2},~~~~~~~~\varepsilon = \pm1.
\fe

In this gauge, the equations of motion is simplified as
\ie
\Bigl( \Box + \partial_z^2 - \frac{1}{z^2}(r-{1\over 2})(r-{3\over 2})\Bigr)\phi_{r}= 0,
\fe
where $\phi_r(x|\vec Y)$ are the components of $\phi(x|Y)$ expanded in $Y^z$ as in (\ref{phiexpand}), and the general solution of this equation is
\ie\label{soleom}
\phi_{r}(\vp,z|\vY)=C^{r}_1(\vp,\vY)\sqrt{z}J_{r-1}(z|\vp|)+C^{r}_2(\vp,\vY)\sqrt{z}Y_{r-1}(z|\vp|),
\fe
where we Fourier transformed $\phi_{r}(x|\vY)$ as
\ie
\phi_{r}(x|\vY)=\int d^2 x ~\phi_{r}(\vp,z|\vY) ~e^{\vp \cdot {\vec x}}.
\fe
Notice that $\vec p$ is imaginary momentum. We can Wick rotate back to the real momentum by $\vec p\rightarrow i\vec p$. For the purpose of computing the boundary-to-bulk propagator, we can simply replace  $J_{r-1}(z|\vp|)$ and $Y_{r-1}(z|\vp|)$ by $i^{-r+1}K_{r-1}(x)$.

Next, let us solve for the functions $C^{r}_1(\vp,\vY)$ and $C^{r}_2(\vp,\vY)$ using the double traceless condition and the gauge condition. Let us first look at the reduced double traceless condition. It is convenient to define
\ie
Y^+=Y^1+iY^2~~~{\rm and}~~~Y^-=Y^1-iY^2.
\fe
The double traceless condition (\ref{dtphi}) can be written as
\ie
\left({\partial\over\partial Y^{+}}{\partial\over\partial Y^{-}}\right)^2C^{r}(\vp,\vY)=0.
\fe
The general solution of it is
\ie\label{doubletraceless}
C^{r}(\vp,\vY)=c^{r}_{++}(\vp)(Y^+)^{r}+c^{r}_{-+}(\vp)Y^-(Y^+)^{r-1}+c^{r}_{+-}(\vp)Y^+(Y^-)^{r-1}+c^{r}_{--}(\vp)(Y^-)^{r}.
\fe
for $r>2$. For the $r=1,2$, we have
\ie\label{doubletracelessa}
C^{1}(\vp,\vY)=c^{1}_{+}Y^++c^{1}_{-}Y^-~~~{\rm and}~~~C^{2}(\vY)=c^{2}_{++}(Y^+)^{2}+c^{2}_{+-}Y^+Y^-+c^{2}_{--}(Y^-)^{2}.
\fe
Next, let us consider the gauge condition (\ref{gauge}).
\ie
&\bar C\phi(x|Y)=\left({\partial\over\partial\vY}\cdot\vp - {1\over 2} \vY\cdot\vp {\partial^2\over\partial\vY^2} +
{1\over 2} e_1 {\partial^2\over\partial\vY^2} - {\bar e}_1\Pi'\right)\sum^s_{r=0}\left(Y^{z}\right)^{s-r}\phi_{r}(\vp,z|\vY)\\
&=\left[ {\partial\over\partial\vY}\cdot\vp - {1\over 2} \vY\cdot\vp {\partial^2\over\partial\vY^2} +
{1\over 2} Y^z \varepsilon
\Bigl(\frac{2s+d-4-N_z}{2s+d-4-2N_z}\Bigr)^{1/2} \left( \partial_z + \frac{2s + d -5
-2N_z}{2z}\right) {\partial^2\over\partial\vY^2} \right.\\
 &~~~~\left.-\Bigl(\partial_z - \frac{2s + d -5
-2N_z}{2z}\Bigr) \varepsilon
\Bigl(\frac{2s+d-4-N_z}{2s+d-4-2N_z}\Bigr)^{1/2}
{\partial\over\partial Y^z}\Pi'\right] \sum^s_{r=0}\left(Y^{z}\right)^{s-r}\phi_{r}(\vp,z|\vY)\\
&=\sum^s_{r=0}\left(Y^{z}\right)^{s-r}\left[ {\partial\over\partial\vY}\cdot\vp - {1\over 2} \vY\cdot\vp {\partial^2\over\partial\vY^2} +
{1\over 2} Y^z \varepsilon
\Bigl(\frac{s+r+d-4}{2r+d-4}\Bigr)^{1/2} \left( \partial_z + \frac{2r+ d -5
}{2z}\right) {\partial^2\over\partial\vY^2} \right.\\
 &~~~~\left.-\varepsilon\Bigl(\partial_z - \frac{2r+ d -3
}{2z}\Bigr) \Bigl(\frac{s+r+d-3}{2r+d-2}\Bigr)^{1/2}
{s-r\over Y^z}\Pi'\right] \phi_{r}(\vp,z|\vY)\\
&=\sum^s_{r=0}\left(Y^{z}\right)^{s-r}\left[{\partial\over\partial\vY}\cdot\vp - {1\over 2} \vY\cdot\vp {\partial^2\over\partial\vY^2} +
{1\over 2} Y^z 
\Bigl(\frac{s+r-2}{2r-2}\Bigr)^{1/2} \left( \partial_z + \frac{2r -3
}{2z}\right) {\partial^2\over\partial\vY^2} \right.\\
 &~~~~\left.-\varepsilon\Bigl(\partial_z - \frac{2r -1
}{2z}\Bigr) \Bigl(\frac{s+r-1}{2r}\Bigr)^{1/2}
{s-r\over Y^z}\Pi'\right]\phi_{r}(\vp,z|\vY).
\fe 
The gauge condition can be written as
\ie\label{ggcd}
&\left({\vp\over p}\cdot{\partial\over\partial\vY} - {1\over 2} {\vp\over p}\cdot\vY {\partial^2\over\partial\vY^2} \right)\phi_{r+1}+{1\over 2} \Bigl(\frac{s+r}{2r+2}\Bigr)^{1/2} \left( \partial_z + \frac{2r +1
}{2z}\right) {\partial^2\over\partial\vY^2}\phi_{r+2}
\\
&~~~~~~~~-\varepsilon\Bigl(\partial_z - \frac{2r -1
}{2z}\Bigr) \Bigl(\frac{s+r-1}{2r}\Bigr)^{1/2}
(s-r)\Pi'\phi_{r}=0.
\fe
with $p\equiv |\vp|$. Plugging (\ref{soleom}) into (\ref{ggcd}), we obtain
\ie
&\left({\vp\over p}\cdot{\partial\over\partial\vY} - {1\over 2} {\vp\over p}\cdot\vY {\partial^2\over\partial\vY^2} \right)C^{r+1}+{1\over 2} \Bigl(\frac{s+r}{2r+2}\Bigr)^{1/2}{\partial^2\over\partial\vY^2}C^{r+2}
\\
&~~~~+\varepsilon\Bigl(\frac{s+r-1}{2r}\Bigr)^{1/2}
(s-r)\left(1 -\vY^2\frac{1}{4(r-1)}{\partial^2\over\partial\vY^2}\right)C^{r}=0,
\fe 
or more explicitly,
\ie
&\left[{p^+\over p}\partial_+ + {p^- \over p}\partial_- - \big({p^+\over p}Y^-+{p^-\over p} Y^+\big)\partial_+\partial_-\right]C^{r+1}+2 \Bigl(\frac{s+r}{2r+2}\Bigr)^{1/2}\partial_+\partial_-C^{r+2}
\\
&~~~~+\varepsilon\Bigl(\frac{s+r-1}{2r}\Bigr)^{1/2}
(s-r)\left(1 -\vY^2\frac{1}{r-1}\partial_+\partial_-\right)C^{r}=0,
\fe
with $\partial_\pm=\partial_{Y^\pm}$. Plugging (\ref{doubletraceless}) and (\ref{doubletracelessa}) into the above equation, we obtain
\ie
r{p^+\over p}c^{r}_{++}(\vp)+\varepsilon\Bigl(\frac{s+r-2}{2(r-1)}\Bigr)^{1/2}
(s-r+1)c^{r-1}_{++}(\vp)+(2-r){p^-\over p}c^{r}_{-+}(\vp)+2 \Bigl(\frac{s+r-1}{2r}\Bigr)^{1/2}rc^{r+1}_{-+}(\vp)=0,
\fe
and
\ie
r{p^-\over p}c^{r}_{--}(\vp)+\varepsilon\Bigl(\frac{s+r-2}{2(r-1)}\Bigr)^{1/2}
(s-r+1)c^{r-1}_{--}(\vp)+(2-r){p^+\over p}c^{r}_{+-}(\vp)+2 \Bigl(\frac{s+r-1}{2r}\Bigr)^{1/2}(r)c^{r+1}_{+-}(\vp)=0,
\fe
for $r>2$, and in the cases $r=1,2$,
\ie
&2{p^+\over p}c^{2}_{++}(\vp)+\varepsilon\Bigl(\frac{s}{2}\Bigr)^{1/2}
(s-1)c^{1}_{+}(\vp)+2 \Bigl(\frac{s+1}{4}\Bigr)^{1/2}2c^{3}_{-+}(\vp)=0,
\\
&2{p^-\over p}c^{2}_{--}(\vp)+\varepsilon\Bigl(\frac{s}{2}\Bigr)^{1/2}
(s-1)c^{1}_{-}(\vp)+2 \Bigl(\frac{s+1}{4}\Bigr)^{1/2}2c^{3}_{+-}(\vp)=0,
\\
&{p^+\over p}c^{1}_{+}(\vp)+{p^-\over p}c^{1}_{-}(\vp)+2 \Bigl(\frac{s}{2}\Bigr)^{1/2}c^{2}_{+-}(\vp)=0.
\fe
We can consistently set $c^{r}_{+-}=0=c^{r}_{-+}$ for $r>2$, and obtain
\ie
r{p^+\over p}c^{r}_{++}(\vp)+\varepsilon\Bigl(\frac{s+r-2}{2(r-1)}\Bigr)^{1/2}
(s-r+1)c^{r-1}_{++}(\vp)+(2-r){p^-\over p}c^{r}_{-+}(\vp)=0,
\fe
and
\ie
r{p^-\over p}c^{r}_{--}(\vp)+\varepsilon\Bigl(\frac{s+r-2}{2(r-1)}\Bigr)^{1/2}
(s-r+1)c^{r-1}_{--}(\vp)+(2-r){p^+\over p}c^{r}_{+-}(\vp)=0,
\fe
for $r>2$, and 
\ie
&2{p^+\over p}c^{2}_{++}(\vp)+\varepsilon\Bigl(\frac{s}{2}\Bigr)^{1/2}
(s-1)c^{1}_{+}(\vp)=0,
\\
&2{p^-\over p}c^{2}_{--}(\vp)+\varepsilon\Bigl(\frac{s}{2}\Bigr)^{1/2}
(s-1)c^{1}_{-}(\vp)=0,
\\
&{p^+\over p}c^{1}_{+}(\vp)+{p^-\over p}c^{1}_{-}(\vp)+2 \Bigl(\frac{s}{2}\Bigr)^{1/2}c^{2}_{+-}(\vp)=0,
\fe
for $r=1,2$.
The solution to the above recursive equations is given by
\ie\label{solcppmm}
&c^{r}_{++}={s!\over (s-r)!r!}\sqrt{2^{s-r}(s-1)!(s+r-2)!\over(r-1)!(2s-2)!}(-\varepsilon{p^+\over p})^{s-r}c^s_{++},
\\
&c^{r}_{--}={s!\over (s-r)!r!}\sqrt{2^{s-r}(s-1)!(s+r-2)!\over(r-1)!(2s-2)!}(-\varepsilon{p^-\over p})^{s-r}c^s_{--},
\fe
and
\ie\label{solc2}
c^{2}_{+-}(\vp)=\sqrt{2^{s-2}s!(s-1)!\over (2s-2)!}(-\varepsilon{p^+\over p})^{s}c^s_{++}+\sqrt{2^{s-2}s!(s-1)!\over (2s-2)!}(-\varepsilon{p^-\over p})^{s}c^s_{--}.
\fe
Starting from here and in what follows, we set $\varepsilon=-1$ and only consider the positively polarized fields by setting $c^s_{--}=0$. Plugging (\ref{solcppmm}) and (\ref{solc2}) back to (\ref{doubletraceless}) and (\ref{doubletracelessa}), then back to (\ref{soleom}), and Wick rotating to the real momenta, we obtain
\ie
&\phi(\vp,z|\vY,Y^z)
\\
&=\sum^s_{r=1}i^{1-r}{s!\over (s-r)!r!}\sqrt{2^{s-r}(s-1)!(s+r-2)!\over(r-1)!(2s-2)!}\left({p^+\over p}\right)^{s-r}(Y^z)^{s-r}(Y^+)^{r}c^s_{++}\sqrt{z}K_{r-1}(p z)\\
&~~~~+i^{-1}\sqrt{2^{s-2}s!(s-1)!\over (2s-2)!}\left({p^+\over p}\right)^{s}c^s_{++}Y^+Y^-(Y^z)^{s-2}\sqrt{z}K_{1}(p z).
\fe
Using the transformation (\ref{transformphi}), we arrive at the expression for the boundary to bulk propagator in momentum space, in the modified de Donder gauge,
\ie
&\Phi^s(\vp,z|Y)
\\
 &= z^{\frac{1}{2}}\Pi^{\Phi\phi}\cN\phi(\vp,z|\vY,Y^z)\\
&=\sum^s_{r=1}\sum_{n=0}^{\infty}  \frac{(-1)^{n} i^{1-r} \Gamma(s-n -
\frac{1}{2})}{ 4^n n!\Gamma(s - \frac{1}{2} ) }{s!\over (s-r-2n)!r!}\left({p^+\over p}\right)^{s-r}Y^{2n}(Y^z)^{s-r-2n} (Y^+)^{r}c^s_{++}zK_{r-1}(p z)\\
&~~~~+\sum_{n=0}^{\infty} \frac{(-1)^n i^{-1}\Gamma(s-n -
\frac{1}{2})}{ 4^n n! \Gamma(s - \frac{1}{2} )}{(s-2)!\over(s-2-2n)!}\left({p^+\over p}\right)^{s}c^s_{++}Y^{2n}(Y^z)^{s-2-2n}Y^+Y^- zK_{1}(p z).
\fe
In terms of the frame-like fields, using (\ref{relation}), we have
\ie
\chi^{(s),+}_{odd}(\vp,z|y)=&c^s_{++}\sum^s_{r=0}{i^{r}}{s!\over (s-r)!r!} p^{r-1} { (p^+)^{s-r}} (y^1)^{s+r} (y^2)^{s-r} zK_{r-1}(z|\vp|),
\\
\chi^{(s),-}_{odd}(\vp,z|y) =&c^s_{++} {s\over 2(2s-1)}\sum^s_{r=0} i^{r}{(s-2)!\over (s-r-2)!r!} {p^{r-1} (p^+)^{s-r}}(y^1)^{s+r-2} (y^2)^{s-r-2}zK_{r-1}(z|\vp|).
\fe

\section{Second order in perturbation theory}

\subsection{A star-product relation}

Let us write the following useful formula for the star-product:
\ie\label{star product expansion}
A(y)*B(y)&=\sum^\infty_{n=0}\left(\sum^n_{m=0}\sum^\infty_{p=0}{(m+p)!(n-m+p)!\over p!m!(n-m)!}A_{\af_1\cdots\af_p(\ul{\beta_{1}\cdots\beta_{m}}}B^{\af_1\cdots\af_p}{}_{\ul{\beta_{m+1}\cdots\beta_{n}})}\right)y^{\beta_{1}}\cdots y^{\beta_{n}}
\fe
where $A(y)$ and $B(y)$ have the expansions:
\ie
A(y)=\sum^{\infty}_{n=0}A_{\A_1\cdots\A_n}y^{\A_1}\cdots y^{\A_n},~~\text{and}~~B(y)=\sum^{\infty}_{n=0}B_{\A_1\cdots\A_n}y^{\A_1}\cdots y^{\A_n}.
\fe
(\ref{star product expansion}) follows from writing the $(\mbox{m-th})*(\mbox{n-th})$ term as
\ie
&\left(A_{\af_1\cdots\af_m}y^{\af_1}\cdots y^{\af_m}\right)*\left(B_{\beta_1\cdots\beta_n}y^{\beta_1}\cdots y^{\beta_n}\right)
\\
&=(-1)^mA^{\A_1\cdots\A_m}(y_{\A_1}+{\partial\over \partial y^{\A_1}})\cdots(y_{\A_m}+{\partial\over \partial y^{\A_m}})B_{\beta_1\cdots\beta_n}y^{\beta_1}\cdots y^{\beta_n}
\\
%&=\sum_{p\le n,m} {(-1)^m m!\over p!(m-p)!} A^{\A_1\cdots\A_m}y_{\A_{p+1}}\cdots y_{\A_m}{\partial\over \partial y^{\A_1}}\cdots{\partial\over \partial y^{\A_p}}B_{\beta_1\cdots\beta_n}y^{\beta_1}\cdots y^{\beta_n}
%\\
&=\sum_{p\le m,n}{n!m!\over (m-p)!(n-p)!p!}A_{\af_1\cdots\af_p(\ul{\af_{p+1}\cdots\af_m}}B^{\af_1\cdots\af_p}{}_{\ul{\beta_{p+1}\cdots\beta_n})}y^{\af_{p+1}}\cdots y^{\af_m}y^{\beta_{p+1}}\cdots y^{\beta_n}.
\fe

\subsection{Derivation of $U^{0,\mu}$ and $U^2_{\mu|\A\B}$}

The purpose of this subsection is to compute the RHS of (\ref{2nd scalar}).

By using the star-product relation (\ref{star product expansion}), we obtain
\ie
&[\Omega^{even},C^{(1)}_{mat}]_*
\\
&~~~~=\sum^\infty_{n=0}\left(\sum^n_{m=0}\sum^\infty_{p=0}{(m+p)!(x-m+p)!\over p!m!(n-m)!}{(1-(-)^p)}\Omega^{even}_{\af_1\cdots\af_p(\ul{\beta_{1}\cdots\beta_{m}}}C^{(1)}_{mat}{}^{\af_1\cdots\af_p}{}_{\ul{\beta_{m+1}\cdots\beta_{n}})}\right)y^{\beta_{1}}\cdots y^{\beta_{n}},
\\
&\{\Omega^{odd},C^{(1)}_{mat}\}_*
\\
&~~~~=\sum^\infty_{n=0}\left(\sum^n_{m=0}\sum^\infty_{p=0}{(m+p)!(n-m+p)!\over p!m!(n-m)!}(1+(-)^p)\Omega^{odd}_{\af_1\cdots\af_p(\ul{\beta_{1}\cdots\beta_{m}}}C^{(1)}_{mat}{}^{\af_1\cdots\af_p}{}_{\ul{\beta_{m+1}\cdots\beta_{n}})}\right)y^{\beta_{1}}\cdots y^{\beta_{n}}.
\fe
The $U^{0}_{\m}$ and $U^{2}_{\m|\A_1\A_2}$ are coefficients of the components in $-[\Omega^{even},C^{(1)}_{mat}]_*+\psi_1\{\Omega^{odd},C^{(1)}_{mat}\}_*$, which are independent and quadratic in $y$. They can be written as
\ie
U^{(0)}_\m=&\psi_1\sum^\infty_{p=0}{p!}(1+(-)^p)\Omega^{odd}_{\m|\af_1\cdots\af_p}C^{(1)}_{mat}{}^{\af_1\cdots\af_p},
\fe
and
\ie
&U^{(2)}_{\m|\af\beta}=-\sum^\infty_{p=0}{(p+1)(p+1)!}(1-(-)^p)\Omega^{even}_{\m|\af_1\cdots\af_p(\af}C^{(1)}_{mat}{}^{\af_1\cdots\af_p}{}_{\beta)}
\\
&+\psi_1\sum^\infty_{p=0}{(p+2)!\over 2}(1+(-)^p)\Omega^{odd}_{\m|\af_1\cdots\af_p}C^{(1)}_{mat}{}^{\af_1\cdots\af_p}{}_{\af\beta}
+\psi_1\sum^\infty_{p=0}{(p+2)!\over 2}(1+(-)^p)\Omega^{odd}_{\m|\af_1\cdots\af_p\af\beta}C^{(1)}_{mat}{}^{\af_1\cdots\af_p}.
\fe
We first compute $\nabla^\m U^{(0)}_\m$:
\ie
\nabla^\m U^{(0)}_\m=&-32\psi_1\sum^\infty_{p=0}{p!}(1+(-)^p)\left(\nabla^{\A\B}\Omega^{odd}_{\A\B|\af_1\cdots\af_p}C^{(1)}_{mat}{}^{\af_1\cdots\af_p}+\Omega^{odd}_{\A\B|\af_1\cdots\af_p}\nabla^{\A\B}C^{(1)}_{mat}{}^{\af_1\cdots\af_p}\right)
\\
=&-32\psi_1\sum^\infty_{p=0}{p!}(1+(-)^p)\left(\nabla^{\A\B} \chi^{p,+,odd}_{\A\B\A_1\cdots\A_p}C^{(1)}_{mat}{}^{\af_1\cdots\af_p}  + \nabla_{\A_1\A_2}  \chi^{p,-,odd}_{\A_3\cdots\A_p}C^{(1)}_{mat}{}^{\af_1\cdots\af_p}\right.
\\
& \left.+ \chi^{p,+,odd}_{\A\B\A_1\cdots\A_p}\nabla^{\A\B}C^{(1)}_{mat}{}^{\af_1\cdots\af_p}  + \chi^{p,-,odd}_{\A_3\cdots\A_p} \nabla_{\A_1\A_2} C^{(1)}_{mat}{}^{\af_1\cdots\af_p}\right)
\\
=&32\psi_1\sum^\infty_{p=0}(1+(-)^p)\left[{C_{mat}^{(1),p}(\partial_y)}\left({\nabla^{-} \chi^{p,+}_{odd} (y)\over (p+2)(p+1)}+ \nabla^+ \chi^{p,-}_{odd}(y)\right)\right.
\\
&\left.+ {(\nabla^{+}C_{mat}^{(1),p})(\partial_y) \chi^{p,+}_{odd}(y)\over (p+2)(p+1)} + (\nabla^- C_{mat}^{(1),p})(\partial_y)\chi^{p,-}_{odd}(y) \right],
\fe
where we have assumed the gauge condition $\chi^{p,0}_{odd}=0$. Using (\ref{eqnscalar}) to express $\nabla^{\pm}C_{mat}^{(1),p}$ in terms of $C_{mat}^{(1),p\pm2}$, we have
\ie\label{ttta}
\nabla^\m U^{(0)}_\m 
=&32\psi_1\sum^\infty_{p=0}(1+(-)^p)\left[{C_{mat}^{(1),p}(\partial_y)}\left({\nabla^{-} \chi^{p,+}_{odd} (y)\over (p+2)(p+1)} + \nabla^+ \chi^{p,-}_{odd}(y)\right)\right.
\\
&\left.+ \psi_1{C_{mat}^{(1),p+2}(\partial_y) \chi^{p,+}_{odd}(y)\over 16}  +  \psi_1{p(p+1)\over 16}C_{mat}^{(1),p-2}(\partial_y)\chi^{p,-}_{odd}(y) \right].
\fe
Next, we compute $(e^\m_0)^{\af\beta}U^{(2)}_{\m|\af\beta}$:
\ie
(e^\m_0)^{\af\beta}U^{(2)}_{\m|\af\beta}=&\sum^\infty_{p=0}{(p+3)(p+1)!\over 2}(1-(-)^p)\chi^{p+1,0,even}_{\af_1\cdots\af_p\beta}C^{(1)}_{mat}{}^{\af_1\cdots\af_p\beta}
\\
&~~~+\psi_1\sum^\infty_{p=0}{(p+2)!\over 2}(1+(-)^p)\chi^{p+1,+,odd}_{\af_1\cdots\af_p\af\beta}C^{(1)}_{mat}{}^{\af_1\cdots\af_p\af\beta}
\\
&~~~+\psi_1\sum^\infty_{p=0}{(p+3)(p+2)p!\over 2}(1+(-)^p)\chi^{p,-,odd}_{\af_1\cdots\af_p}C^{(1)}_{mat}{}^{\af_1\cdots\af_p}
\\
=&\sum^\infty_{p=0}{(p+3)(1-(-)^p)\over 2}C^{(1),p+1}_{mat}(\partial_y)\chi^{p+1,0}_{even}(y)+\psi_1\sum^\infty_{p=0}{(1+(-)^p)\over 2}C^{(1),p+2}_{mat}(\partial_y)\chi^{p,+}_{odd}(y)
\\
&~~~+\psi_1\sum^\infty_{p=0}{(p+3)(p+2)(1+(-)^p)\over 2}C^{(1),p}_{mat}(\partial_y)\chi^{p+2,-}_{odd}(y),
\fe
where we have assumed the gauge $\chi^{p,0}_{odd}=0$. Using (\ref{odd to even}) to express $\chi^{p+1,0}_{even}$ in terms of $\chi^{p+1,+}_{odd}$ and $\chi^{p+1,-}_{odd}$, we have
\ie\label{tttb}
&(e^\m_0)^{\af\beta}U^{(2)}_{\m|\af\beta}=\sum^\infty_{p=0}{(1-(-)^p)}C^{(1),p+1}_{mat}(\partial_y)\left[{4(p+1)\over (p+3)(p+2)}\nabla^- \chi_{odd}^{p+1,+}(y) - {4}(p+3)\nabla^+ \chi_{odd}^{p+1,-}(y)\right]
\\
&+\psi_1\sum^\infty_{p=0}{(1+(-)^p)\over 2}C^{(1),p+2}_{mat}(\partial_y)\chi^{p,+}_{odd}(y) 
+ \psi_1\sum^\infty_{p=0}{(p+3)(p+2)(1+(-)^p)\over 2}C_{mat}^{(1),p}(\partial_y)\chi^{p+2,-}_{odd}(y),
\fe
Adding the two terms (\ref{ttta}) and (\ref{tttb}), we obtain
\ie
&\nabla^\m U^{(0)}_\m + 4\psi_1 (e^\m_0)^{\af\beta}U^{(2)}_{\m|\af\beta}
\\
&=4\sum^\infty_{p=0}(1+(-)^p) \left[{C_{mat}^{(1),p+2}(\partial_y) \chi^{p,+}_{odd}(y)}+{(p+1)p}C_{mat}^{(1),p-2}(\partial_y)\chi^{p,-}_{odd}(y)\right]
\\
&~~+ 16\psi_1 \sum^\infty_{p=2}{(1+(-)^p)}C_{mat}^{(1),p}(\partial_y)\left[{1\over (p+1)}\nabla^- \chi_{odd}^{p,+}(y) - p\nabla^+ \chi_{odd}^{p,-}(y)\right].
\fe

\subsection{Computation of the three point function}

In this subsection, we compute the three point function of a higher spin current with two scalars by explicitly evaluating the integral (\ref{vaction}).

To begin with, let us turn on boundary sources only for the $C_{even}$ component of the scalars in (\ref{vaction}). It is convenient to decompose $\Xi_s$ as $\Xi_s=\Xi^+_s + \Xi^0_s + \Xi^-_s $, with $\Xi^{\pm/0}_s$ being the homogeneous polynomials in $y$ of degree $2s$, $2s-2$, and $2s-4$, respectively. The action (\ref{vaction}) splits into three terms. The terms with $\Xi^{\pm}_s$ have already been of the form (\ref{action1}). For the term with $\Xi^0_s$, we need to perform an integration by part:
\ie
&\int {dx^2}\left({dz\over z^3}\right)\Xi^0_s(\partial_y)\delta C_{mat}^{(1),0} C_{mat}^{(1),2s-2}
\\
&=\int {dx^2}\left({dz\over z^3}\right)32\psi_1 \left( {1\over (2s-1)} \nabla^- \chi_{odd}^{(s),+}(\partial_y) - ( 2s-2) \nabla^+ \chi_{odd}^{(s),-}(\partial_y) \right) \delta C_{mat}^{(1),0} C_{mat}^{(1),2s-2}
%\\
%&=\int {dx^2}\left({dz\over z^3}\right)\bigg[ - 32\psi_1 {1\over (2s-1)}  \chi_{odd}^{(s),+}(\partial_y) \nabla^+ \delta C_{mat}^{(1),0} C_{mat}^{(1),2s-2} 
%\\
%&~~~~- 32\psi_1 {1\over (2s-1)}  \chi_{odd}^{(s),+}(\partial_y)  \delta C_{mat}^{(1),0} \nabla^+ C_{mat}^{(1),2s-2} + 32\psi_1  ( 2s-2)  \chi_{odd}^{(s),-}(\partial_y) \nabla^{-}\delta C_{mat}^{(1),0}C_{mat}^{(1),2s-2}
%\\
%&~~~~ + 32\psi_1  ( 2s-2)  \chi_{odd}^{(s),-}(\partial_y)\delta  C_{mat}^{(1),0}\nabla^- C_{mat}^{(1),2s-2} \bigg]
\\
&= \int {dx^2}\left({dz\over z^3}\right)\bigg[- 4 {1\over (2s-1)}  \chi_{odd}^{(s),+}(\partial_y) \delta C_{mat}^{(1),2} C_{mat}^{(1),2s-2} - 4s \chi_{odd}^{(s),+}(\partial_y)  \delta C_{mat}^{(1),0} C_{mat}^{(1),2s}
\\
&~~~~+ 4 ( 2s_{mat}-2)  \chi_{odd}^{(s),-}(\partial_y) \delta C_{mat}^{(1),2}(\partial_y) C_{mat}^{(1),2s-2} + 2 ( 2s-2)^2(2s-1)  \chi_{odd}^{(s),-}(\partial_y)\delta  C_{mat}^{(1),0} C_{mat}^{(1),2s-4} \bigg],
\fe
where we have used (\ref{eqnscalar}) to express $\nabla^{\pm}C_{mat}^{(1),p}$ in terms of $C_{mat}^{(1),p\pm2}$. The variation of the action $\delta S$ is then given by
\ie
&\delta S=\int d^2 x \left({dz\over z^3}\right)\bigg[\chi_{odd}^{(s),+}(\partial_y)\left((8-4s) \delta C_{mat}^{(1),0} C_{mat}^{(1),2s}- 4 {1\over (2s-1)} \delta C_{mat}^{(1),2} C_{mat}^{(1),2s-2}\right)
\\
&~~~~+4\chi_{odd}^{(s),-} (\partial_y)\left( ( 2s-2)  \delta C_{mat}^{(1),2}(\partial_y) C_{mat}^{(1),2s-2} + 2 ( s-1)(s+1)(2s-1)\delta  C_{mat}^{()1,0} C_{mat}^{(1),2s-4}\right)\bigg]
\\
&=-\int d^2 x \left({dz\over z^3}\right)\Bigg[\nabla^+\lambda(\partial_y)\left((8-4s) \delta C_{mat}^{(1),0} C_{mat}^{(1),2s}- 4 {1\over (2s-1)} \delta C_{mat}^{(1),2} C_{mat}^{(1),2s-2}\right)
\\
&~~~~-4\nabla^-\lambda(\partial_y)\bigg( {1\over (2s-1)} \delta C_{mat}^{(1),2}(\partial_y) C_{mat}^{(1),2s-2} + (s+1)  \delta  C_{mat}^{(1),0} C_{mat}^{(1),2s-4}\bigg)\Bigg]
\\
&=-\int d^2 x dz\partial_z \bigg[{1\over z^2}\lambda(\partial_y)\partial_{y^1}\partial_{y^2}\left((2-s) \delta C_{mat}^{(1),0} C_{mat}^{(1),2s}-  {1\over (2s-1)} \delta C_{mat}^{(1),2} C_{mat}^{(1),2s-2}\right)
\\
&~~~~-{1\over z^2}\left(\partial_{y^1}\partial_{y^2}\lambda\right)(\partial_y)\left( {1\over 2s-1} \delta C_{mat}^{(1),2}(\partial_y) C_{mat}^{(1),2s-2} + (s+1)\delta  C_{mat}^{(1),0} C_{mat}^{(1),2s-4}\right)\bigg]
\\
&=\lim_{z\rightarrow 0}\int d^2 x {1\over z^2}\bigg[\lambda(\partial_y)\partial_{y^1}\partial_{y^2}\left((2-s) \delta C_{mat}^{(1),0} C_{mat}^{(1),2s}-  {1\over (2s-1)} \delta C_{mat}^{(1),2} C_{mat}^{(1),2s-2}\right)
\\
&~~~~+\left(\partial_{y^1}\partial_{y^2}\lambda\right)(\partial_y)\left( {1\over 2s-1} \delta C_{mat}^{(1),2}(\partial_y) C_{mat}^{(1),2s-2} + (s+1) \delta  C_{mat}^{(1),0} C_{mat}^{(1),2s-4}\right)\bigg]
\\
&=4\lim_{z\rightarrow 0}\int d^2 x \sum^{2s-1}_{r=1} {z^{r-s-2}\over (x^- - x_3^-)^{r}} \Bigg[(\partial_{y^2})^{2s-r}(-\partial_{y^1})^{r} \left((2-s) \delta C_{mat}^{(1),0} C_{mat}^{(1),2s}-  {1\over (2s-1)} \delta C_{mat}^{(1),2} C_{mat}^{(1),2s-2}\right)
\\
&~~~~-(2s-r-1)(r-1)(\partial_{y^2})^{2s-r-2}(-\partial_{y^1})^{r-2}\left({1\over 2s-1} \delta C_{mat}^{(1),2}(\partial_y) C_{mat}^{(1),2s-2} + (s+1)  \delta  C_{mat}^{(1),0} C_{mat}^{(1),2s-4}\right)\Bigg]
\\
&\equiv\delta S_{1}+\delta S_{2}+\delta S_{3}+\delta S_{4},
\fe
where we substituted the boundary to bulk propagator for $\chi_{odd}^{(s),+}$ and $\chi_{odd}^{(s),-}$ in the ``pure gauge" form, and we also performed the similar step as illustrated in (\ref{intbypt}), and we used (\ref{eqnscalar}) again to express $\nabla^{\pm}C_{mat}^{(1),p}$ in terms of $C_{mat}^{(1),p\pm2}$. For the convenience of the later computation, we have split $\delta S$ into four terms $\delta S=\delta S_{1}+\delta S_{2}+\delta S_{3}+\delta S_{4}$. We will compute these four terms  one by one in the following.
%\ie
%\delta S_2&=4\int d^2 x \left({dz\over z^3}\right)\chi_{odd}^{(s),-} (\partial_y)\left( ( 2s-2)  \delta C_{mat}^{(1),2}(\partial_y) C_{mat}^{(1),2s-2} + 2 ( s-1)(s+1)(2s-1)\delta  C_{mat}^{()1,0} C_{mat}^{(1),2s-4}\right)
%\\
%&=-4\int d^2 x \left({dz\over z^3}\right){1\over (2s-2)(2s-1)}\nabla^-\lambda(\partial_y)\big( ( 2s-2) \delta C_{mat}^{(1),2}(\partial_y) C_{mat}^{(1),2s-2} 
%\\
%&~~~~+ 2 ( s-1)(s+1)(2s-1)  \delta  C_{mat}^{(1),0} C_{mat}^{(1),2s-4}\big)
%\\
%&=-\int d^2 x dz\partial_z{1\over z^2}\left(\partial_{y^1}\partial_{y^2}\lambda\right)(\partial_y)\left( {1\over 2s-1} \delta C_{mat}^{(1),2}(\partial_y) C_{mat}^{(1),2s-2} + (s+1)\delta  C_{mat}^{(1),0} C_{mat}^{(1),2s-4}\right)
%\\
%&=\lim_{z\rightarrow 0}\int d^2 x {1\over z^2}\left(\partial_{y^1}\partial_{y^2}\lambda\right)(\partial_y)\left( {1\over 2s-1} \delta C_{mat}^{(1),2}(\partial_y) C_{mat}^{(1),2s-2} + (s+1) \delta  C_{mat}^{(1),0} C_{mat}^{(1),2s-4}\right)
%\\
%&=-4\lim_{z\rightarrow 0}\int d^2 x \sum^{2s-1}_{r=1}(2s-r-1)(r-1)(\partial_{y^2})^{2s-r-2}(-\partial_{y^1})^{r-2}{z^{r-s-2}\over (x^- - x_3^-)^{r}}
%\\
%&~~~~\times\left({1\over 2s-1} \delta C_{mat}^{(1),2}(\partial_y) C_{mat}^{(1),2s-2} + (s+1)  \delta  C_{mat}^{(1),0} C_{mat}^{(1),2s-4}\right)
%\\
%&\equiv\delta S_{2a}+\delta S_{2b}.
%\fe
%The three point function is given by evaluating $\delta S$ on the boundary-to-bulk propagators of the higher spin gauge fields and scalars. 
The next step is to substitute the boundary-to-bulk propagator for the master field $C^{(1)}_{mat}$. We first expand $C^{(1)}_{mat}$ as
\ie
C^{(1)}_{mat}(y)&=\left(1+\psi_1{1+\tilde \psi_1\over 2} y\Sigma y\right)e^{{\psi_1\over 2}y\Sigma y}K^{1+{\tilde\psi_1\over 2}}
\\
&=\sum^\infty_{s=0}{1\over s!}\left(1+s(1+\tilde\psi_1)\right)\left({\psi_1\over 2}\right)^s (y\Sigma y)^s K^{1+{\tilde\psi_1\over 2}}
\\
&=\sum^\infty_{s=0}{\psi_1^s\over s!}\left(1+s(1+\tilde\psi_1)\right) \left[ \big(z-{x^+x^-\over z}\big)y^1 y^2 - (y^1)^2 x^- +( y^2)^2 x^+ \right]^s K^{1+{\tilde\psi_1\over 2}+s}
\\
&=\sum^\infty_{s=0}{\psi_1^s\over s!}\left(1+s(1+\tilde\psi_1)\right)\sum^s_{u=0}\sum^u_{w=0}\sum^{u-w}_{v=0}  {s!\over (s-u)!(u-w-v)! w!v!}\\
&~~~~\times z^{u-w-2v} (-x^-)^{w+v}(x^+)^{s-u+v} (y^1)^{u+w} (y^2)^{2s-u-w} K^{1+{\tilde\psi_1\over 2}+s}.
\fe
In particular, the piece of homogeneous degree $2s$ is given by
\ie
C_{mat}^{(1),2s} (y)&={\psi_1^s\over s!}\left(1+s(1+\tilde\psi_1)\right)\sum^s_{u=0}\sum^u_{w=0}\sum^{u-w}_{v=0}  {s!\over (s-u)!(u-w-v)! w!v!}\\
&~~~~\times z^{u-w-2v} (-x^-)^{w+v}(x^+)^{s-u+v} (y^1)^{u+w} (y^2)^{2s-u-w} K^{1+{\tilde\psi_1\over 2}+s}.
\fe
where $K={z\over z^2+x^2}$ is the scalar boundary-to-bulk propagator. Near the boundary, $K^{1+{\tilde\psi_1\over 2}+s}$ has the following expansion
\ie\label{z limit}
K^{1+{\tilde\psi_1\over 2}+s}\rightarrow \pi\sum^{s}_{q=0}  {\Gamma(s-q+{\tilde\psi_1\over 2})\over q!\Gamma(1+s+{\tilde\psi_1\over 2})}z^{2q+1-{\tilde\psi_1\over 2}-s}(\partial_{x^+}\partial_{x^-})^q\delta^2(x) + z^{1+{\tilde\psi_1\over 2}+s}{1\over x^{2+\tilde \psi_1+2s}}+\cdots,
\fe
where we keep only the leading analytic term and the first $s$ contact terms. The subleading terms will not contribute to the three point function.

Let us first compute $\delta S_{1}$.
\ie
&\delta S_{1}
\\
&=4\lim_{z\rightarrow 0}\int d^2 x_0 \sum^{2s-1}_{r=1}(2-s){1\over (x^-_{03})^r}z^{r-s-2}(\partial_{y^2})^{2s-r}(-\partial_{y^1})^{r}\delta C_{mat}^{(1),0}(x_{01}) C_{mat}^{(1),2s} (x_{02}|y)
\\
&=4\lim_{z\rightarrow 0}\int d^2 x_0 \sum^{2s-1}_{r=1}\psi_1^s\left(1+s(1+\tilde\psi_1)\right)\sum^s_{u=0}\sum^{2u-r}_{v=0}{(2-s) r!(2s-r)!(-1)^{-u+v}\over  (s-u)!(r-u)!(2u-r-v)!v!}
\\
&~~~~\times z^{2u-2v-s-2}(x_{02}^-)^{r-u+v}(x_{02}^+)^{s-u+v}{1\over (x^-_{03})^r} K^{1+{\tilde\psi_1\over 2}}_{01} K^{1+{\tilde\psi_1\over 2}+s}_{02}
\\
&=4\int d^2 x_0 \sum^{2s-1}_{r=1}\psi_1^s\left(1+s(1+\tilde\psi_1)\right)\sum^s_{u=0}\sum^{2u-r}_{v=0}{(2-s) r!(2s-r)!(-1)^{-u+v}\over  (s-u)!(r-u)!(2u-r-v)!v!}
\\
&~~\times \bigg[ \pi^{3\over 2} {\Gamma({1\over 2}\tilde\psi_1)\over \Gamma({1\over 2})\Gamma(1+{\tilde\psi_1\over 2})}\delta^2(x_{01}) {1\over x_{02}^{2+\tilde\psi_1+2s}}(x_{02}^-)^{r}(x_{02}^+)^{s}\delta_{u,v}{1\over (x^-_{03})^r}
\\
&~~~~+\delta_{v,u+q-s}   \pi \sum^{s}_{q=0}  {\Gamma(s-q+{\tilde\psi_1\over 2})\over \Gamma(1+s+{\tilde\psi_1\over 2})}\delta^2(x_{02})\sum^q_{n=0}{q!(q+r-s)!\over (q-n)!n!(r-s+n)!}(x^-_{02})^{r-s+n}\partial_{x_0^-}^n\left({1\over (x^-_{03})^r}{1\over x_{01}^{2+\tilde\psi_1}}\right)\bigg],
\\
%&=4\int d^2 x_0 \sum^{2s-1}_{r=1}(2-s)\psi_1^s\left(1+s(1+\tilde\psi_1)\right) \bigg[2\pi \tilde\psi_1 { (2s-r)!\over  (s-r)!}\delta^2(x_{01}) {1\over x_{02}^{2+\tilde\psi_1}}{1\over (x_{02}^-)^{s-r}}{1\over (x^-_{03})^r}
%\\
%&~~+\sum^s_{u=0} \sum^{s}_{q=0} { r!(2s-r)!\Gamma(s-q+{\tilde\psi_1\over 2}) q! (-1)^{q-s}\over  (s-u)!(r-u)!(u-r-q+s)!(u+q-s)!\Gamma(1+s+{\tilde\psi_1\over 2})(s-r)!}\pi \delta^2(x_{02})\partial_{x^-}^{s-r}\left({1\over (x^-_{03})^r}{1\over x_{01}^{2+\tilde\psi_1}}\right)\bigg],
\fe
where we have substituted the boundary-to-bulk propagator for $\delta C_{mat}^{(1),0}(x_{01})$ and $C_{mat}^{(1),2s} (x_{02}|y)$, and the $K_{ij}$ stands for $K\big|_{x\rightarrow x_{ij}}$, and we have substituted the expansion (\ref{z limit}) for $K_{ij}$. Integrating out the delta functions gives
\ie
&\delta S_1
=4 \sum^{2s-1}_{r=1}(2-s)\psi_1^s\left(1+s(1+\tilde\psi_1)\right) \bigg[2\pi \tilde\psi_1 { (2s-r)!\over  (s-r)!}  {1\over x_{12}^{2+\tilde\psi_1} (x_{12}^-)^{s-r}  (x^-_{13})^r}
\\
&+\sum^s_{u=0} \sum^{s}_{q=0} { r!(2s-r)!\Gamma(s-q+{\tilde\psi_1\over 2}) q! (-1)^{q-s}\over  (s-u)!(r-u)!(u-r-q+s)!(u+q-s)!\Gamma(1+s+{\tilde\psi_1\over 2})(s-r)!}\pi \partial_{x_2^-}^{s-r}\left({1\over (x^-_{23})^r x_{21}^{2+\tilde\psi_1}}\right)\bigg].
\fe

Similarly, let us compute $\delta S_{2}$ and $\delta S_{3}$ as follows. Substituting the boundary-to-bulk propagator for the master field $C_{mat}^{(1)}$, we have
\ie
&\delta S_{2}
=-  4\lim_{z\rightarrow 0}\int d^2 x_0 \sum^{2s-1}_{r=1}{z^{r-s-2}\over (2s-1)}{1\over (x^-_{03})^r} (\partial_{y^2})^{2s-r}(-\partial_{y^1})^{r}\delta C_{mat}^{(1),2}(x_{01}) C_{mat}^{(1),2s-2}(x_{02}|y)
\\
&=-  4\lim_{z\rightarrow 0}\int d^2 x_0 \sum^{2s-1}_{r=1}{1\over (2s-1)} {1\over (x^-_{03})^r}\psi_1^{s}\left(1+(s-1)(1+\tilde\psi_1)\right)(2+\tilde\psi_1) K_{01}^{2+{\tilde\psi_1\over 2}}K_{02}^{{\tilde\psi_1\over 2}+s}
\\
&~~\times\bigg[\sum^{s-1}_{u=0}\sum^{2u-r+1}_{v=0}{r!(2s-r)!(-1)^r\over (s-u-1)!(2u-r+1-v)!(r-u-1)!v!}
\\
&~~~~~~\times\left(z-{x_{01}^+x_{01}^-\over z}\right)z^{2u-2v-s-1} (-x_{02}^-)^{r-u+v-1}(x_{02}^+)^{s-u+v-1} 
\\
&~~+\sum^{s-1}_{u=0} \sum^{2u-r+2}_{v=0}{r!(2s-r)!(-1)^r\over (s-u-1)!(2u-r+2-v)!(r-u-2)!v!}(-x^-_{01})z^{2u-2v-s} (-x_{02}^-)^{r-u+v-2}(x_{02}^+)^{s-u+v-1} 
\\
&~~+\sum^{s-1}_{u=0}\sum^{2u-r}_{v=0} {r!(2s-r)!(-1)^r\over (s-u-1)!(2u-r-v)!(r-u)!v!}(x^+_{01})z^{2u-2v-s-2} (-x_{02}^-)^{r-u+v}(x_{02}^+)^{s-u+v-1} \bigg],
\fe
and
\ie
&\delta S_{3}
=-  4\lim_{z\rightarrow 0}\int d^2 x_0 \sum^{2s-1}_{r=1}{z^{r-s-2}\over (2s-1)}{1\over (x^-_{03})^r} (2s-r-1)(r-1)
\\
&~~~~~~~\times (\partial_{y^2})^{2s-r-2}(-\partial_{y^1})^{r-2}\delta C_{mat}^{(1),2}(x_{01}|\partial_y) C_{mat}^{(1),2s-2}(x_{02}|y)
\\
&=- 4\lim_{z\rightarrow 0}\int d^2 x_0 \sum^{2s-1}_{r=1} {1\over (2s-1)}{1\over (x^-_{03})^r} (2s-r-1)(r-1)\psi_1^{s}\left(1+(s-1)(1+\tilde\psi_1)\right) (2+\tilde\psi_1)
\\
&~~~\times K_{01}^{2+{\tilde\psi_1\over 2}} K_{02}^{{\tilde\psi_1\over 2}+s}\bigg[\sum^{s-1}_{u=0} \sum^{2u-r+1}_{v=0}{(r-1)!(2s-r-1)!(-1)^{r-1}\over (s-u-1)!(2u-r+1-v)!(r-u-1)!v!}
\\
&~~~~~~~~\times\left(z-{x_{01}^+x_{01}^-\over z}\right)z^{2u-2v-s-1}  (-x_{02}^-)^{r-u+v-1}(x_{02}^+)^{s-u+v-1} 
\\
&~~~~+\sum^{s-1}_{u=0}\sum^{2u-r+2}_{v=0} {(r-2)!(2s-r)!(-1)^{r-1}\over (s-u-1)!(2u-r+2-v)!(r-u-2)!v!}(x^-_{01})z^{2u-2v-s} (-x_{02}^-)^{r-u+v-2}(x_{02}^+)^{s-u+v-1}
\\
&~~~~+\sum^{s-1}_{u=0}\sum^{2u-r}_{v=0} {r!(2s-r-2)!\over (s-u-1)!(2u-r-v)!(r-u)!v!}(-1)^r (x^+_{01}) z^{2u-2v-s-2} (-x_{02}^-)^{r-u+v}(x_{02}^+)^{s-u-1+v} \bigg].
\fe
These two terms can be combined as
\ie
&\delta S_{2}+\delta S_{3}
%\\
%&=-4\lim_{z\rightarrow 0}\int d^2 x_0 \sum^{2s-1}_{r=1}\bigg[  {z^{r-s-2}\over (2s-1)}{1\over (x^-_{03})^r} (\partial_{y^2})^{2s-r}(-\partial_{y^1})^{r}\delta C_{mat}^{(1),2}(x_{01}) C_{mat}^{(1),2s-2}(x_{02}|y)
%\\
%&~~~~+ {z^{r-s-2}\over (2s-1)}{1\over (x^-_{03})^r}(2s-r-1)(r-1)(\partial_{y^2})^{2s-r-2}(-\partial_{y^1})^{r-2}\delta C_{mat}^{(1),2}(x_{01}|\partial_y) C_{mat}^{(1),2s-2}(x_{02}|y)\bigg]
\\
&=- 4\lim_{z\rightarrow 0}\int d^2 x_0 \sum^{2s-1}_{r=1}\psi_1^{s}\left(1+(s-1)(1+\tilde\psi_1)\right)(2+\tilde\psi_1) K_{01}^{2+{\tilde\psi_1\over 2}}K_{02}^{{\tilde\psi_1\over 2}+s}{1\over (x^-_{03})^r}
\\
&~~\times\bigg[\sum^{s-1}_{u=0}\sum^{2u-r+1}_{v=0}{(r-1)!(2s-r-1)!(-1)^r\over (s-u-1)!(2u-r+1-v)!(r-u-1)!v!}
\\
&~~~~~~\times\left(z-{x_1^+x_1^-\over z}\right)z^{2u-2v-s-1} (-x_{02}^-)^{r-u+v-1}(x_{02}^+)^{s-u+v-1} 
\\
&~~+\sum^{s-1}_{u=0}\sum^{2u-r+2}_{v=0} {(r-1)!(2s-r)!(-1)^r\over (s-u-1)!(2u-r+2-v)!(r-u-2)!v!}(-x^-_{01})z^{2u-2v-s} (-x_{02}^-)^{r-u+v-2}(x_{02}^+)^{s-u+v-1} 
\\
&~~+\sum^{s-1}_{u=0}\sum^{2u-r}_{v=0} {r!(2s-r-1)!(-1)^r\over (s-u-1)!(2u-r-v)!(r-u)!v!}(x^+_{01})z^{2u-2v-s-2} (-x_{02}^-)^{r-u+v}(x_{02}^+)^{s-u+v-1} \bigg]
\\
&\equiv U_1+U_2+U_3,
\fe
where we have split $\delta S_{2}+\delta S_{3}$ into three terms $U_1,U_2,U_3$. These are computed as follows.
\ie
& U_1 =- 4\int d^2 x_0 \sum^{2s-1}_{r=1} \psi_1^{s}\left(1+(s-1)(1+\tilde\psi_1)\right)(2+\tilde\psi_1)
\\
&~~\times\sum^{s-1}_{u=0} \bigg[-{2\pi\over 2+\tilde\psi_1}\delta^2(x_{01}) {1\over x_{02}^{\tilde\psi_1+2}} {1\over (x_{02}^-)^{s-r}}{1\over (x^-_{03})^r}{(r-1)!(2s-r-1)!\over (s-u-1)!(u-r+1)!(r-u-1)!u!}
\\
&~~ + {4\pi\over 2\tilde\psi_1+1}\delta^2(x_{01})  {1\over x_{02}^{\tilde\psi_1+2}} {1\over (x_{02}^-)^{s-r}}{1\over (x^-_{03})^r} {(r-1)!(2s-r-1)!\over (s-u-1)!(u-r+1)!(r-u-1)!u!}
\\
&~~+\sum^{s-1}_{q=0}  {(r-1)!(2s-r-1)!\Gamma(s-1-q+{\tilde\psi_1\over 2})q!(-1)^{s+q+1}\over (s-u-1)!(u-r-q+s)!(r-u-1)!(q+u-s+1)! \Gamma(s+{\tilde\psi_1\over 2})(s-r)!}
\\
&~~~~~~~~\times\pi\delta^2(x_{02})\partial_{x_0^-}^{s-r}\left( {1\over x_{01}^{2+\tilde\psi_1}}{1\over (x^-_{03})^r}\right)\bigg]
\\
%&=-  4\int d^2 x_0 \sum^{2s-1}_{r=1}\psi_1^{s}\left(1+(s-1)(1+\tilde\psi_1)\right)(2+\tilde\psi_1)
%\\
%&~~\times\bigg[{10\tilde\psi_1-8\over 3}\pi{(2s-r-1)!\over (s-r)!}\delta^2(x_{01}) {1\over x_{02}^{\tilde\psi_1+2}} {1\over (x_{02}^-)^{s-r}}{1\over (x^-_{03})^r}
%\\
%&~~+\sum^{s-1}_{u=0} \sum^{s-1}_{q=0}  {(r-1)!(2s-r-1)!\Gamma(s-1-q+{\tilde\psi_1\over 2})q!(-1)^{s+q+1}\over (s-u-1)!(u-r-q+s)!(r-u-1)!(q+u-s+1)! \Gamma(s+{\tilde\psi_1\over 2})(s-r)!}
%\\
%&~~~~~~~~\times\pi\delta^2(x_{02})\partial_{x_0^-}^{s-r}\left( {1\over x_{01}^{2+\tilde\psi_1}}{1\over (x^-_{03})^r}\right) \bigg]
&=-  4 \sum^{2s-1}_{r=1}\psi_1^{s}\left(1+(s-1)(1+\tilde\psi_1)\right)(2+\tilde\psi_1)
\bigg[{10\tilde\psi_1-8\over 3}\pi{(2s-r-1)!\over (s-r)!} {1\over x_{12}^{\tilde\psi_1+2} (x_{12}^-)^{s-r} (x^-_{13})^r}
\\
&~~+\sum^{s-1}_{u=0} \sum^{s-1}_{q=0}  {(r-1)!(2s-r-1)!\Gamma(s-1-q+{\tilde\psi_1\over 2})q!(-1)^{s+q+1}\over (s-u-1)!(u-r-q+s)!(r-u-1)!(q+u-s+1)! \Gamma(s+{\tilde\psi_1\over 2})(s-r)!}
\\
&~~~~~~~~\times\pi \partial_{x_2^-}^{s-r}\left( {1\over x_{21}^{2+\tilde\psi_1} (x^-_{23})^r}\right) \bigg],
\fe
\ie
U_2&=- 4\lim_{z\rightarrow 0}\int d^2 x_0 \sum^{2s-1}_{r=1}\psi_1^{s}\left(1+(s-1)(1+\tilde\psi_1)\right)(2+\tilde\psi_1) {1\over (x^-_{03})^r}
\\
&~~\times\sum^{s-1}_{u=0}\sum^{2u-r+2}_{v=0} {(r-1)!(2s-r)!\over (s-u-1)!(2u-r+2-v)!(r-u-2)!v!}(-1)^r(-x^-_{01}) (-x_{02}^-)^{r-u+v-2}(x_{02}^+)^{s-u+v-1} 
\\
&~~\times\bigg[\pi \sum^{1}_{q=0}  {\Gamma(1-q+{\tilde\psi_1\over 2})\over q!\Gamma(2+{\tilde\psi_1\over 2})}(\partial_{x_{0}^+}\partial_{x_{0}^-})^q\delta^2(x_{01})  {1\over x_{02}^{\tilde\psi_1+2s}}z^{2u-2v+2q}
\\
&~~ {1\over x_{01}^{2+\tilde\psi_1+4}}\pi \sum^{s-1}_{q=0}  {\Gamma(s-1-q+{\tilde\psi_1\over 2})\over q!\Gamma(s+{\tilde\psi_1\over 2})}z^{2u-2v+2q+4-2s}(\partial_{x_0^+}\partial_{x_0^-})^q\delta^2(x_{02}) \bigg],
\\
&=0,
\fe
and
\ie
U_3%&=-4\int d^2 x_0 \sum^{2s-1}_{r=1} \psi_1^{s}\left(1+(s-1)(1+\tilde\psi_1)\right)(2+\tilde\psi_1) 
%\\
%&\times\bigg[ {4\pi\over 1+2\tilde\psi_1}{(2s-r-1)!\over (s-r-1)!} \delta^2(x_{01})\partial_{x_0^-}\left({1\over x_{02}^{2+\tilde\psi_1}}{1\over (x_{02}^-)^{s-r-1}}{1\over (x^-_{03})^r}\right)
%\\
%&+\sum^{s-1}_{q=0}  {\over }\sum^{s-1}_{u=0} {\Gamma(s-1-q+{\tilde\psi_1\over 2})r!(2s-r-1)!q! \pi (-1)^{1+s+q}\over\Gamma(s+{\tilde\psi_1\over 2}) (s-u-1)!(u-r-q+s-1)!(r-u)!(q+1+u-s)!(s-r-1)!}
%\\
%&~~~~~~\times\delta^2(x_{02})\partial_{x_0^-}^{s-r-1}\left({1\over x_{01}^{2+\tilde\psi_1}}{1\over (x^-_{01})(x^-_{03})^r}\right)\bigg]
&=-4 \sum^{2s-1}_{r=1} \psi_1^{s}\left(1+(s-1)(1+\tilde\psi_1)\right)(2+\tilde\psi_1) 
\left[ {4\pi\over 1+2\tilde\psi_1}{(2s-r-1)!\over (s-r-1)!} \partial_{x_1^-}\left({1\over x_{12}^{2+\tilde\psi_1}  (x_{12}^-)^{s-r-1}  (x^-_{13})^r}\right)\right.
\\
&+\sum^{s-1}_{q=0}  {\over }\sum^{s-1}_{u=0} {\Gamma(s-1-q+{\tilde\psi_1\over 2})r!(2s-r-1)!q! \pi (-1)^{1+s+q}\over\Gamma(s+{\tilde\psi_1\over 2}) (s-u-1)!(u-r-q+s-1)!(r-u)!(q+1+u-s)!(s-r-1)!}
\\
&~~~~~~\left.\times \partial_{x_2^-}^{s-r-1}\left({1\over x_{21}^{2+\tilde\psi_1} (x^-_{21})(x^-_{23})^r}\right)\right].
\fe
where we have substituted the expansion (\ref{z limit}) and taken the $z\rightarrow 0$ limit.
Finally, let us compute $\delta S_{4}$:
\ie
&\delta S_{4}
=-4\lim_{z\rightarrow 0}\int d^2 x_0 \sum^{2s-1}_{r=1} (2s-r-1)(r-1){1\over (x^-_{03})^r} z^{r-s-2}(s+1) 
\\
&~~~~\times(\partial_{y^2})^{2s-r-2}(-\partial_{y^1})^{r-2} \delta  C_{mat}^{(1),0}(x_{01}) C_{mat}^{(1),2s-4}(x_{02}|y)
\\
&=-4\lim_{z\rightarrow 0}\int d^2 x_0 \sum^{2s-1}_{r=1}(-1)^{r-2}   {1\over (x^-_{03})^r} K_{01}^{1+{\tilde\psi_1\over 2}}K_{02}^{{\tilde\psi_1\over 2}+s-1}{\psi_1^s\over (s-2)!}\left(1+(s-2)(1+\tilde\psi_1)\right)
\\
&~~\times\sum^{s-2}_{u=0}\sum^{2u-r+2}_{v=0} {(s-2)! (r-1)! (2s-r-1)!\over (s-u-2)!(2u-r+2-v)!(r-u-2)!v!}z^{2u-2v-s} (-x_{02}^-)^{r-u+v-2}(x_{02}^+)^{s-u+v-2}.
\fe
After substituting the boundary to bulk propagators and taking the $z\to 0$ limit, we obtain
\ie
\delta S_4 %&=-4\int d^2 x_0 \sum^{2s-1}_{r=1}(s+1)  \psi_1^s\left(1+(s-2)(1+\tilde\psi_1)\right)
%\\
%&~~\times\bigg[\pi   {\Gamma({\tilde\psi_1\over 2})\over \Gamma(1+{\tilde\psi_1\over 2})}\delta^2(x_{01})  {1\over x_{02}^{\tilde\psi_1+2s-2}} { (r-1) (2s-r-1)!\over (s-r)!}(x_{02}^-)^{r-2}(x_{02}^+)^{s-2} {1\over (x^-_{03})^r}
%\\
%&~~+  \pi \sum^{s-2}_{q=0} \sum^{s-2}_{u=0} {\over }  {\Gamma(s-2-q+{\tilde\psi_1\over 2})(r-1)! (2s-r-1)!q!\over \Gamma(s-1+{\tilde\psi_1\over 2})(s-u-2)!(u-r-q+s)!(r-u-2)!(q+u-s+2)!(s-r)!} 
%\\
%&~~~~~~\times\delta^2(x_{02})(-1)^{q-s}\partial_{x_0^-}^{s-r}\left({1\over x_{01}^{2+\tilde\psi_1}}{1\over (x^-_{03})^r}\right)\bigg]
&=-4 \sum^{2s-1}_{r=1}(s+1)  \psi_1^s\left(1+(s-2)(1+\tilde\psi_1)\right)
\\
&~~\times\bigg[\pi   {\Gamma({\tilde\psi_1\over 2})\over \Gamma(1+{\tilde\psi_1\over 2})}  {1\over x_{12}^{\tilde\psi_1+2s-2}} { (r-1) (2s-r-1)!\over (s-r)!} {(x_{12}^-)^{r-2}(x_{12}^+)^{s-2}\over (x^-_{13})^r}
\\
&~~+  \pi \sum^{s-2}_{q=0} \sum^{s-2}_{u=0} {\over }  {\Gamma(s-2-q+{\tilde\psi_1\over 2})(r-1)! (2s-r-1)!q!\over \Gamma(s-1+{\tilde\psi_1\over 2})(s-u-2)!(u-r-q+s)!(r-u-2)!(q+u-s+2)!(s-r)!} 
\\
&~~~~~~\times (-1)^{q-s}\partial_{x_2^-}^{s-r}\left({1\over x_{21}^{2+\tilde\psi_1}}{1\over (x^-_{23})^r}\right)\bigg].
\fe

The three point function is proportional to $\delta S=\delta S_1+U_1+U_3+\delta S_4$. One can simplify the above expressions and compute the full three point function directly, but since we are only interested in the overall coefficient whereas the position dependence is completely fixed by the conformal symmetry, we can take the limit in which one of the two scalar operators collides with the higher spin current, and extract the overall coefficient.

Let us define the variables $y_1^{\pm}=x_1^{\pm}-x_3^{\pm}$ and $y_2^{\pm}=x_2^{\pm}-x_3^{\pm}$, and consider the limit $y_1\ll y_2$. The various pieces of contributions are given in this limit by
\ie
\delta S_{1}\rightarrow&4(2-s)\psi_1^s\left(1+s(1+\tilde\psi_1)\right) 2\pi \tilde\psi_1  s! {1\over y_{2}^{2+\tilde\psi_1}}{1\over (y^-_1)^s},
\\
U_1\rightarrow&- 4\psi_1^{s}\left(1+(s-1)(1+\tilde\psi_1)\right)(2+\tilde\psi_1){10\tilde\psi_1-8\over 3}\pi (s-1)! {1\over y_{2}^{\tilde\psi_1+2}} {1\over (y^-_1)^s},
\\
U_3\rightarrow&-4 \psi_1^{s}\left(1+(s-1)(1+\tilde\psi_1)\right)(2+\tilde\psi_1)   {4\pi\over 1+2\tilde\psi_1}s! {1\over y_{2}^{2+\tilde\psi_1}}\,{-s+1\over (y^-_1)^{s}},
\\
\delta S_{4}\rightarrow&-4(s+1)  \psi_1^s\left(1+(s-2)(1+\tilde\psi_1)\right) \pi   {\Gamma({\tilde\psi_1\over 2})\over \Gamma(1+{\tilde\psi_1\over 2})}  (s-1) (s-1)! {1\over (y^-_1)^s} {1\over y_{2}^{\tilde\psi_1+2}}.
\fe
Summing these four terms, and recovering the full position dependence using the conformal symmetry, we obtain the three point function of one higher spin current and two scalar operators:
\ie\label{tmaa}
&\vev{\left(\cO+\overline{\cO}\right)(x_1)\left(\cO+\overline{\cO}\right)(x_2)J^s(x_3)}=8\pi (s+\tilde\psi_1(s-1))(1+(-)^s)\Gamma(s) {1\over |x_{12}|^{2+\tilde\psi_1}}\left({x_{12}^-\over x^-_{13} x^-_{23}}\right)^s.
\fe
Note that since we have turned on the sources for $C_{even}$ so far, the dual scalar operator is ${\cal O}+\overline{\cal O}$. The three point function involving an insertion of ${\cal O}-\overline{\cal O}$, dual to the bulk field $C_{odd}$, can be computed analogously by turning on a source for $C_{odd}$. Note that $C_{odd}$ is a purely imaginary field; in other words, if we write $C_{odd}=i\varphi$, then $\varphi$ is a real field with the ``right sign" kinetic term. A computation similar to the above gives
\ie\label{tmbb}
&\vev{\left(\cO-\overline{\cO}\right)(x_1)\left(\cO+\overline{\cO}\right)(x_2)J^s(x_3)}=%-i
8\pi (s+\tilde\psi_1(s-1))(1-(-)^s)\Gamma(s) {1\over |x_{12}|^{2+\tilde\psi_1}}\left({x_{12}^-\over x^-_{13} x^-_{23}}\right)^s.
\fe
%Adding these two terms and expressing the result in terms of the normalized higher spin current $j_s$ (which differs from $J_s$ by a factor of $i$ when $s$ is odd), we have 
Adding (\ref{tmaa}) and (\ref{tmbb}), we obtain
\ie
&\vev{\overline{\cO}(x_1)\cO(x_2)J^s(x_3)}=-4\pi (s+\tilde\psi_1(s-1))\Gamma(s) {1\over |x_{12}|^{2+\tilde\psi_1}}\left({x_{12}^-\over x^-_{13} x^-_{23}}\right)^s.
\fe

\section{The deformed vacuum solution}

In this section, we discuss the formulation of the three dimensional Vasiliev system as originally written in \cite{Vasiliev:1999ba}, which amounts to an extension of the equations (\ref{Vasiliev's eq}) by introducing two additional auxiliary variables $k$ and $\rho$, as described below, and the 1-parameter family of ``deformed" vacuum solutions. The deformed vacuum solution of the system (\ref{Vasiliev's eq}) can be obtain by a simple projection on the extended system. We will also present the boundary to bulk propagator for the $B$ master field, which contains the bulk ``matter" scalar field, in the deformed vacua, by solving the linearized equations.

To describe the deformed vacuum, it is useful to introduce two additional auxiliary variables $k$ and $\rho$. They obey the following (anti-)commutation relations with one another and with the twistor variables $(y,z)$:
\ie
k^2=\rho^2=1,~~~~\{k,\rho\}=\{k,y_\A\}=\{k,z_\A\}=0,~~~~[\rho,y_\A]=[\rho,z_\A]=0.
\fe
It will be also convenient to define the variable
\ie
w_\A =(z_\A+y_\A)\int^1_0 dt \,t e^{tzy}.
\fe
It is straightforward to show that $w_\A$ satisfy the following star commutation relations:
\ie\label{w relations}
&[w_\A,w_\B]_*=0,
\\
&\left[w_\A,y_\B\right]_*+\left[y_\A,w_\B\right]_*=2\epsilon_{\af\beta} K,
\\
&\left[w_\A,z_\B\right]_*+\left[z_\A,w_\B\right]_*=-2\epsilon_{\af\beta} K,
\\
&\{w_\A , z_\B\}_* *K- \{y_\A,w_\B\}_*=0.
\fe
Next, let us define
\ie
&\tz_\A (\n)=z_\A+\n w_\A k,
\\
&\ty_\A(\n)=y_\A+\n w_\A *Kk.
\fe
Using the relations (\ref{w relations}), it is easy to show that
\ie
&\left[\tilde y_\af,\tilde y_\beta\right]_*=2\epsilon_{\af\beta}(1+\n k),
\\
&\left[\rho \tz_\af,\rho \tz_\beta\right]_*=-2\epsilon_{\af\beta}\left(1+\n Kk\right),
\\
&\left[\rho\tz_\af,\tilde y_\beta\right]_*=0.
\fe
Under the star algebra, $\ty_\A$ generate the (deformed) three dimensional higher spin algebra $hs(\lambda)$ with $\lambda ={1\over 2} (1+ \nu k)$. Later we will make the projection onto the eigenspace of $k=1$ or $k=-1$, in which case $\lambda={1\over 2}(1+\nu)$ or $\lambda={1\over 2}(1-\nu)$. The higher spin algebra $hs(\lambda)$ is an associative algebra, whose general element can be represented by an even analytic star-function in $\ty_\A$. In particular, it has an $sl(2)$-subalgebra whose generator can be written as $T_{\A\B}=\ty_{(\A}*\ty_{\B)}$.

The deformed vacuum solution is given by
\ie
&B={1\over 4}\n,~~~~S_\A={1\over 2}\rho (\tz_\A -z_\A),
\\
&W=W_0=w_0+\psi_1 e_0=\left(w_0^{\A\B}(x) +\psi_1 e_0^{\A\B}(x)\right) T_{\A\B}.
\fe
They satisfy the $(k,\rho)$-extended Vasiliev equations:\footnote{Note that the form of these equations differs from the system (\ref{Vasiliev's eq}) only in the RHS of the third equation.}
\ie
&d_x W + W*W = 0,
\\
&d_x S + d_z W + \{W,S\}_* = 0,
\\
&d_zS + S*S = B*K k dz^2,
\\
&d_z B + [S,B]_* = 0,
\\
&d_x B + [W,B]_* = 0,
\fe
We can go back to the system (\ref{Vasiliev's eq}) by simply multiplying a projector ${1\over 2}(1+k)$ on the left of every equation. Given any solution of the extended Vasiliev equations, by acting on it with the projector we obtain a solution of the equations (\ref{Vasiliev's eq}). It follows that the deformed vacuum solution of (\ref{Vasiliev's eq}) is
\ie
&B={1\over 4}\n,~~~~S_\A={1\over 2}\left( \tz_\A(-\n) -z_\A\right),
\\
&W=\left(w_0^{\A\B}(x) +\psi_1 e_0^{\A\B}(x)\right) \ty_\A(\n)*\ty_\B(-\n).
\fe

Next, we will solve the linearize equation on the deformed vacua, and derive the boundary to bulk propagator for $B$ (the scalar and corresponding auxiliary fields). For simplicity of the notation, we will work in the extended Vasiliev system. The boundary to bulk propagator for fields in the system (\ref{Vasiliev's eq}) can be obtained simply by applying the projector ${1\over 2}(1+k)$. The linearized equations for $B$ are
\ie\label{lin sc de}
&\left[\rho\tilde z_\af,B^{(1)}\right]_*=0,
\\
&D_0 B^{(1)}=0.
\fe
where $D_0$ is defined by $D_0\equiv d+[W_0,\cdot]$. The first equation of (\ref{lin sc de}) immediately implies $B^{(1)}(x|y,z,\psi)=B^{(1)}_{*}(x|\ty,\psi)$, where the subscript $*$ of a function means that it is a star-function.

Decomposing $B^{(1)}_{*}(x|\ty,\psi)$ as $B^{(1)}_{*}(x|\ty,\psi)=C^{(1)}_{aux*}(x|\ty,\psi_1)+\psi_2 C^{(1)}_{mat*}(x|\ty,\psi_1)$, the second equation of (\ref{lin sc de}) gives
\ie\label{deq}
&dC^{(1)}_{aux*}+[w_0,C^{(1)}_{aux*}]_*+\psi_1[e_0,C^{(1)}_{aux*}]_*=0,
\\
&dC^{(1)}_{mat*}+[w_0,C^{(1)}_{mat*}]_*-\psi_1\{e_0,C^{(1)}_{mat*}\}_*=0.
\fe
As in the case of equations in the undeformed vacuum analyzed in section 3.1 and Appendix A.1, the equation for $C^{(1)}_{aux*}$ is over-constraining, and eliminates all dynamical degrees of freedom of $C^{(1)}_{aux*}$. We will simply set $C^{(1)}_{aux*}=0$, and only study the equation of the ``matter" component $C^{(1)}_{mat*}$ in the following. Let us expand $C^{(1)}_{mat*}$ in the form
\ie
C^{(1)}_{mat*}(\ty)=\sum_{n=0}^\infty C^{(1)}_{mat*,}{}_{\A_1\cdots\A_n}\tilde y^{(\af_1}*\cdots*\tilde y^{\af_n)}.
\fe
To compute the (anti-)commutators in (\ref{deq}), let us first consider the star product of $\ty^\A$ with $\tilde y^{(\af_1}*\cdots*\tilde y^{\af_n)}$:
\ie
&\tilde y^\af*\tilde y^{(\af_1}*\cdots*\tilde y^{\af_n)}\\
&=\tilde y^{(\af}*\tilde y^{\af_1}*\cdots*\tilde y^{\af_n)}+{1\over n+1}\sum^n_{i=1}(n-i+1)\ty^{(\underline{\af_1}}*\cdots*[\ty^\af,\ty^{\underline{\af_i}}]_**\cdots*\ty^{\underline{\af_n})}
\\
&=\tilde y^{(\af}*\tilde y^{\af_1}*\cdots*\tilde y^{\af_n)}+{1\over n+1}\sum^n_{i=1}(n-i+1)(1+(-)^{i-1}\n k)2\epsilon^{\af(\af_i}\ty^{\af_1}*\cdots*{\slash\!\!\!\ty^{{\slash\!\!\!\A_i}}}*\cdots*\ty^{\af_n)}.
\fe
Contracting the above with $e_\af C_{\af_1\cdots\af_n}$ (here and in what follows, $e$ and $C$ are used to denote arbitrary totally symmetric tensors), we obtain
\ie
&e_\af\tilde y^\af*C_{\af_1\cdots\af_n}\tilde y^{\af_1}*\cdots*\tilde y^{\af_n}\\
&=e_{(\underline{\af}}C_{\underline{\af_1\cdots\af_n})}\tilde y^{\af}*\tilde y^{\af_1}*\cdots*\tilde y^{\af_n}- a(n,\n k)e^\af C_{\af\af_1\cdots\af_{n-1}}\ty^{\af_1}*\cdots*\ty^{\af_{n-1}},
\fe
where
\ie
a(n,\n k)=2\sum^n_{i=1}{1\over (n+1)}(n-i+1)(1+(-)^{i-1}\n k).
\fe
Applying a similar operation, staring $\tilde y^{(\af}*\tilde y^{\B)}$ with $\tilde y^{(\af_1}*\cdots*\tilde y^{\af_n)}$ and contracting with $e_{\beta\af}C_{\af_1\cdots\af_n}$, we get
\ie\label{star yyl}
&e_{\beta\af}\ty^\beta*\tilde y^\af*C_{\af_1\cdots\af_n}\tilde y^{\af_1}*\cdots*\tilde y^{\af_n}
=e_{(\underline{\beta\af}}C_{\underline{\af_1\cdots\af_n})}\ty^\beta*\tilde y^{\af}*\tilde y^{\af_1}*\cdots*\tilde y^{\af_n}\\
&~~~~-{n\over n+1}a(n+1,\n k)e^\beta{}_{(\underline{\af}}C_{\beta\underline{\af_1\cdots\af_{n-1}})}\tilde y^{\af}*\tilde y^{\af_1}*\cdots*\tilde y^{\af_{n-1}}\\
&~~~~-a(n,-\n k)e_{(\underline{\beta}}{}^\af C_{\af\underline{\af_1\cdots\af_{n-1}})}\ty^\beta*\ty^{\af_1}*\cdots*\ty^{\af_{n-1}}\\
&~~~~+a(n,-\n k)a(n-1,\n k)e^{\af\beta} C_{\af\beta\af_1\cdots\af_{n-2}}\ty^{\af_1}*\cdots*\ty^{\af_{n-2}}.
\fe
Now, starring $\ty^\af$ with $\tilde y^{(\af_1}*\cdots*\tilde y^{\af_n)}$ from the right side,
\ie
&\tilde y^{(\af_1}*\cdots*\tilde y^{\af_n)}*\tilde y^\af\\
&=\tilde y^{(\af}*\tilde y^{\af_1}*\cdots*\tilde y^{\af_n)}+{1\over n+1}\sum^n_{i=1}(-i)\ty^{(\underline{\af_1}}*\cdots*[\ty^\af,\ty^{\underline{\af_i}}]_**\cdots*\ty^{\underline{\af_n})}\\
&=\tilde y^{(\af}*\tilde y^{\af_1}*\cdots*\tilde y^{\af_n)}+{1\over n+1}\sum^n_{i=1}(-i)(1+(-)^{i-1}\n k)2\epsilon^{\af(\af_i}\ty^{\af_1}*\cdots*\ty^{\not\af_i}*\cdots*\ty^{\af_n)}.
\fe
Contracting this formula with $e_\af C_{\af_1\cdots\af_n}$, we have
\ie
&C_{\af_1\cdots\af_n}\tilde y^{\af_1}*\cdots*\tilde y^{\af_n}*e_\af\tilde y^\af\\
&=e_{(\underline{\af}}C_{\underline{\af_1\cdots\af_n})}\tilde y^{\af}*\tilde y^{\af_1}*\cdots*\tilde y^{\af_n}-b(n,\n k)e^\af C_{\af\af_1\cdots\af_{n-1}}\ty^{\af_1}*\cdots*\ty^{\af_{n-1}},
\fe
where
\ie
b(n,\n k)=2\sum^n_{i=1}{1\over (n+1)}(-i)(1+(-)^{i-1}\n k).
\fe
Performing a similar operation with $\ty^{(\A}*\tilde y^{\B)}$, we obtain
\ie\label{star yyr}
&C_{\af_1\cdots\af_n}\tilde y^{\af_1}*\cdots*\tilde y^{\af_n}*e_{\beta\af}\ty^\beta*\tilde y^\af
=e_{(\underline{\beta\af}}C_{\underline{\af_1\cdots\af_n})}\ty^\beta*\tilde y^{\af}*\tilde y^{\af_1}*\cdots*\tilde y^{\af_n}\\
&~~~~ -{n\over n+1}b(n+1,\n k)e^\beta{}_{(\underline{\af}}C_{\beta\underline{\af_1\cdots\af_{n-1}})}\tilde y^{\af}*\tilde y^{\af_1}*\cdots*\tilde y^{\af_{n-1}}\\
&~~~~ -b(n,\n k)e_{(\underline{\beta}}{}^\af C_{\af\underline{\af_1\cdots\af_{n-1}})}\ty^\beta*\ty^{\af_1}*\cdots*\ty^{\af_{n-1}}\\
&~~~~ +b(n,\n k)b(n-1,\n k)e^{\af\beta} C_{\af\beta\af_1\cdots\af_{n-2}}\ty^{\af_1}*\cdots*\ty^{\af_{n-2}}.
\fe
Adding (\ref{star yyl}) and (\ref{star yyr}), we obtain the anticommutator:
\ie
&\{e_{\beta\af}\ty^\beta*\tilde y^\af,C_{\af_1\cdots\af_n}\tilde y^{\af_1}*\cdots*\tilde y^{\af_n}\}_* =2e_{(\underline{\beta\af}}C_{\underline{\af_1\cdots\af_n})}\ty^\beta*\tilde y^{\af}*\tilde y^{\af_1}*\cdots*\tilde y^{\af_n}\\
&~~+f(n,\n k)e^\beta{}_{(\underline{\af}}C_{\beta\underline{\af_1\cdots\af_{n-1}})}\tilde y^{\af}*\tilde y^{\af_1}*\cdots*\tilde y^{\af_{n-1}}+g(n,\n k)e^{\af\beta}C_{\af\beta\af_1\cdots\af_{n-2}}\ty^{\af_1}*\cdots*\ty^{\af_{n-2}},
\fe
where
\ie
f(n,\n k)&=-{n\over n+1}a(n+1,\n k)-a(n,-\n k)-{n\over n+1}b(n+1,\n k)-b(n,\n k),
\\
g(n,\n k)&=a(n,-\n k)a(n-1,\n k)+b(n,\n k)b(n-1,\n k).
\fe
If $n$ is even, $f(n,\n k)$ and $g(n,\n k)$ can be further simplified to
\ie
&f(2j,\n k)=0,
\\
&g(2j,\n k)=4j{(1+2j-\n k)(-1+2j+\n k)\over 1+2j}.
\fe
Subtracting (\ref{star yyl}) from (\ref{star yyr}), we obtain the commutator:
\ie
\left[w_{\beta\af}\ty^\beta*\tilde y^\af,C_{\af_1\cdots\af_n}\tilde y^{\af_1}*\cdots*\tilde y^{\af_n}\right]_*=-4nw^\beta{}_{(\underline{\af}}C_{\beta\underline{\af_1\cdots\af_{n-1}})}\tilde y^{\af}*\tilde y^{\af_1}*\cdots*\tilde y^{\af_{n-1}}.
\fe
The linearized equation (\ref{deq}) for the matter field, therefore, can be written as
\ie\label{mt de eq}
&\partial_\m C^{(1),n}_{mat}{}_{\af_1\cdots\af_n}-4n(w_{0\m})_{(\underline\af_1}{}^{\beta}C^{(1),n}_{mat}{}_{\beta\underline{\af_2\cdots\af_{n}})}-2\psi_1(e_{0\m})_{(\underline{\af_1\af_2}}C^{(1),n-2}_{mat}{}_{\underline{\af_3\cdots\af_{n}})}
\\
&~~~~~~~~-g(n+2,\n k)\psi_1(e_{0\m})^{\af\beta}C^{(1),n+2}_{mat}{}_{\af\beta\af_1\cdots\af_{n}}=0.
\fe
After contracting with $(e_0^\m)_{\A\B}$, this equation is written as
\ie\label{mt de eq 2}
&\nabla_{\A\B} C_{mat}^{(1),n}{}_{\af_1\cdots\af_n}+{1\over 16}\psi_1\epsilon_{(\A(\ul{\A_1}}\epsilon_{\B)\ul{\A_2}} C_{mat}^{(1),n-2}{}_{\underline{\af_3\cdots\af_{n}})}
+{1\over 32}g(n+2,\n k)\psi_1C_{mat}^{(1),n+2}{}_{\A\B\A_1\cdots\A_{n}}=0.
\fe
We follow the same procedure used in analyzing the undeformed vacuum, decomposing the above equation according to the action of permutation group on the indices. Contracting (\ref{mt de eq 2}) with $\epsilon^{\A\A_1}$ gives
\ie\label{mt de eq 3}
\nabla^{\A}{}_{\B} C_{mat}^{(1),n}{}_{\af\A_2\cdots\af_n}-{n+1\over 16n}\psi_1\epsilon_{\B(\ul{\A_2}} C_{mat}^{(1),n-2}{}_{\underline{\af_3\cdots\af_{n}})}=0.
\fe
Further contracting (\ref{mt de eq 3}) with $\epsilon^{\B\A_2}$ gives
\ie\label{mt de eq 4}
\nabla^{\A\B} C_{mat}^{(1),n}{}_{\A\B\af_3\cdots\af_n}+{n+1\over 16(n-1)}\psi_1 C_{mat}^{(1),n-2}{}_{\af_3\cdots\af_{n}}=0.
\fe
As in the analysis of undeformed vacuum, now contracting the indices of the equations (\ref{mt de eq 2}), (\ref{mt de eq 3}), and (\ref{mt de eq 4}) with the $y^{\A}$'s, we obtain
\ie\label{eqn de scalar}
&\nabla^+ C_{mat}^{(1),n}(y)-{1\over 32}g(n+2,\n k)\psi_1C_{mat}^{(1),n+2}(y)=0,
\\
&\nabla^0 C_{mat}^{(1),n}(y)=0,
\\
&\nabla^{-} C_{mat}^{(1),n}{}(y)-{1\over 16}(n+1)n\psi_1 C_{mat}^{(1),n-2}(y)=0,
\fe
where
\ie
C_{mat}^{(1),n}(y) \equiv C_{mat}^{(1),n}{}_{\af_1\cdots\af_n}y^{\A_1}\cdots y^{\A_n}.
\fe
Iterating the first equation of (\ref{eqn de scalar}), we obtain
\ie\label{iterate de}
C_{mat}^{(1),2s}(y)=\left(\prod^s_{j=1}{1\over g(2j,\n k)}\right)(32\psi_1\nabla^+)^s C_{mat}^{(1),0}.
\fe
Since $C^{(1)}_{mat}(y)$ is restricted to be even in $y^\A$, it is entirely determined by the bottom component $C_{mat}^{(1),0}$ via the above relation. After some simple manipulations of (\ref{eqn de scalar}) using (\ref{y-algebra}), we derive the second order form linearized equation
\ie
\Box_{AdS} C_{mat}^{(1),n}&=-{1\over 8}\left(4n+8+{n+1\over n}g(n,\n k)\right)C_{mat}^{(1),n}.
\fe
For $n=0$, the equation is just the usual Klein-Gordon equation on $AdS_3$, and can be rewritten in a more familiar form:
\ie\label{df Klein-Gordon}
\left(\nabla^\mu \partial_\mu - m^2 \right) C^{(1),0}_{mat} = 0,~~~m^2 = -{1\over 4}(3-\n k)(1+\n k).
\fe
Depending on the choice of AdS boundary condition, this scalar field is dual to an operator of dimension 
\ie
\Delta_{\pm}=1\pm{1-\n k\over 2}={1+\n k\over 2}~~\text{or}~~{3-\n k\over 2}.
\fe
It is convenient to package the choice of boundary condition into a variable $\tilde\psi_1$, obeying $\tilde \psi_1^2=1$, so that the scaling dimension of the dual operator can be written as
\ie
\Delta=1+\tilde\psi_1\left({1-\n k\over 2}\right).
\fe
The boundary to bulk propagator for the scalar field is a solution of (\ref{df Klein-Gordon}), which up to normalization is given by
\ie
C^{(1),0}_{mat}=K^\Delta,~~~~\text{where}~~~~K={z\over \vec x^2 + z^2}.
\fe
Here $(\vec x,z)$ are Poincar\'e coordinates of the $AdS_3$ (not to be confused with the twistor variable $z_\A$).
Using (\ref{nsK}) and (\ref{iterate de}), we obtain
\ie
C_{mat}^{(1)}(y)&=\sum^{\infty}_{s=0}C_{mat}^{(1),2s}(y)
\\
&=\sum^{\infty}_{s=0}\left(\prod^s_{j=1}{\Delta +j-1\over g(2j,\n k)}\right)(4\psi_1)^s(y\Sigma y)^s K^\Delta
\\
&=\sum^{\infty}_{s=0}\left(\prod^s_{j=1}{(\Delta +j-1)(1+2j)\over j(1+2j-\n k)(-1+2j+\n k)}\right) \psi_1^s(y\Sigma y)^s K^\Delta
\\
&={}_1F_1\left({3\over 2},1-\tilde\psi_1\left({1-\n k\over 2}\right),{1\over 2}\psi_1 y\Sigma y\right)K^{1+\tilde\psi_1\left({1-\n k\over 2}\right)}.
\fe
In the actual master field, the above expression should be understood as a star-function, with $y$ replaced by $\tilde y$. More concretely, we can transform the ordinary function $C_{mat}^{(1)}(y)$ to the star-function $C_{mat*}^{(1)}(\ty)$ via the formula
\ie
C_{mat*}^{(1)}(\ty)={1\over (2\pi)^2}\int d^2 y d^2 u \,C_{mat}^{(1)}(y) e^{i uy}\exp_*(-iu\ty).
\fe

\end{document}